\documentclass[%
 aip,
 amsmath,amssymb,
preprint,%
]{revtex4-1}

\usepackage{graphicx}
\usepackage{dcolumn}
\usepackage{bm}
\usepackage[mathlines]{lineno}
\usepackage[normalem]{ulem}

\usepackage[utf8]{inputenc}
\usepackage[T1]{fontenc}
\usepackage{mathptmx}
\usepackage{etoolbox}
\usepackage{xcolor}
\graphicspath{{./}}
\usepackage{comment}
\usepackage{subfig}
\usepackage{ulem}

\newcommand{\subfigthreeB}{0.23}

\newcommand{\Df}{Df}

\newcommand{\NS}{NS}
\newcommand{\WS}{WS}

\newcommand{\ES}{ES}

\newcommand{\ida}[1]{{\color{black} #1}}
\newcommand{\idaa}[1]{{\color{black} #1}}
\definecolor{trackgreen}{RGB}{0,120,60}

\makeatletter
\def\@email#1#2{%
 \endgroup
 \patchcmd{\titleblock@produce}
  {\frontmatter@RRAPformat}
  {\frontmatter@RRAPformat{\produce@RRAP{*#1\href{mailto:#2}{#2}}}\frontmatter@RRAPformat}
  {}{}
}%
\makeatother
\begin{document}

\preprint{AIP/123-QED}

\title[]{Mixing dynamics and transport mechanisms during laminar stirring flows}
\author{Mohammad Reza Daneshvar Garmroodi}
\author{Ida Karimfazli$^*$}%
 \email{Ida.karimfazli@concordia.ca}
\affiliation{ 
Department of Mechanical, Industrial and Aerospace Engineering, Concordia University, 1515 St. Catherine W., Montreal, QC H3G 2W1, Canada
}%

\date{\today}

\begin{abstract}
\idaa{We investigate laminar mixing of a passive dye in an infinite, two-dimensional domain filled with a Newtonian fluid by simulating a cylindrical stirrer that rotates at constant speed along a circular path, stirring an initially quiescent fluid. The fluid is marked by a passive dye in the lower half of the domain, enabling a systematic analysis of dye-interface evolution and mixing dynamics.} \ida{By systematically varying the stirring Reynolds number within the laminar regime, we identify how transitions in flow topology govern the evolution of mixing and distinguish three mixing regimes spanning diffusion-dominated and advective mixing.} In the diffusion-dominated regime, mixing is characterized by the formation of a well-mixed central region and the development of a spiral dye pattern that evolves in an approximately self-similar manner. Advective mixing is marked by substantial deformation of the dye interface beyond the central region.

We provide a mechanistic interpretation of advective mixing by relating mixing events to specific flow features: (i) direct interaction between the stirrer and the dye interface, and (ii) vortex shedding near the stirrer and away from the central region. Enhanced mixing occurs when vortical structures are able to escape the central region and transport scalar gradients nonlocally. When vortical activity remains confined near the stirrer’s path, \ida{mixing per stirrer period exhibits only weak dependence on stirring speed despite increasing stirring intensity. Significant enhancement of mixing per period occurs only when escaping vortices introduce a new transport mechanism that carries scalar gradients far beyond the stirrer's path.} 

\idaa{Overall, we present a mechanistic link between flow topology and mixing by identifying the specific transport mechanisms through which flow features govern scalar evolution, thereby providing a bridge between the kinematics of mixing and the underlying fluid dynamics.}

\end{abstract}

\maketitle

\section{Introduction}

Mixing is the process by which an initially heterogeneous mixture of fluids or additives evolves toward a homogeneous state. Mixing is ubiquitous in both natural and industrial environments \cite{paul2004handbook}. Examples range from everyday processes such as mixing milk and coffee, food preparation \cite{cullen2009food}, and household cleaning, to industrial applications including oil transportation \cite{hermoso2012high}, pharmaceutical production \cite{yu2018mixing}, and wastewater treatment \cite{singh2019state}.

During a mixing process, there is often a discrepancy between the rheological properties of the initial fluid and those of the final product, for example, when an additive such as flour or powder is added to a base fluid like water. The process is complicated and difficult to model due to the wide variety of initial conditions, the properties of the initial products, and the changes in the mixture properties throughout the process, along with the extensive choices of stirrer and reactor shape, size, and stirring protocol. Given the vast range of parameters and the complexity of the problem, most studies have focused on Newtonian fluids, and mixing remains an active research area (see \cite{spencer1951mixing, ottino1990mixing,warhaft2000passive,peltier2003mixing,wunsch2004vertical,villermaux2019mixing,caulfield2021layering} and references therein). 

\ida{In the present study, we restrict our attention to laminar mixing of Newtonian fluids.} Within this reduced scope extensive simplifications have been made to arrive at tractable model problems that can be studied numerically or experimentally. In the simplest approach, mixing is probed by introducing a passive dye into a fluid. The dye is assumed not to influence the fluid properties in any way. Mixing then becomes the homogenization of a passive dye in a fluid with uniform and constant properties, referred to here as \textit{Level-1} (see, e.g., \cite{bacsbuug2018reduced, el2021active}). At the next level of complexity, the dye concentration modifies the physical and rheological properties of the base fluid (\textit{Level-2}), thereby altering the flow dynamics during mixing (see, e.g., \cite{garmroodi2024mixing, mirfasihi2025numerical, derksen2013simulations}). A further step involves chemical reactions occurring during mixing, introducing additional complexity (\textit{Level-3}) (see, e.g., \cite{guo2023structure, moon2021determination}).

In its most simplified form, a prescribed and often periodic flow field is imposed to promote mixing (see, e.g., \cite{gubanov2010towards}) (\textit{Level-1-A}). This approach allows the study of mixing as a geometric problem \cite{ottino1992chaos} and facilitates mixing analysis using dynamical systems-based approaches, including characterization of the time evolution of the dye pattern \cite{meunier2003how}, optimization of mixing \cite{gubanov2010towards, lin2011optimal, gubanov2012cost}, and the proposal of new measures to quantify mixing \cite{mathew2005multiscale, thiffeault2012using}. \idaa{These approaches have provided a powerful kinematic description of mixing through particle trajectories, stretching and folding, and the emergence of chaotic advection.}

\ida{Rather than prescribing the flow field directly, a second class of studies generates mixing through the prescribed motion of one or more stirrers (Level-1-B). This class of problems occupies an intermediate position between prescribed-flow studies and realistic mixer simulations}. \idaa{Unlike prescribed-flow approaches, the flow field is generated by the interaction between the moving stirrer and the fluid, providing an opportunity to relate scalar transport not only to the resulting kinematics, but also to the fluid-dynamical processes through which the flow develops. At the same time, the simplified geometry enables the underlying transport mechanisms to be isolated without the additional complexity associated with practical impeller designs.} For example, Celik et al.\ \cite{celik2009mixing} numerically examined laminar channel mixing in which an oscillating cylinder enhances downstream transport of a passive scalar, highlighting the role of confinement and wall-induced modifications of the wake and vortex dynamics. Eggl \textit{et al.}\ \cite{eggl2020mixing} optimized laminar mixing in a circular container driven by two cylindrical stirrers traversing concentric circular paths, demonstrating that vortex interactions play a central role in mixing efficiency within an optimization framework. Related studies have further shown that boundaries and no-slip walls can strongly influence mixing by constraining flow dynamics and limiting global scalar transport \cite{gouillart2007walls, gouillart2008slow, shi2024mutual, celik2008flow, eggl2022mixing}.

A third subset of studies is dedicated to simulating the flow and mixing of a passive dye in more realistic reactor geometries and with stirrer shapes resembling impeller geometries (\textit{Level-1-C}). Here, the analyses focus on the effect of stirring speed \cite{zalc2002using, alvarez2002mechanisms}, stirrer blade design \cite{ameur2020newly, ameur2015energy}, stirrer shape \cite{bacsbuug2018reduced}, and rheological properties of the fluid \cite{noble2023blending, russell2019mixing}.

In summary, on the one hand, there exists an extensive body of work on the emergence of dye patterns from prototypical flows and on the optimization of such flows for rapid mixing. On the other hand, numerical simulations of near-real reactor and impeller designs have been conducted to explore the influence of stirring speed or impeller design on flow features and passive scalar transport. \idaa{Between these perspectives, comparatively less attention has been given to systematically connecting specific Eulerian flow features and events to the transport mechanisms responsible for the observed evolution of the scalar field. Establishing such a connection provides a bridge between the kinematic description of mixing and the underlying fluid dynamics, since the emergence and evolution of these flow structures can in turn be related to the forces, stresses, and momentum transport that generate the flow}. 

\idaa{In our recent study on yield-stress fluids \cite{garmroodi2025yield}, we introduced an Eulerian framework aimed at establishing this connection by relating scalar evolution to identifiable features of the evolving flow field. Here, we apply this perspective to the Newtonian counterpart using the same stirring configuration, where the simpler rheology allows the underlying transport mechanisms to be isolated more clearly}. We consider stirring a quiescent Newtonian fluid with a cylinder moving at constant speed along a circular path in an \idaa{infinite} two-dimensional domain. \ida{This simplified configuration serves as an archetypal model of laminar stirring. Although practical impellers differ widely in shape and operate within confined vessels, they often generate mixing through the periodic motion of solid bodies relative to the surrounding fluid. The present configuration isolates the fundamental fluid-mechanical processes underlying laminar stirring by considering the simplest idealized representation of a rotating impeller. In doing so, it avoids the additional complexity introduced by specific impeller geometries, stirring protocols, and reactor designs}, \idaa{allowing the relationship between flow evolution, transport mechanisms, and scalar mixing to be examined directly.} By exploring the full range of laminar stirring speeds, \idaa{we aim to identify the transport mechanisms responsible for changes in scalar evolution and mixing rate, and to relate the emergence of these mechanisms to specific features and events in the evolving Eulerian flow field.}

The remainder of this paper is organized as follows. In Section~\ref{sec:III_problem}, we present the model problem, governing equations, and numerical methods. Section~\ref{sec:III_results} presents the flow dynamics and mixing behavior across different stirring speeds, leading to the identification of distinct flow regimes. Section~\ref{sec:III_results} also examines mixing from an energetic perspective. Finally, Section~\ref{sec:III_summary} summarizes the results.

\section{Problem setup}
\label{sec:III_problem}

\subsection{Model problem}

\begin{figure}
	\centering
	\subfloat{
		\includegraphics[trim=26cm 17cm 27cm 8cm, clip=true,height=.35\textwidth]{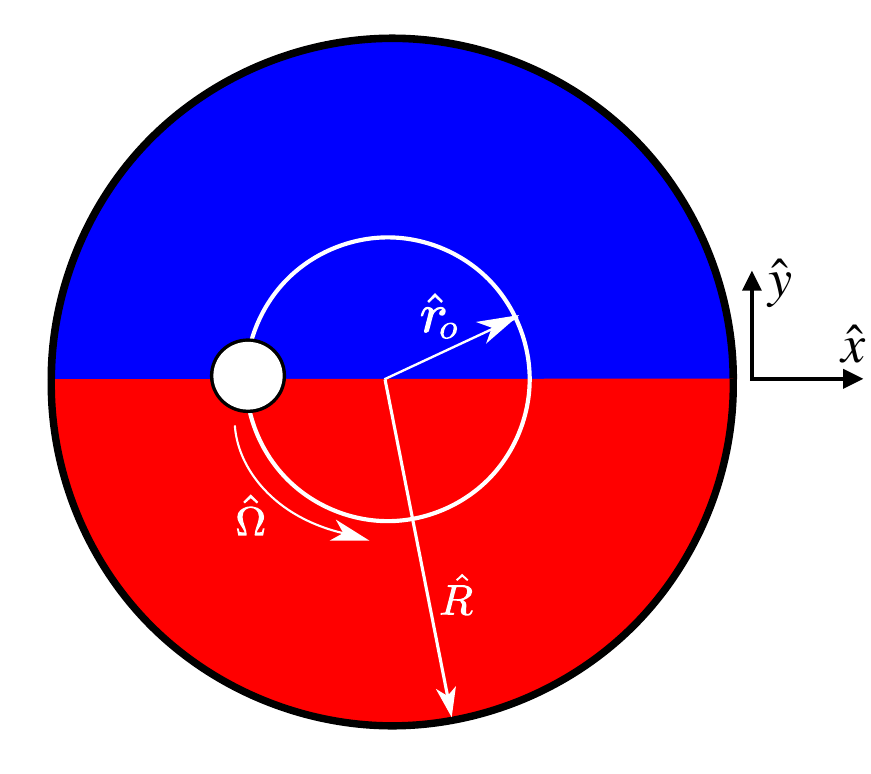}
	}
	\caption{Schematic of the domain geometry and the initial conditions. The white solid line represents the stirrer's path. The red and blue colors indicate the dyed and dye-free regions. Note that the figure is not to scale. }
	\label{fig:geometry}
\end{figure}

We consider an archetypal stirring strategy \ida{in which an idealized circular impeller, represented by a cylindrical stirrer of diameter $\hat{d}_s$, rotates about the center of the domain with constant angular velocity $\hat{\Omega}$. The velocity prescribed on the stirrer boundary is $\hat{\boldsymbol{\Omega}}\times \hat{\boldsymbol{r}}$ where $\hat{\boldsymbol{r}}$ is the position vector measured from the center of the domain.  Consequently, the center of the stirrer follows a circular path of radius \idaa{$\hat{r}_o$}, shown by the white solid line in Figure~\ref{fig:geometry}. {\color{black}The geometric ratio is defined as $c = \hat{r}_o/\hat{d}_s$ and is fixed at $c = 2$ for all simulations.} Throughout the manuscript, this cylindrical impeller is referred to simply as the stirrer. The present geometry is deliberately chosen as the simplest representation of a rotating impeller capable of generating laminar stirring, allowing the influence of fluid inertia on transport and mixing to be examined independently of blade-specific geometric effects.}

The quantities with and without $\hat{.}$ refer to dimensional and non-dimensional parameters, respectively. To conduct numerical simulations of mixing in an infinite domain, we consider a circular domain with radius $\hat{R}\gg \hat{r}_o$. \idaa{In all simulations reported here, the outer radius is set to $\hat{R}=34\,\hat{r}_o$, and the results are analyzed only over subdomains and time intervals for which wall effects are negligible.} \idaa{This allows the present work to isolate the effect of the stirrer on the resulting flow and mixing dynamics. $\hat{R}_{sd}$ indicates the radius of the subdomain, over which the results are presented.} The no-slip boundary condition is applied on the vessel and stirrer's walls.

The material initially contained in the bottom half of the domain is marked by a passive dye, described by a concentration field, $\alpha$ such that at $t=0$, 
\begin{align*}
\left\{
\begin{array}{lr}
	\alpha = 0 & \text{at}~ y\ge 0 \\
	\alpha=1 & \text{at}~ y<0
\end{array}
\right.	
\end{align*} 

\ida{The dye is treated as a passive scalar and therefore influences neither the fluid properties nor the flow field.}

{\color{black}The fluid is initially at rest. At $t=0$, the prescribed stirrer motion is imposed and maintained thereafter. The results reported below therefore include the transient flow development that follows the onset of stirring.}

The dimensionless Cauchy's laws of motion, continuity and advection-diffusion are,

\begin{equation}
\label{eq:Cauchy}
\begin{aligned}
&\frac{\partial \boldsymbol{u}}{\partial {t}} + \ \boldsymbol{u} \cdot \boldsymbol{\nabla} \boldsymbol{u} + \  \boldsymbol{\nabla} P = \idaa{\frac{1}{\, c~Re}} \boldsymbol{\nabla} \cdot \boldsymbol{\tau} \\
&\boldsymbol{\nabla} \cdot \boldsymbol{u} = 0 \\
&\frac{\partial \alpha}{\partial t} + \ \boldsymbol{\nabla} \cdot (\boldsymbol{u} \alpha) = \idaa{\frac{1}{\, c~Pe}} \nabla^2 \alpha
\end{aligned}
\end{equation}

Here $\boldsymbol{u}$, $P$ and $\boldsymbol{{\tau}}$ are the non-dimensional velocity, pressure and  deviatoric stress tensor, respectively.  $\hat{r}_o$,  $\hat{\Omega}^{-1}$, $\hat{r}_o \hat{\Omega}$, $\hat{\mu} \hat{\Omega}$ and $\hat{\rho}\hat{r}_o^2\hat{\Omega}^2$ are used as the scales for length, time, velocity, shear stress and pressure. {\color{black}We present the results in terms of the stirrer's period, $\hat{T}_{stirrer}= 2\pi/\hat{\Omega}$,
	\begin{align}
	T = \frac{\hat{t}}{\hat{T}_{stirrer}} = \frac{t}{2\pi}
	\end{align}
}

{\color{black}$Re$ and $Pe$ are the Reynolds and Peclet numbers,
	
	\begin{align}
	& Re=\frac{\hat{\rho} \hat{\Omega} \hat{r}_o \hat{d}_s}{\hat{\mu}}
	\\
	& Pe=\displaystyle \frac{\hat{\Omega} \hat{r}_{o} \hat{d}_s}{ \hat{D}_{m}} 
	\label{eq:Peh}
	\end{align}

	\noindent where, $\hat{D}_m$ is the diffusion coefficient. We maintain a constant Peclet number throughout the study, $Pe=500$, to isolate the effect of Reynolds number on the mixing topology. \ida{Sensitivity tests at higher $Pe$ suggest that molecular diffusion remains dominant over numerical diffusion at the adopted value of $Pe=500$.} This value is sufficiently large to preserve advection-driven scalar structures, while allowing molecular diffusion to be adequately resolved at the adopted spatial resolution. It also permits measurable homogenization over the simulated time interval.} 
	
	\subsection{Quantifying mixing}
	
	To characterize the rate of mixing, we define a normalized variance of the dye concentration,
	\begin{equation}
	\label{eq:variance}
	\sigma^{2}_{R_{sd}} = \frac{1}{A_{R_{sd}}} \int_{A_{R_{sd}}} \left(1 - \frac{\alpha}{\bar{\alpha}}\right)^{2} dA
	\end{equation}
	
	Here, $R_{sd}$ indicates the radius of the subdomain over which the variance is measured, $r\le R_{sd}$. The normalized variance, $\sigma^{2}_{r}$, is defined over a subdomain of finite size, $\Omega_r=\pi r^2$, to avoid division by infinity and integration over an \idaa{infinite domain}. For given values of the governing dimensionless parameters, $r_{sd}$ is chosen to ensure the subdomain captures the region where dye concentration is affected by the stirring (within the timeframe of interest). $\bar{\alpha}$ is the average concentration over the subdomain.
	
	{\color{black}To quantify the kinetic energy, $KE$ is defined as,
		\begin{equation}
		\label{eq:velNorm}
		KE
		= \sqrt{ 
			\int_{A} |\boldsymbol{u}|^{2} dA}
		\end{equation}
		where $KE$ is the kinetic energy, $|\boldsymbol{u}|$ is the speed, and $A$ represents the flow domain.
	}


	\subsection{\ida{Numerical implementation}}
	\label{sec:numerical}

	\ida{Numerical simulations were performed using OpenFOAM \cite{OpenFOAM} version 7.} The \textit{twoLiquidMixingFoam} solver, which employs the PIMPLE or PISO-SIMPLE algorithm to decouple pressure and velocity in the governing equations, was utilized. For temporal discretization, we applied the second-order Crank-Nicolson scheme, while spatial discretization was second-order as well. Adaptive time stepping was implemented based on a constant Courant-Friedrichs-Lewy number, set to 0.05. \ida{The numerical methodology, including the implementation of the governing equations, moving boundary treatment, and passive scalar transport, is identical to that validated by Garmroodi \textit{et al.} \cite{garmroodi2024mixing}. Therefore, only the mesh-independence assessment for the present simulations is summarized here.}

\begin{figure}
		\centering
		\subfloat[\label{fig:gridVariance}]{
			\includegraphics[trim=0cm 0cm 0cm 0cm, clip=true,height=.30\textwidth]{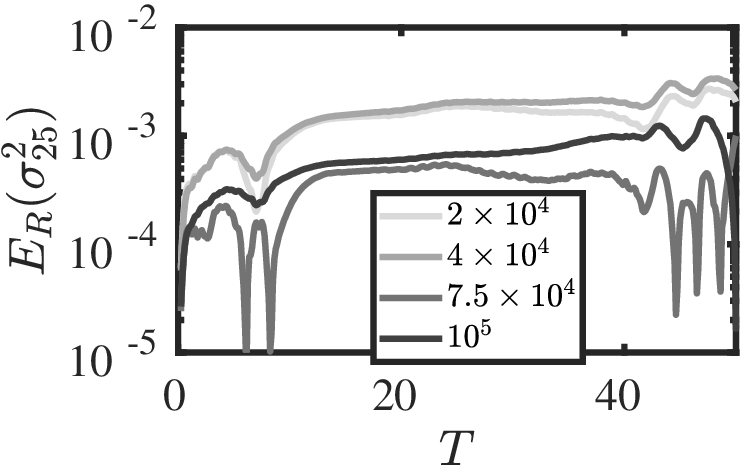}
		}
		\subfloat[\label{fig:gridKE}]{
			\includegraphics[trim=0cm 0cm 0cm 0cm, clip=true,height=.30\textwidth]{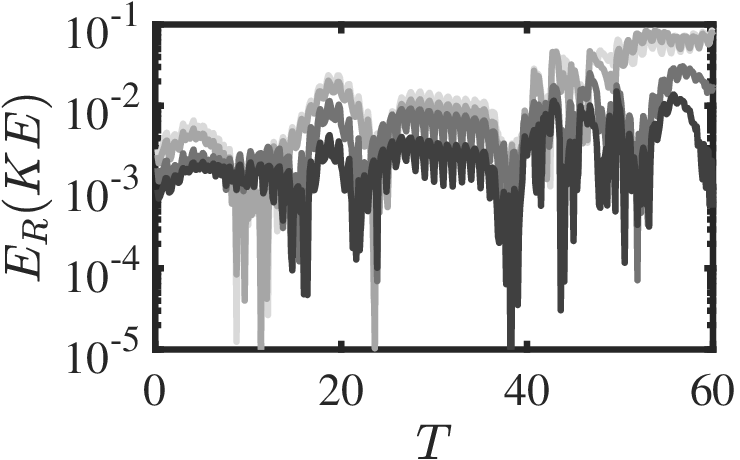}
		}

		\caption{Estimated numerical error in normalized variance of dye concentration, $\sigma^2_{25}$, and {\color{black}kinetic} energy, $KE$, (a) relative error, $\displaystyle E_R (\sigma^2_{25})= \left|\frac{\sigma^2_{25}-\sigma^2_{25, 1.5\times10^5}}{\sigma^2_{25, 1.5\times10^5}}\right|$ and (b) relative error, $\displaystyle E_R (KE)= \left|\frac{KE-KE_{1.5\times10^5}}{KE_{1.5\times10^5}}\right|$. $Re=50$.}
		\label{fig:grid}

	\end{figure}

	To verify grid independence, five different grid sizes were tested. \ida{Figures~\ref{fig:gridVariance} and \ref{fig:gridKE} present the relative errors in kinetic energy and normalized variance for the different mesh sizes at $Re=50$. The finest mesh ($1.5\times10^5$ cells) was taken as the reference solution for estimating the discretization error.}
	
	For the remainder of the simulations, \ida{a mesh containing $10^5$ cells was adopted, resulting} in a relative error of less than 1\% for both the normalized variance and kinetic energy. Further details on the benchmarking and validation of the numerical solver can be found in Garmroodi \textit{et al.}\cite{garmroodi2024mixing}. \ida{Additional sensitivity tests at higher $Pe$ suggest that molecular diffusion remains dominant over numerical diffusion at the adopted value of $Pe=500$. Together, these analyses indicate that the chosen spatial resolution is sufficient for the physical analyses presented in this study.}

\section{Results and discussion}
\label{sec:III_results}
	
	\subsection{\ida{Evolution of mixing mechanisms}}
	\label{sec:mixingEv}
	
	\subsubsection{Diffusive mixing}
	
\begin{figure}
		\centering
		\subfloat[$R_{sd}=4$\label{fig:alpha_Re=0.05_A}]{
			\includegraphics[trim=2cm 0.5cm 6cm 1.25cm, clip=true,height=\subfigthreeB\textwidth]{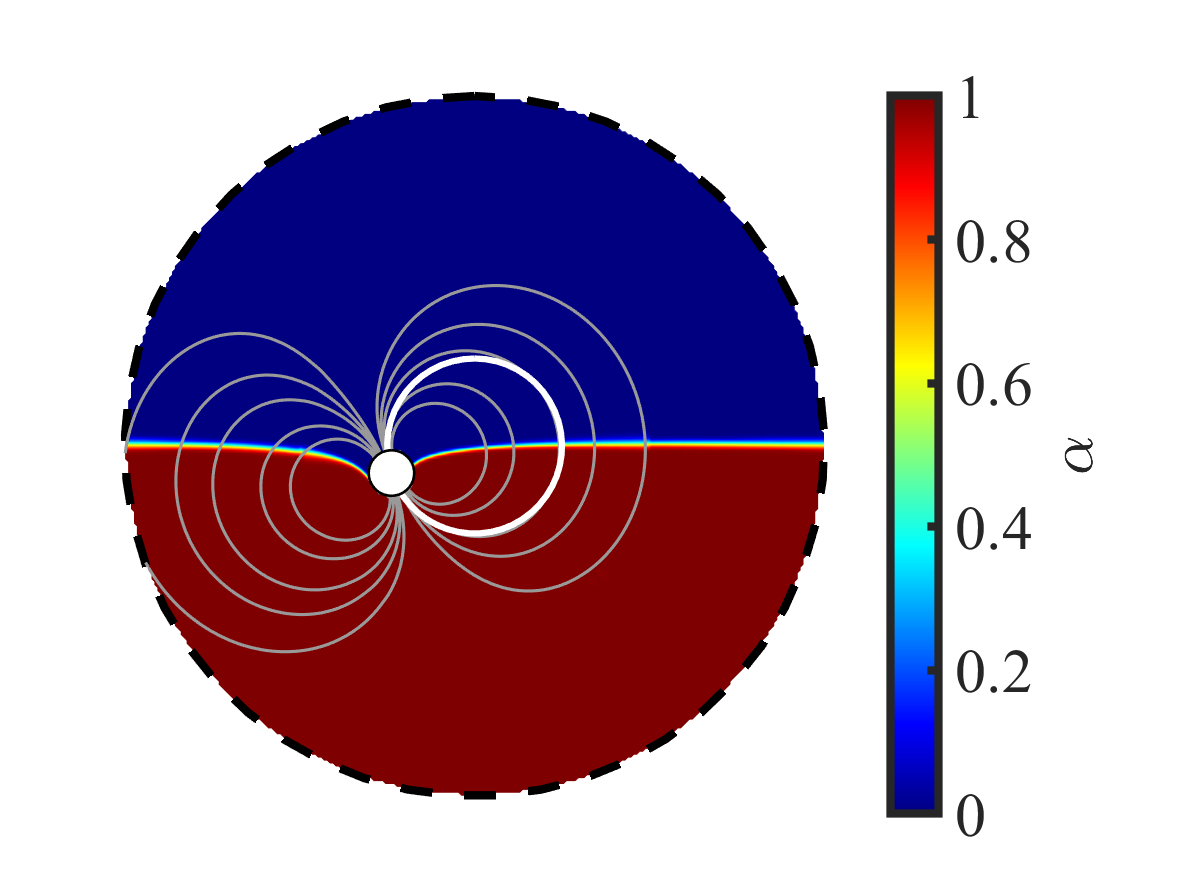}
		}~
		\hspace{-0.3cm}
		\subfloat[$R_{sd}=4$\label{fig:alpha_Re=0.05_B}]{
			\includegraphics[trim=2cm 0.5cm 6cm 1.25cm, clip=true,height=\subfigthreeB\textwidth]{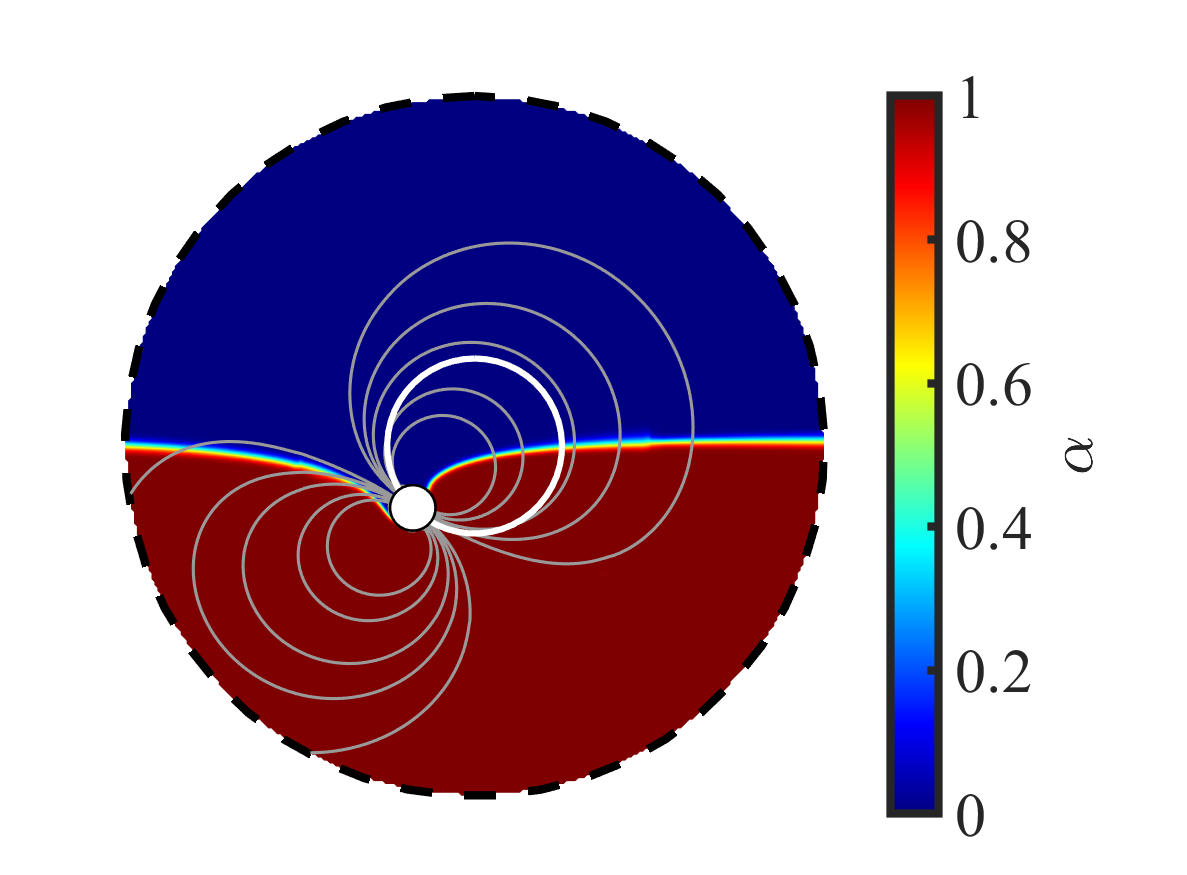}
		}~
		\hspace{-0.3cm}
		\subfloat[$R_{sd}=4$\label{fig:alpha_Re=0.05_C}]{
			\includegraphics[trim=2cm 0.5cm 6cm 1.25cm, clip=true,height=\subfigthreeB\textwidth]{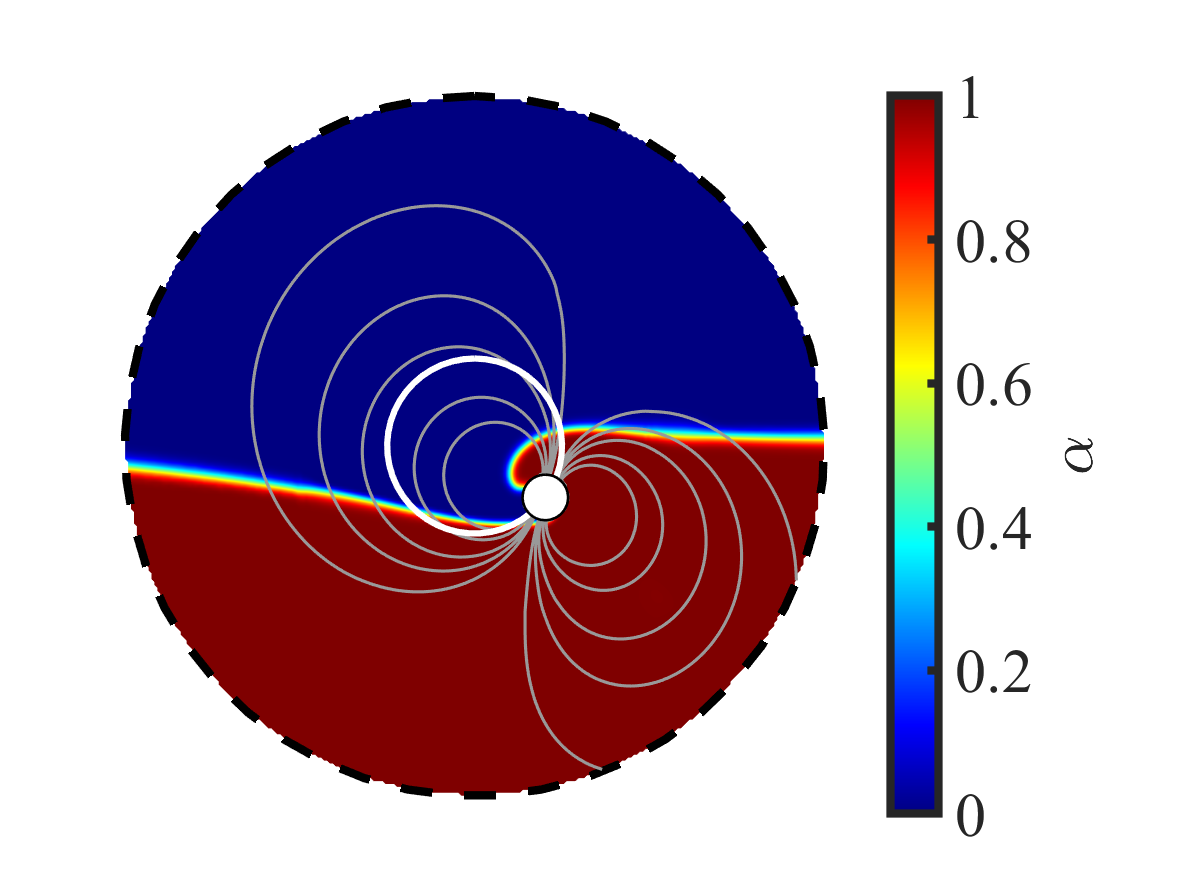}
		}~
		\hspace{-0.3cm}
		\subfloat[$R_{sd}=4$\label{fig:alpha_Re=0.05_D}]{
			\includegraphics[trim=2cm 0.5cm 1.5cm 1.25cm, clip=true,height=\subfigthreeB\textwidth]{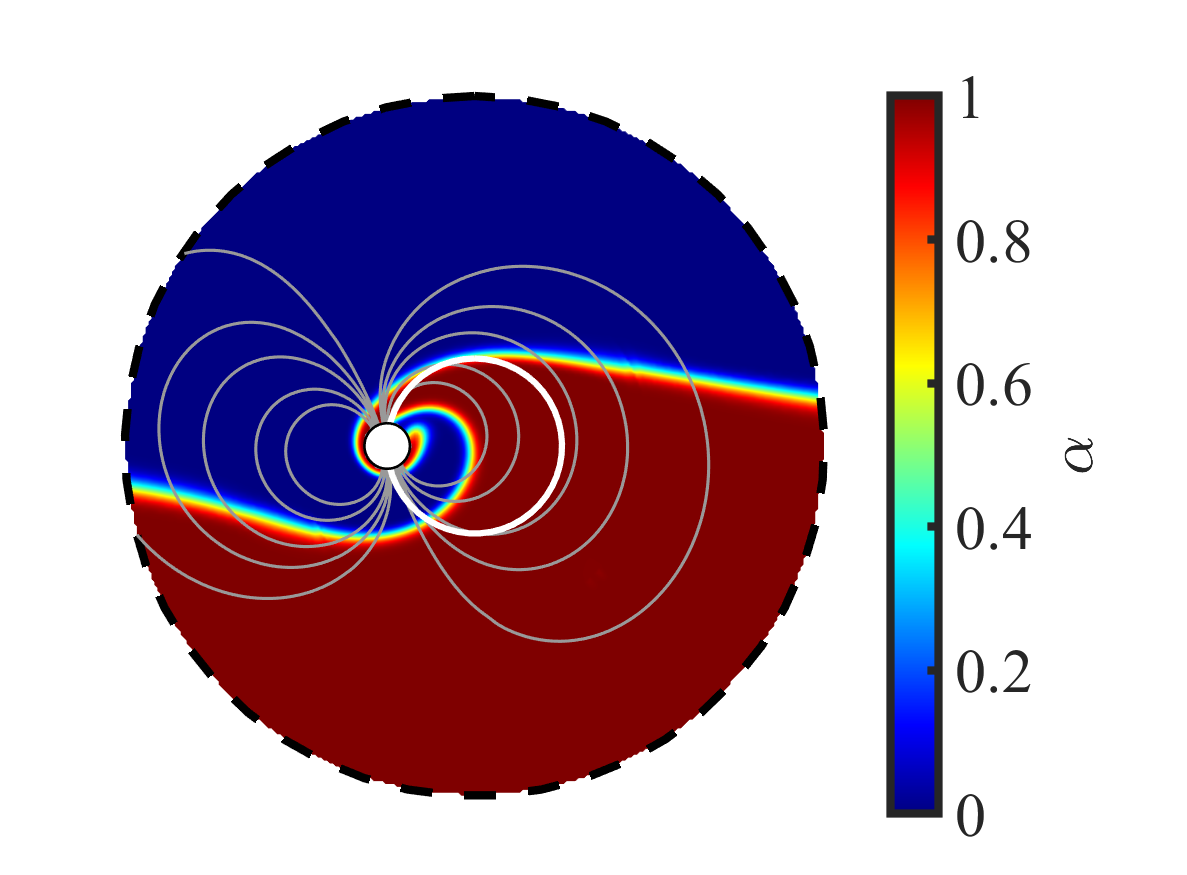}
		}\\
		\vspace{-0.3cm}
		\subfloat[$R_{sd}=4$\label{fig:alpha_Re=0.05_E}]{
			\includegraphics[trim=2cm 0.5cm 6cm 1.25cm, clip=true,height=\subfigthreeB\textwidth]{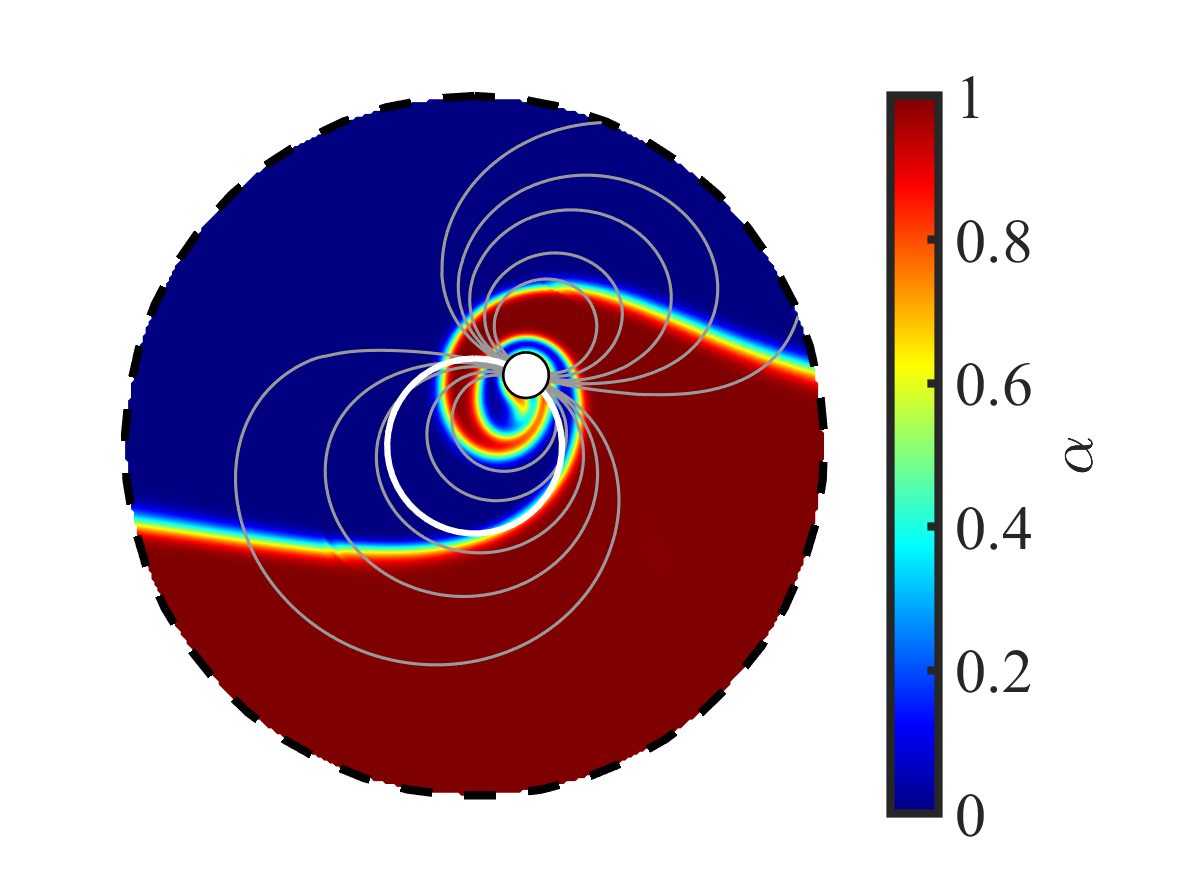}
		}~
		\hspace{-0.3cm}
		\subfloat[$R_{sd}=4$\label{fig:alpha_Re=0.05_F}]{
			\includegraphics[trim=2cm 0.5cm 6cm 1.25cm, clip=true,height=\subfigthreeB\textwidth]{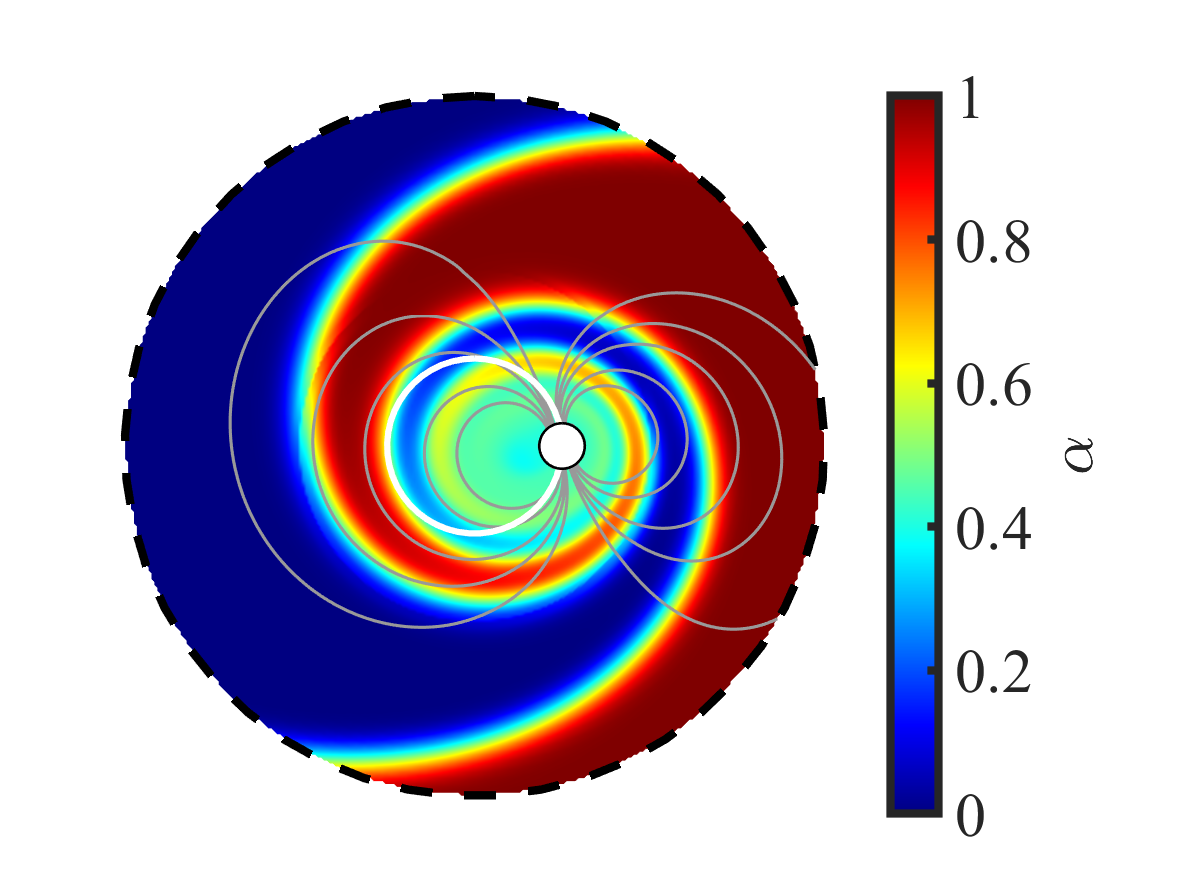}
		}~
		\hspace{-0.3cm}
		\subfloat[$R_{sd}=10$\label{fig:alpha_Re=0.05_G}]{
			\includegraphics[trim=2cm 0.5cm 6cm 1.25cm, clip=true,height=\subfigthreeB\textwidth]{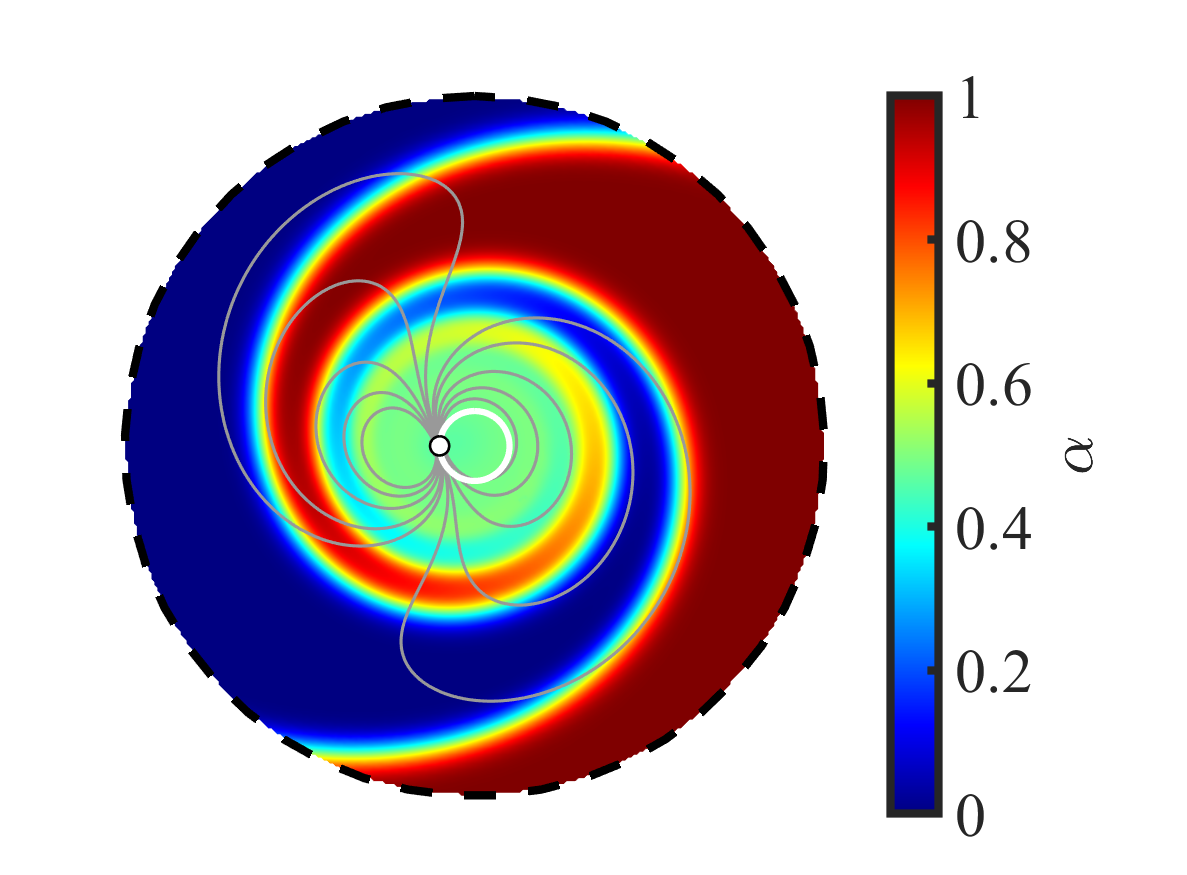}
		}~
		\hspace{-0.3cm}
		\subfloat[$R_{sd}=20$\label{fig:alpha_Re=0.05_H}]{
			\includegraphics[trim=2cm 0.5cm 1.5cm 1.25cm, clip=true,height=\subfigthreeB\textwidth]{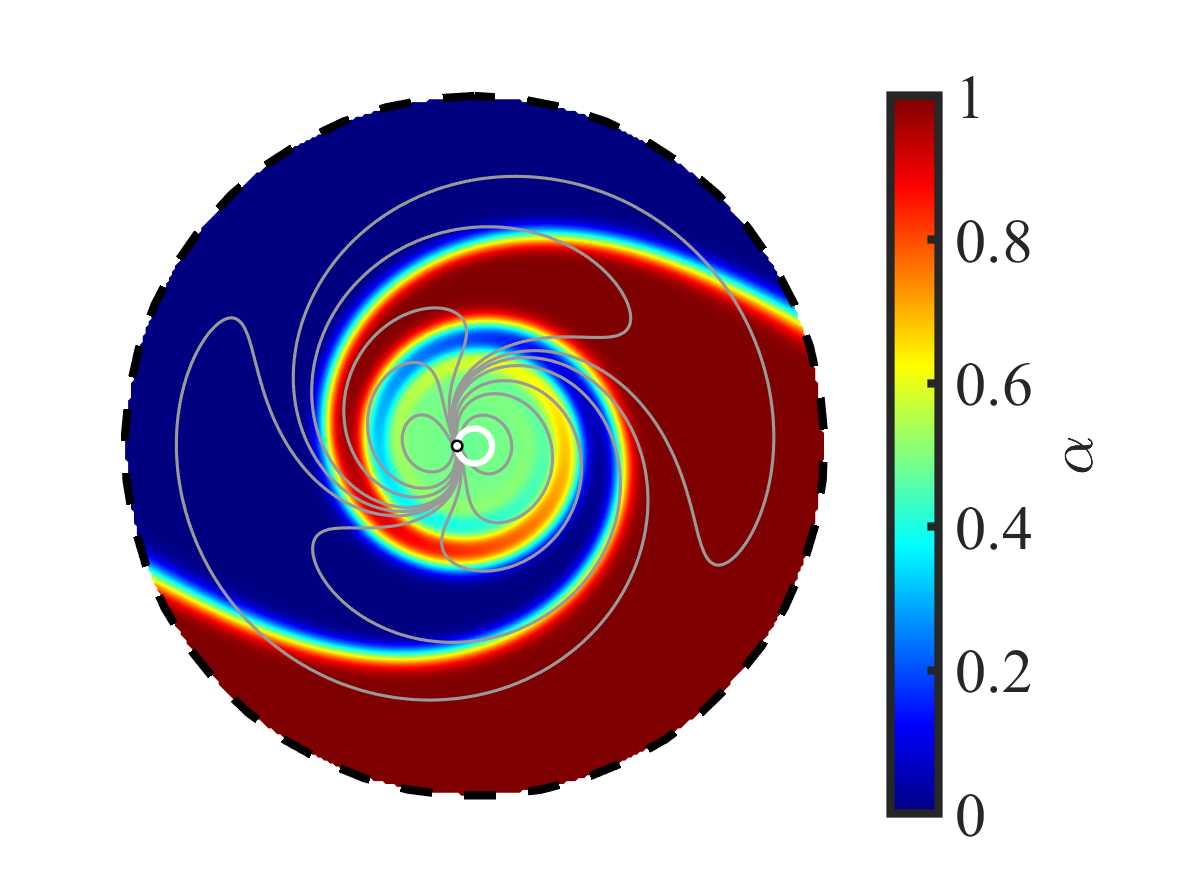}
		}
		\\
		\subfloat[$R_{sd}=5$\label{fig:vorticity_Re=0.05_A}]{
			\includegraphics[trim=2cm 0.5cm 6cm 1.25cm, clip=true,height=\subfigthreeB\textwidth]{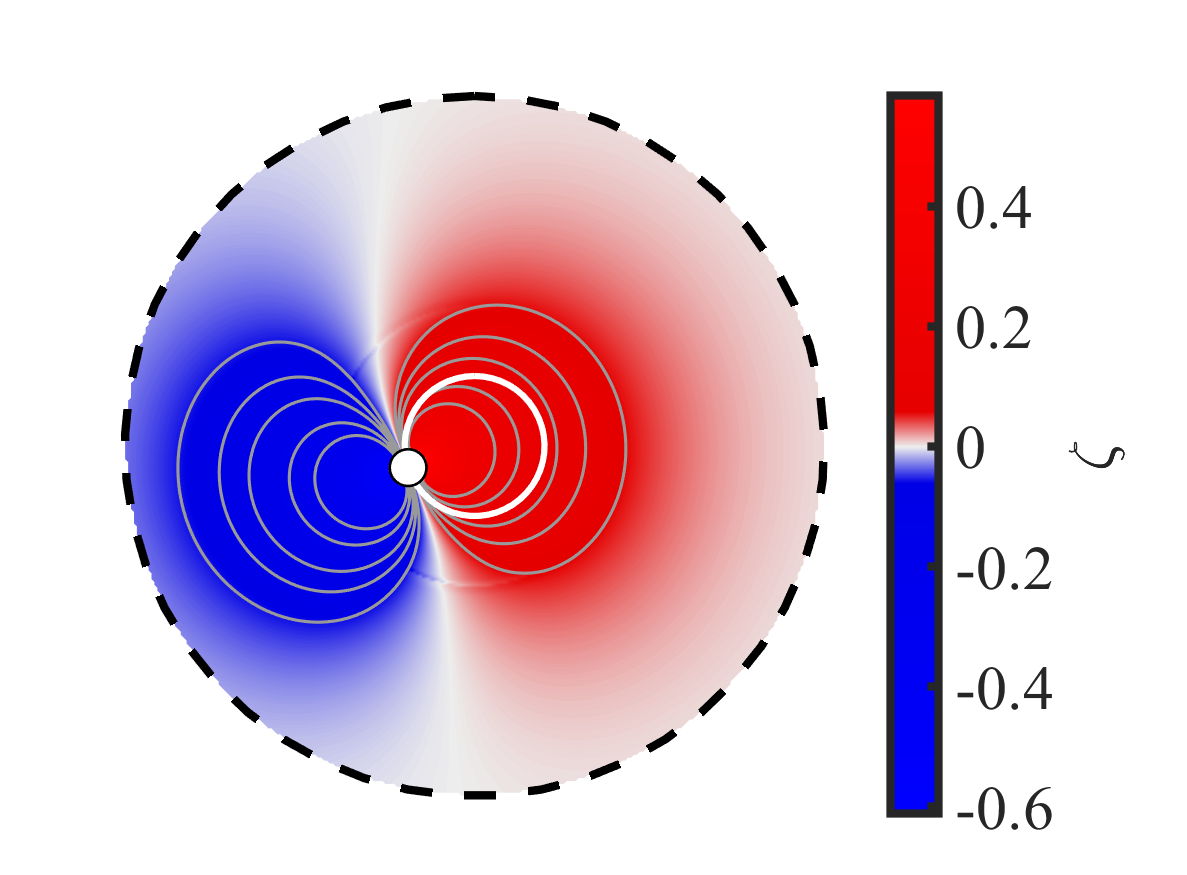}
		}~
		\hspace{-0.3cm}
		\subfloat[$R_{sd}=5$\label{fig:vorticity_Re=0.05_B}]{
			\includegraphics[trim=2cm 0.5cm 6cm 1.25cm, clip=true,height=\subfigthreeB\textwidth]{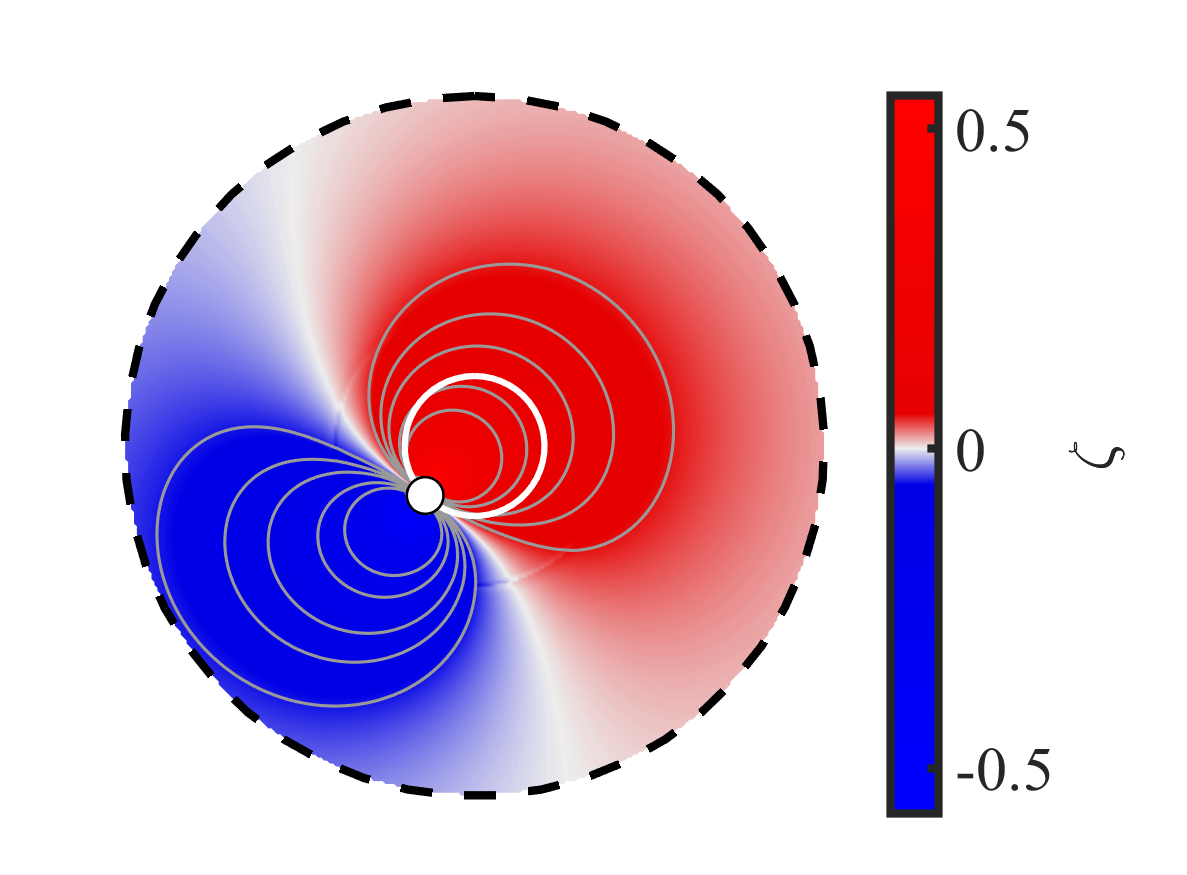}
		}~
		\hspace{-0.3cm}
		\subfloat[$R_{sd}=10$\label{fig:vorticity_Re=0.05_C}]{
			\includegraphics[trim=2cm 0.5cm 6cm 1.25cm, clip=true,height=\subfigthreeB\textwidth]{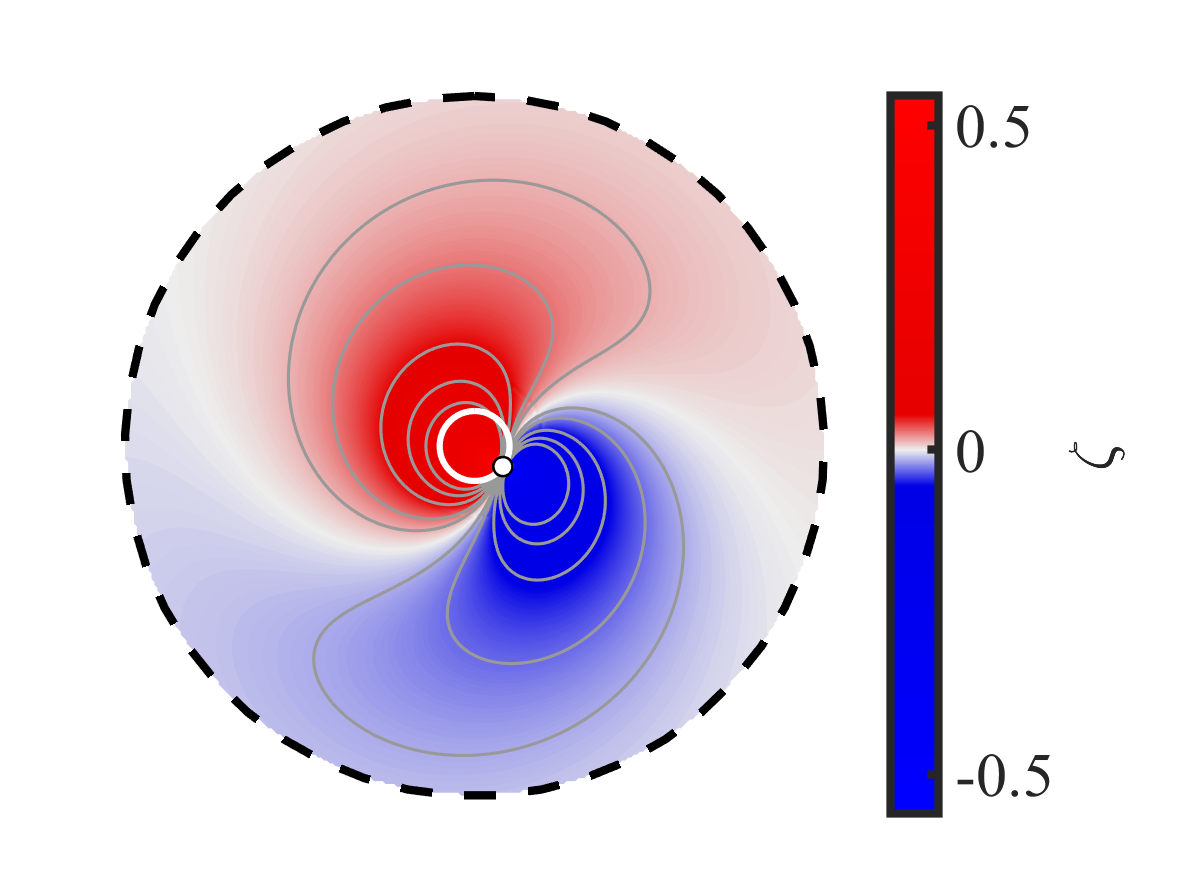}
		}~
		\hspace{-0.3cm}
		\subfloat[$R_{sd}=10$\label{fig:vorticity_Re=0.05_D}]{
			\includegraphics[trim=2cm 0.5cm 6cm 1.25cm, clip=true,height=\subfigthreeB\textwidth]{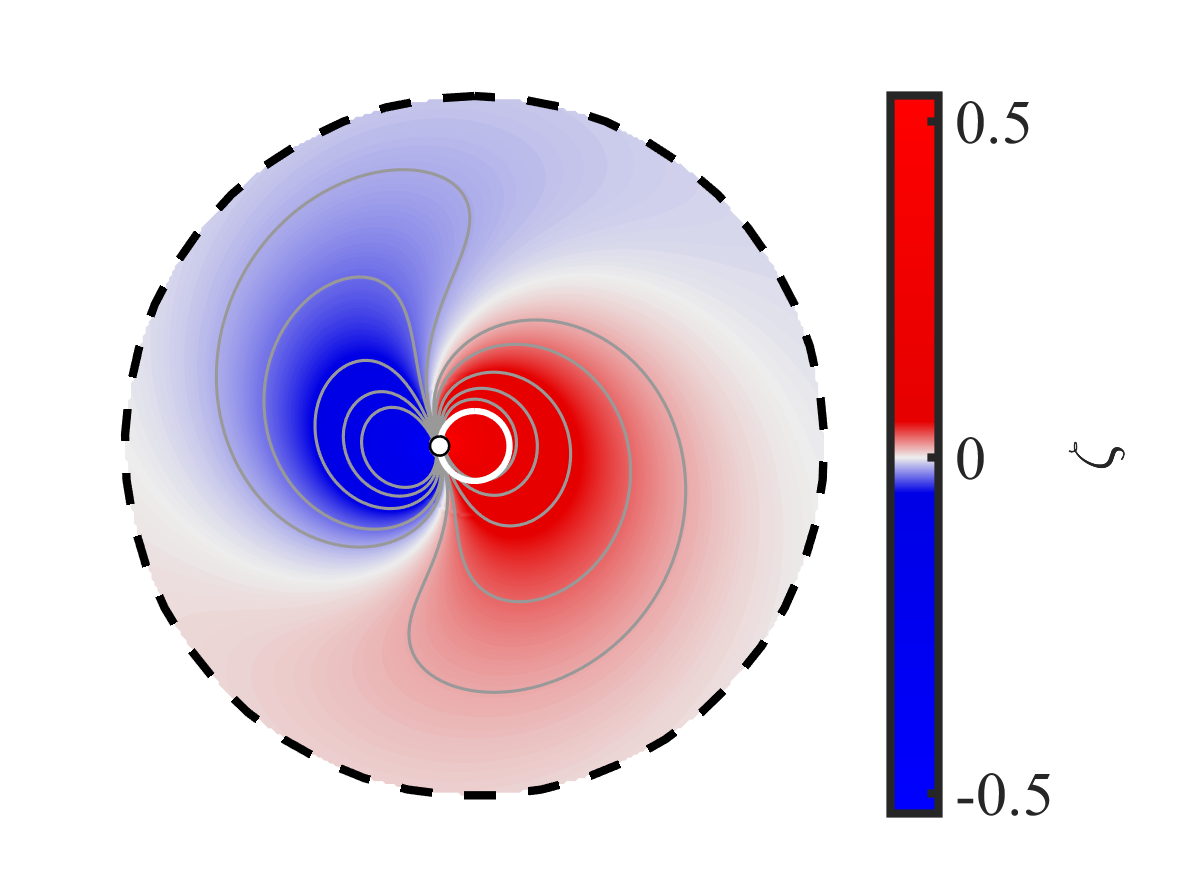}
		}\\
		\vspace{-0.3cm}
		\subfloat[$R_{sd}=10$\label{fig:vorticity_Re=0.05_E}]{
			\includegraphics[trim=2cm 0.5cm 6cm 1.25cm, clip=true,height=\subfigthreeB\textwidth]{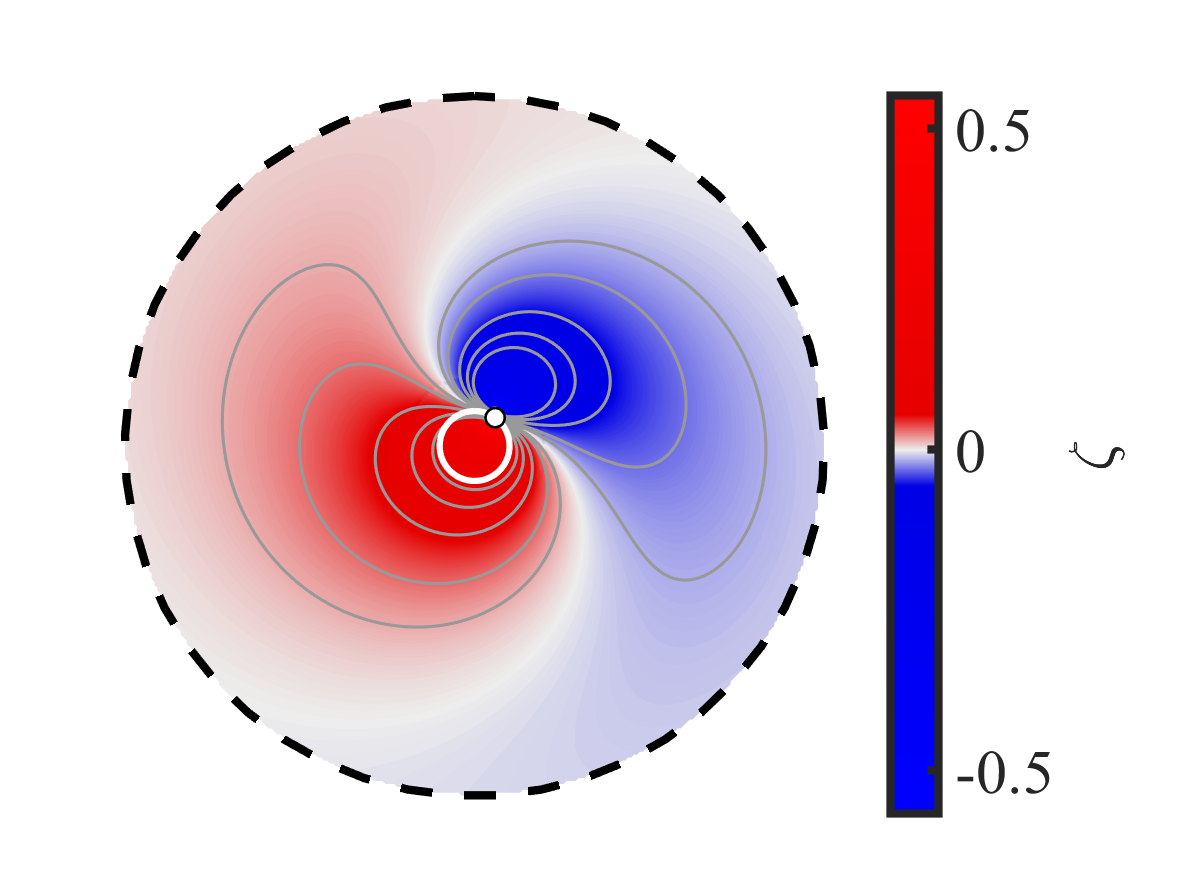}
		}~
		\hspace{-0.3cm}
		\subfloat[$R_{sd}=10$\label{fig:vorticity_Re=0.05_F}]{
			\includegraphics[trim=2cm 0.5cm 6cm 1.25cm, clip=true,height=\subfigthreeB\textwidth]{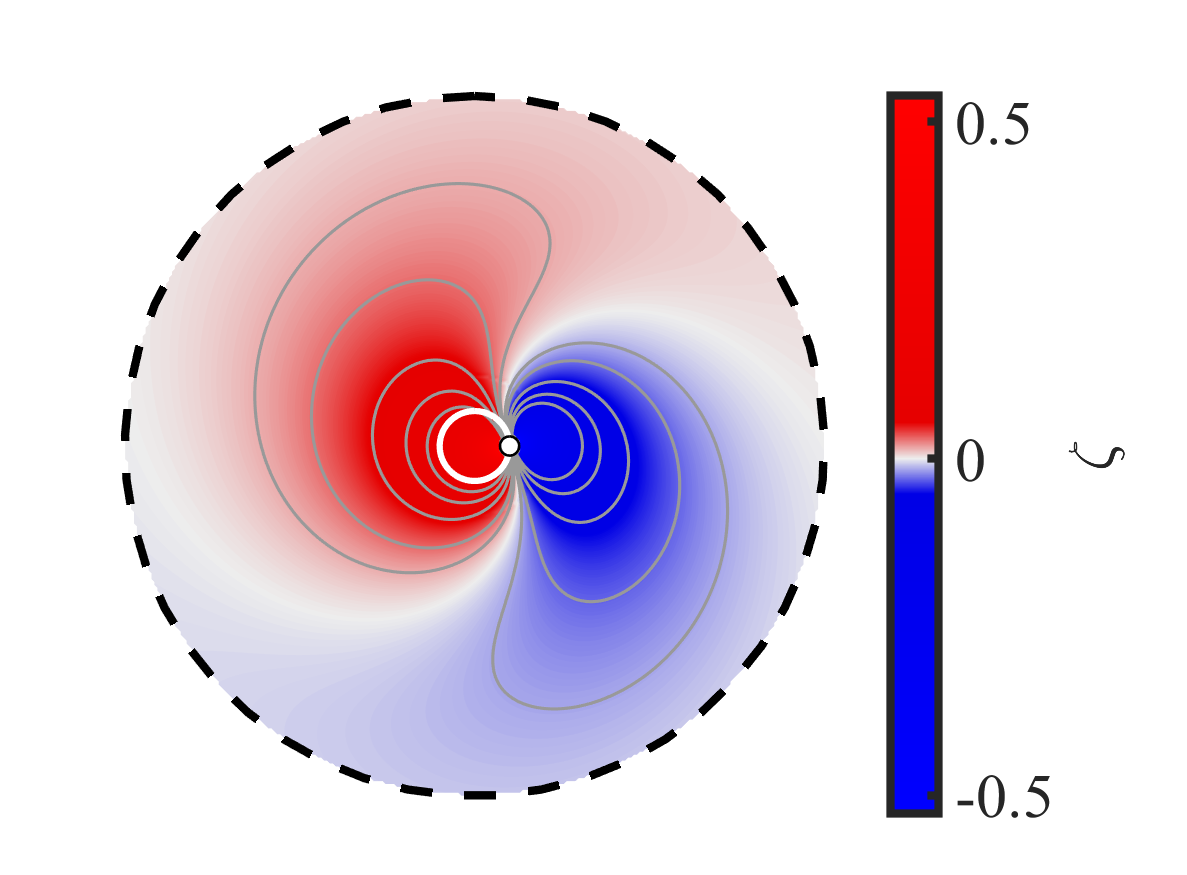}
		}~
		\hspace{-0.3cm}
		\subfloat[$R_{sd}=10$\label{fig:vorticity_Re=0.05_G}]{
			\includegraphics[trim=2cm 0.5cm 6cm 1.25cm, clip=true,height=\subfigthreeB\textwidth]{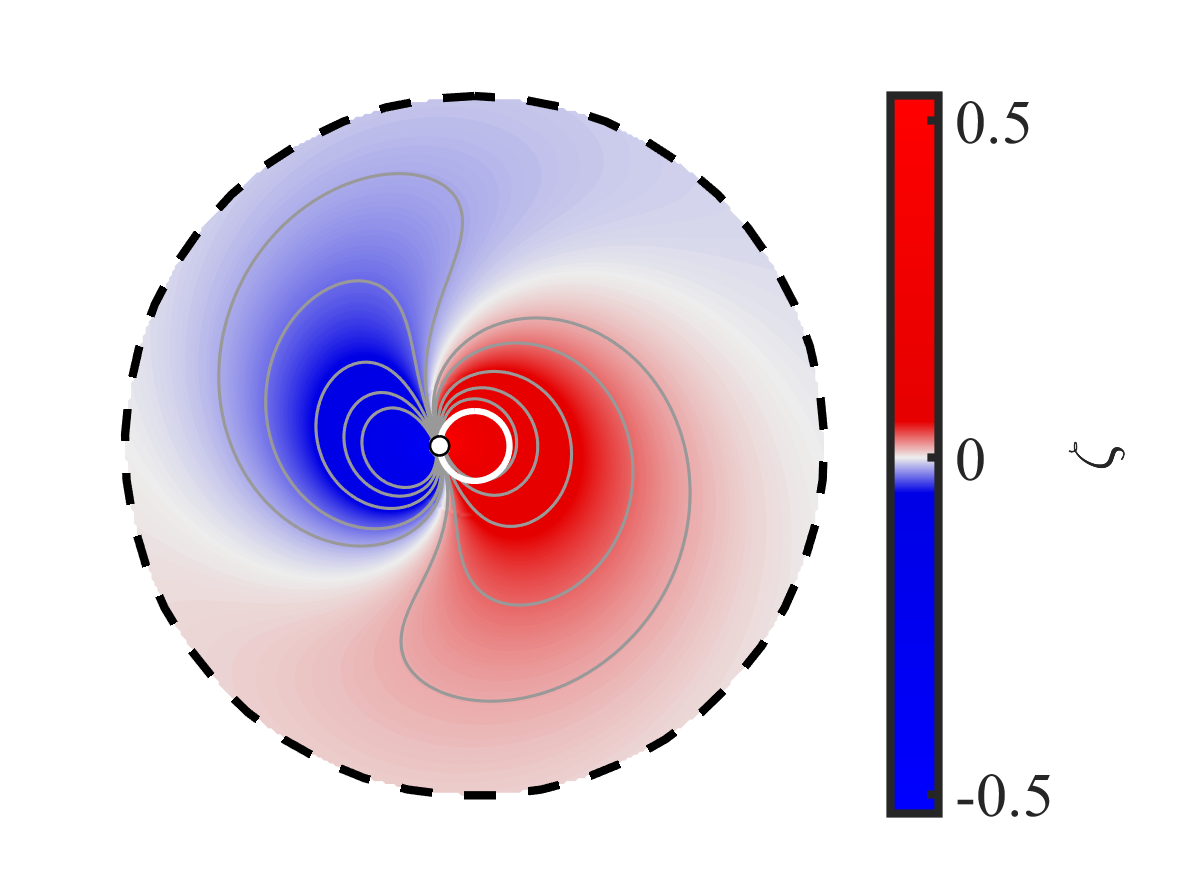}
		}~
		\hspace{-0.3cm}
		\subfloat[$R_{sd}=20$\label{fig:vorticity_Re=0.05_H}]{
			\includegraphics[trim=2cm 0.5cm 6cm 1.25cm, clip=true,height=\subfigthreeB\textwidth]{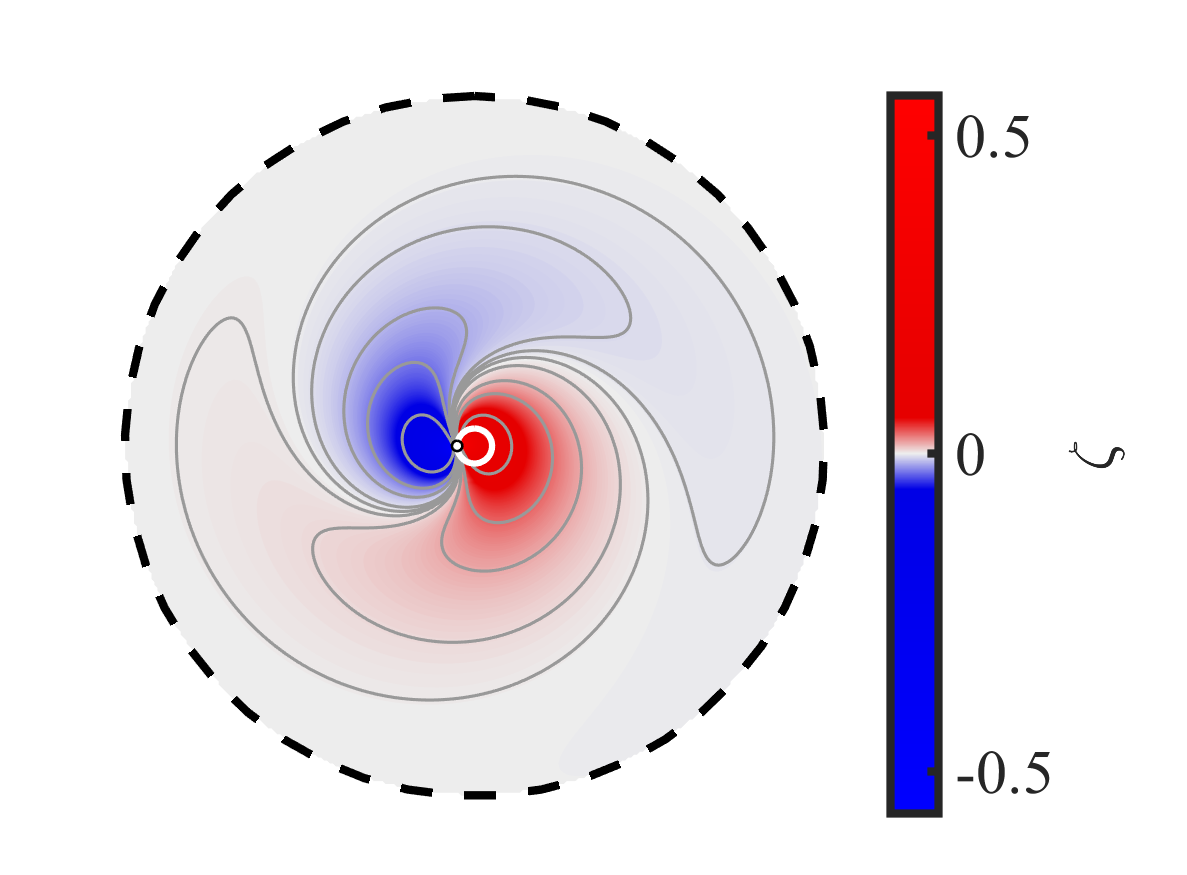}
		}
		\caption{\color{black}Snapshots of the dye concentration field (a-h) and vorticity field (i-p) at times $T= 0.05, 0.125, 0.4, 1, 1.65{\color{black},} 7.5, 45, \text{and}~90$. In the bottom two rows, the red, blue, and grey colors denote counterclockwise (CCW), clockwise (CW), and near-zero vorticity values, respectively. The white and grey lines indicate the stirrer’s path and the streamlines, respectively. The radius of the field of view is provided in the caption. $Re=0.05$.
		}
		\label{fig:alpha_Re=0.05}
	\end{figure}

	Figure \ref{fig:alpha_Re=0.05} illustrates the dynamics at a very low stirring speed, $Re \ll 1$. The first two rows display snapshots of the dye concentration, while the bottom two rows show snapshots of the vorticity field. In the bottom two rows, the red, blue, and grey colors denote counterclockwise (CCW), clockwise (CW), and near-zero vorticity values, respectively. The white and grey lines indicate the stirrer's path and the streamlines, respectively. To facilitate \ida{the interpretation of the scalar evolution}, snapshots of the two fields are displayed at the same time instances. Furthermore, the snapshots are shown within subdomains of varying radii, with the dimensionless subdomain radius, $R_{sd}$, indicated in each subfigure caption.
	
	At the onset of stirring, the stirrer crosses the dye interface, stretching and folding it near its path, an area we will refer to as the ‘\emph{central region}’ (see Figures \ref{fig:alpha_Re=0.05_A} to \ref{fig:alpha_Re=0.05_E}). This stretching enhances local diffusion, leading to more efficient mixing in the immediate vicinity. As a result, a well-mixed zone develops \ida{as the dye interface is stretched into progressively thinner lamellae whose width eventually} approaches the diffusion length scale, \ida{$\delta_D\approx \sqrt{\hat{t}\hat{D}_m}/\hat{r}_o = \sqrt{t/Pe}$} (see Figure \ref{fig:alpha_Re=0.05_F}).
	
	Beyond this stage, the dye concentration field forms a spiral pattern with a circular, well-mixed region at its center (see Figures \ref{fig:alpha_Re=0.05_F} to \ref{fig:alpha_Re=0.05_H}). This pattern appears to evolve in an approximately self-similar manner over time. From this point onward, mixing is primarily driven by diffusion across streamlines, with only gradual stretching of the interface, \ida{which preserves} the self-similar spiral structure.

	The evolution of the dye concentration is influenced by momentum transfer through one-way coupling; i.e., the rheology and density remain independent of the dye concentration. The development of the vorticity field, displayed in Figure \ref{fig:alpha_Re=0.05}, \ida{provides insight into this scalar evolution. As} expected in Stokes flow ($Re \ll 1$), the local streamlines exhibit an approximate fore-and-aft symmetry. The flow field around the stirrer is established shortly after the onset of stirring and remains approximately steady from the perspective of an observer moving with the stirrer. The flow becomes nearly periodic, with a period matching that of the stirrer. The closed streamlines on either side of the stirrer indicate the presence of two attached vortices, which dominate the flow dynamics (see, e.g., Figure \ref{fig:vorticity_Re=0.05_B}). These vortices exhibit approximate antisymmetry with respect to the stirrer's path. 
	
	In all instances shown, streamlines originating within the stirrer's path remain largely confined to this region. This explains why, once the dye concentration becomes uniform along the stirrer's path, \ida{subsequent mixing is no longer driven directly by the stirrer-interface interaction.}
	
	Streamlines that connect to the stirrer but lie outside its path form progressively larger closed trajectories (see $T \gtrsim 1$ in Figure \ref{fig:alpha_Re=0.05}). These outer streamlines drive the expansion of the self-similar spiral dye pattern observed at later times (see Figures \ref{fig:alpha_Re=0.05_F} to \ref{fig:alpha_Re=0.05_H}). {\color{black} Moving away from the stirrer, the streamlines expand \ida{primarily in the} azimuthal direction, indicating negligible radial flow (see Figures \ref{fig:vorticity_Re=0.05_F} to \ref{fig:vorticity_Re=0.05_H}).}
	
\begin{figure}
		\centering
		\subfloat[$R_{sd}=4$\label{fig:alpha_Re=5_A}]{
			\includegraphics[trim=2cm 0.5cm 6cm 1.25cm, clip=true,height=\subfigthreeB\textwidth]{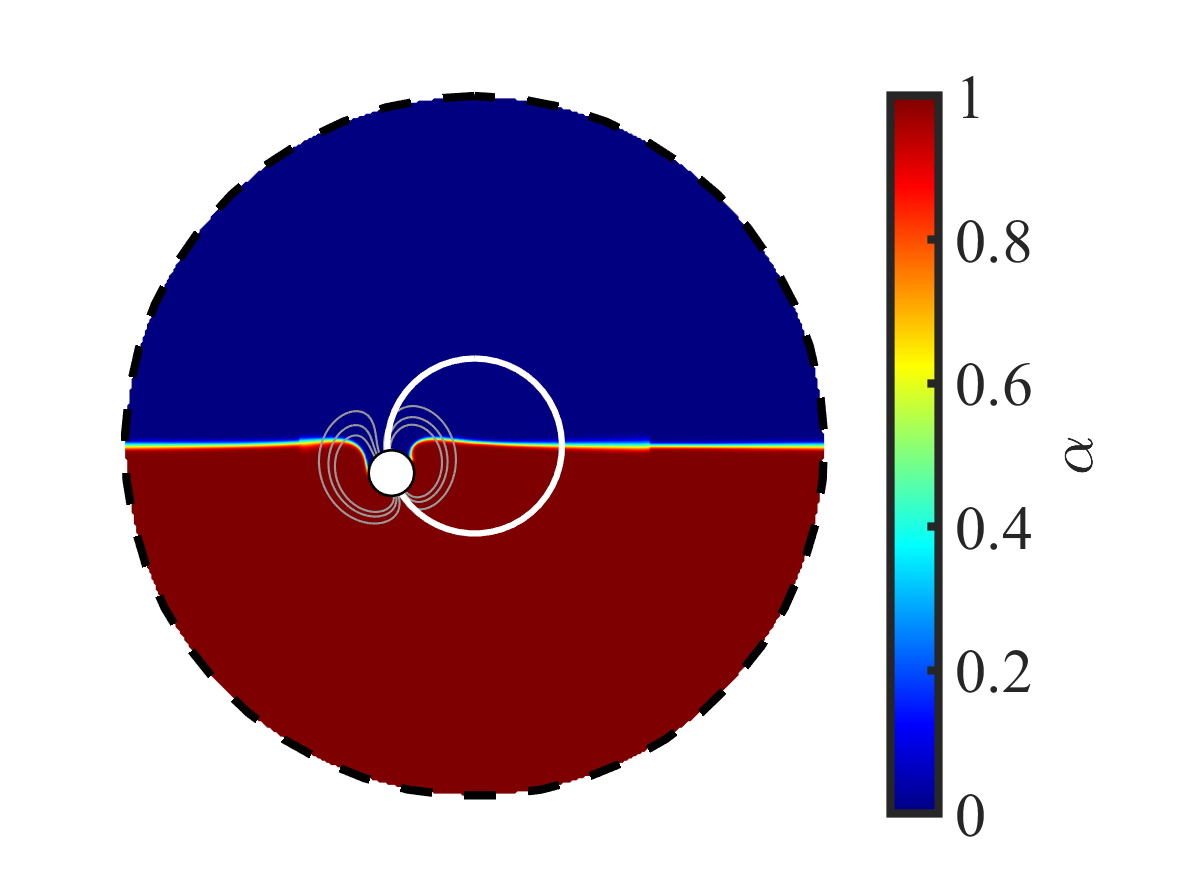}
		}~
		\hspace{-0.3cm}
		\subfloat[$R_{sd}=4$\label{fig:alpha_Re=5_B}]{
			\includegraphics[trim=2cm 0.5cm 6cm 1.25cm, clip=true,height=\subfigthreeB\textwidth]{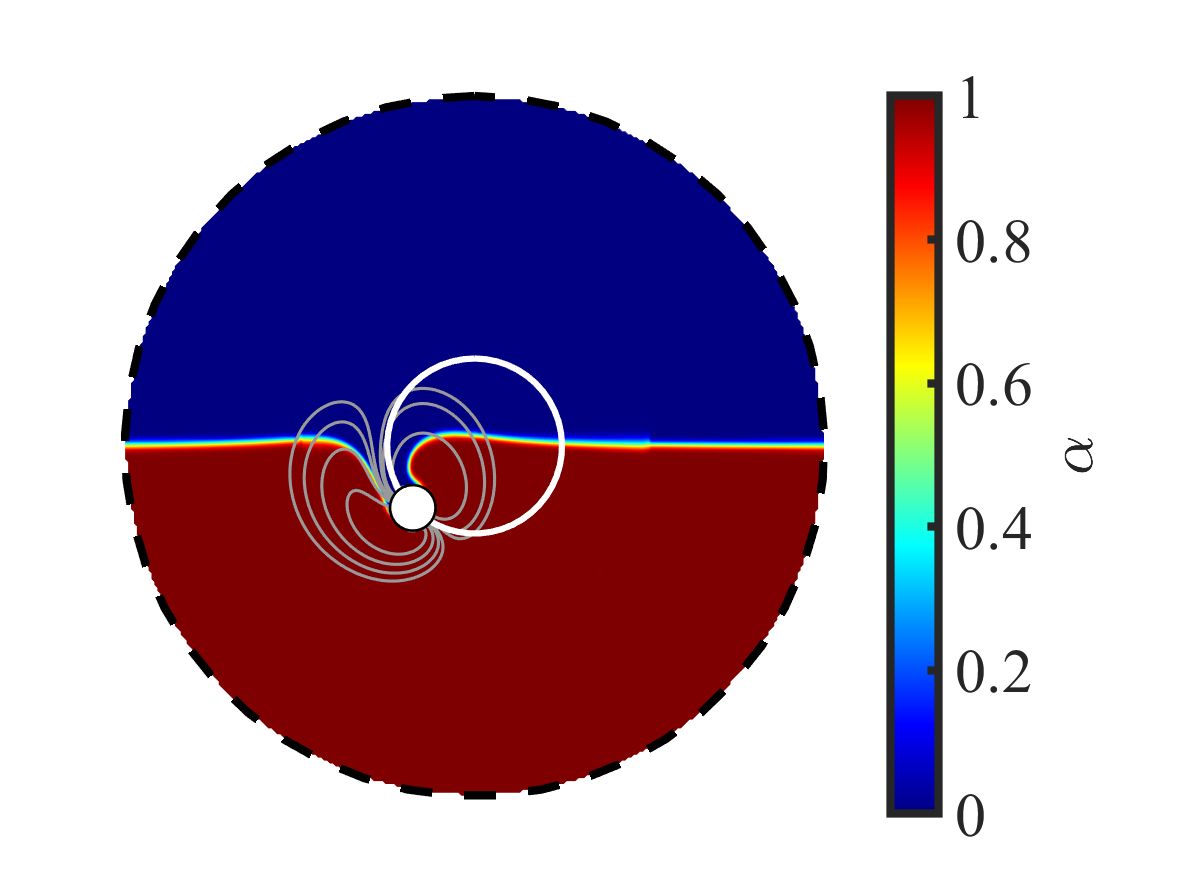}
		}~
		\hspace{-0.3cm}
		\subfloat[$R_{sd}=4$\label{fig:alpha_Re=5_C}]{
			\includegraphics[trim=2cm 0.5cm 6cm 1.25cm, clip=true,height=\subfigthreeB\textwidth]{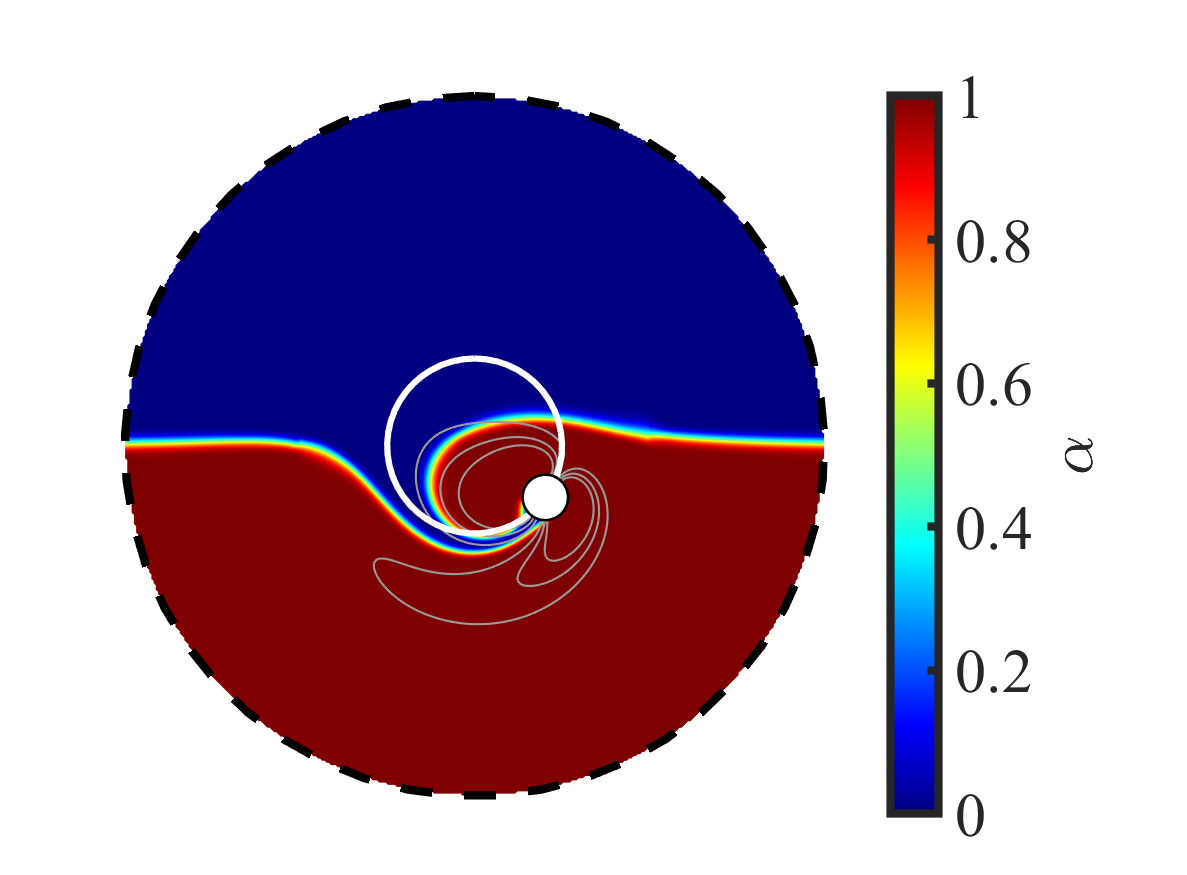}
		}~
		\hspace{-0.3cm}
		\subfloat[$R_{sd}=4$\label{fig:alpha_Re=5_D}]{
			\includegraphics[trim=2cm 0.5cm 1.5cm 1.25cm, clip=true,height=\subfigthreeB\textwidth]{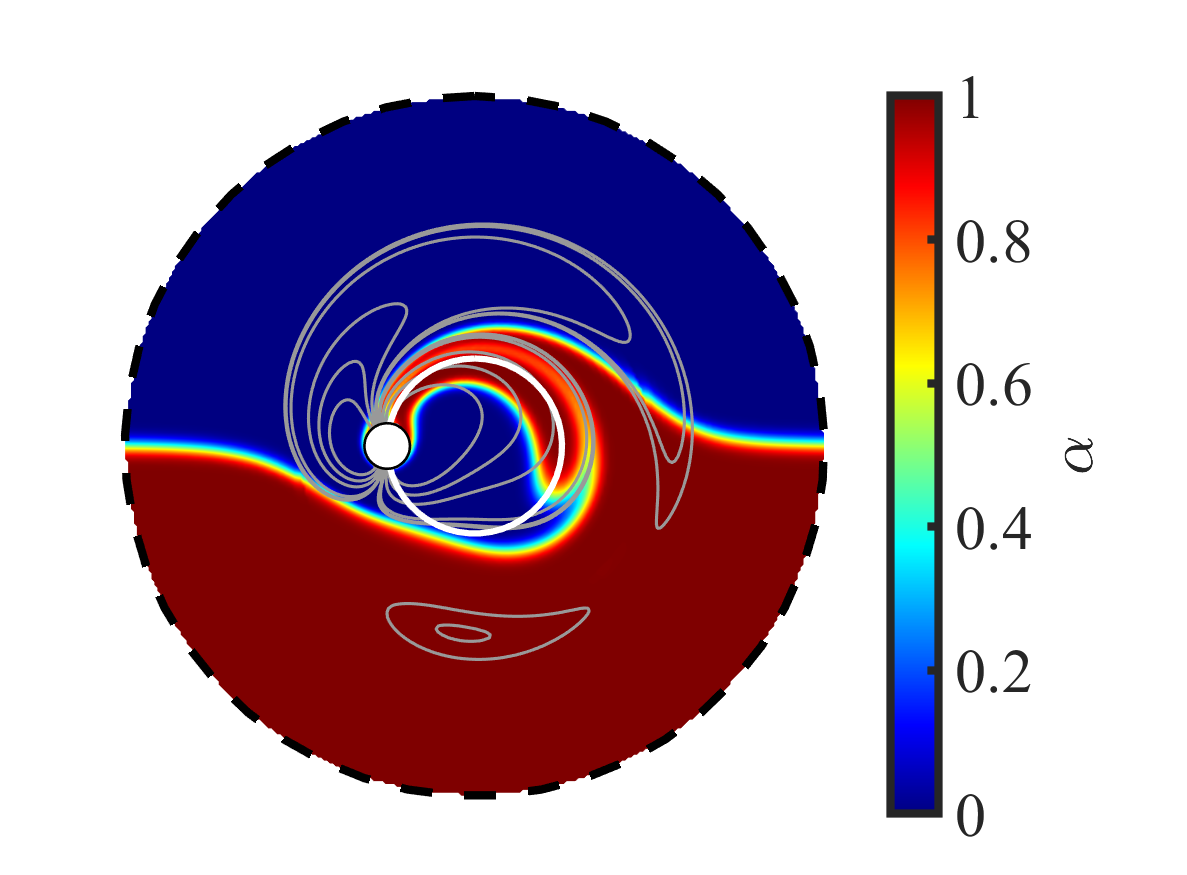}
		}\\
		\vspace{-0.3cm}
		\subfloat[$R_{sd}=4$\label{fig:alpha_Re=5_E}]{
			\includegraphics[trim=2cm 0.5cm 6cm 1.25cm, clip=true,height=\subfigthreeB\textwidth]{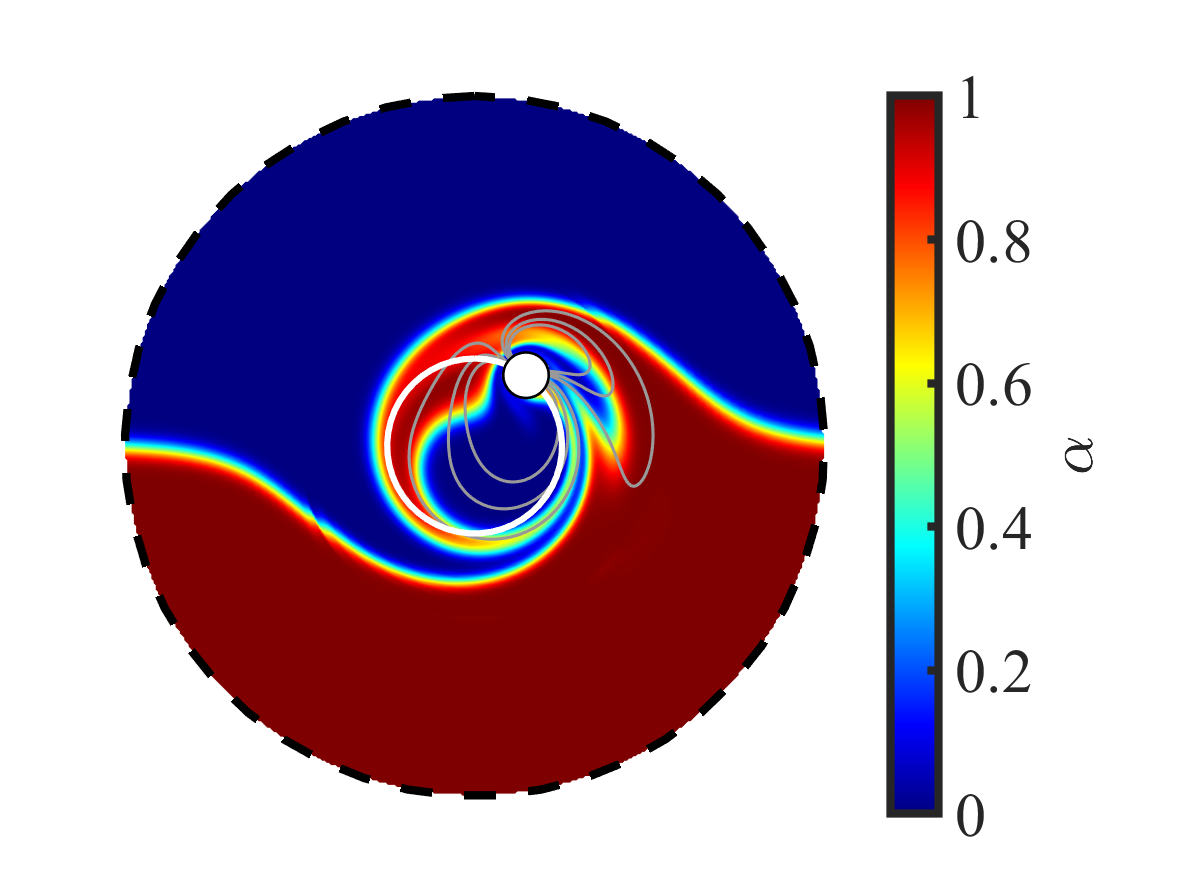}
		}~
		\hspace{-0.3cm}
		\subfloat[$R_{sd}=4$\label{fig:alpha_Re=5_F}]{
			\includegraphics[trim=2cm 0.5cm 6cm 1.25cm, clip=true,height=\subfigthreeB\textwidth]{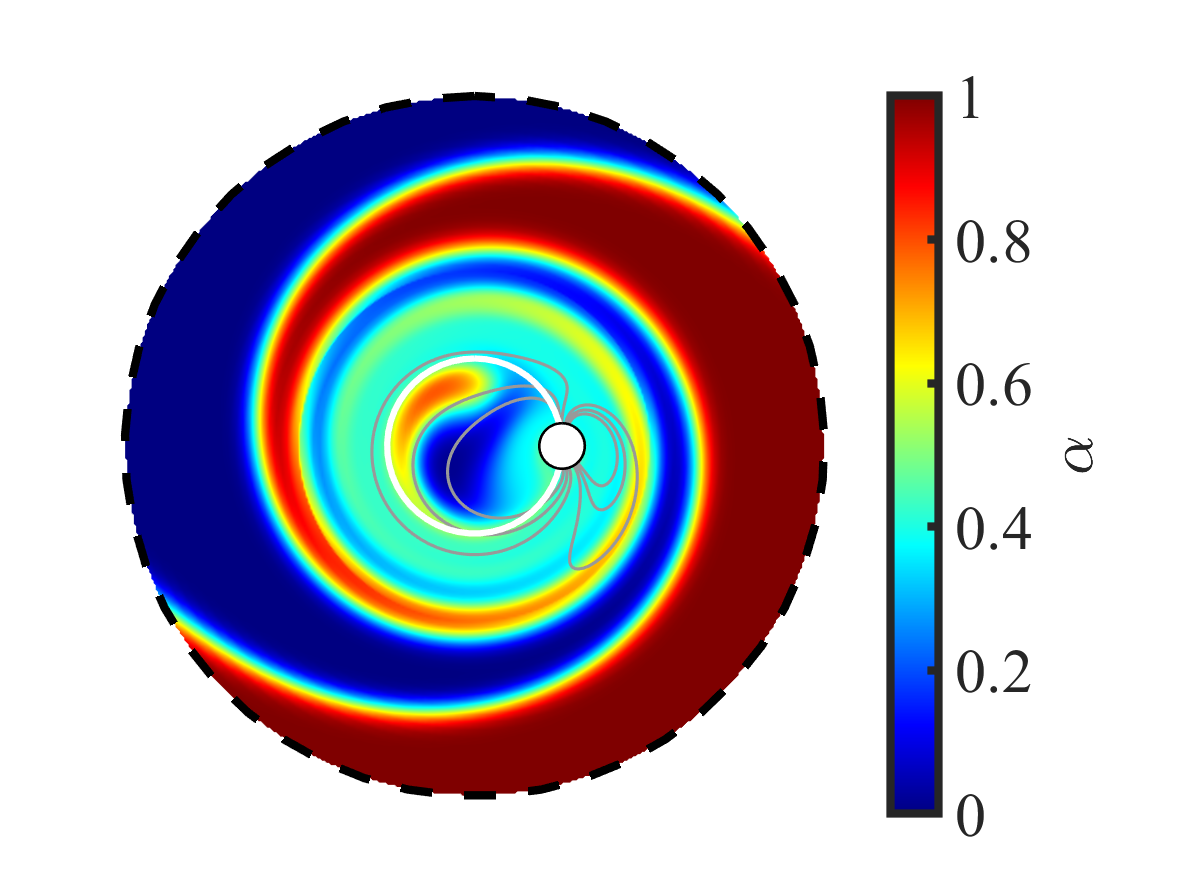}
		}~
		\hspace{-0.3cm}
		\subfloat[$R_{sd}=10$\label{fig:alpha_Re=5_G}]{
			\includegraphics[trim=2cm 0.5cm 6cm 1.25cm, clip=true,height=\subfigthreeB\textwidth]{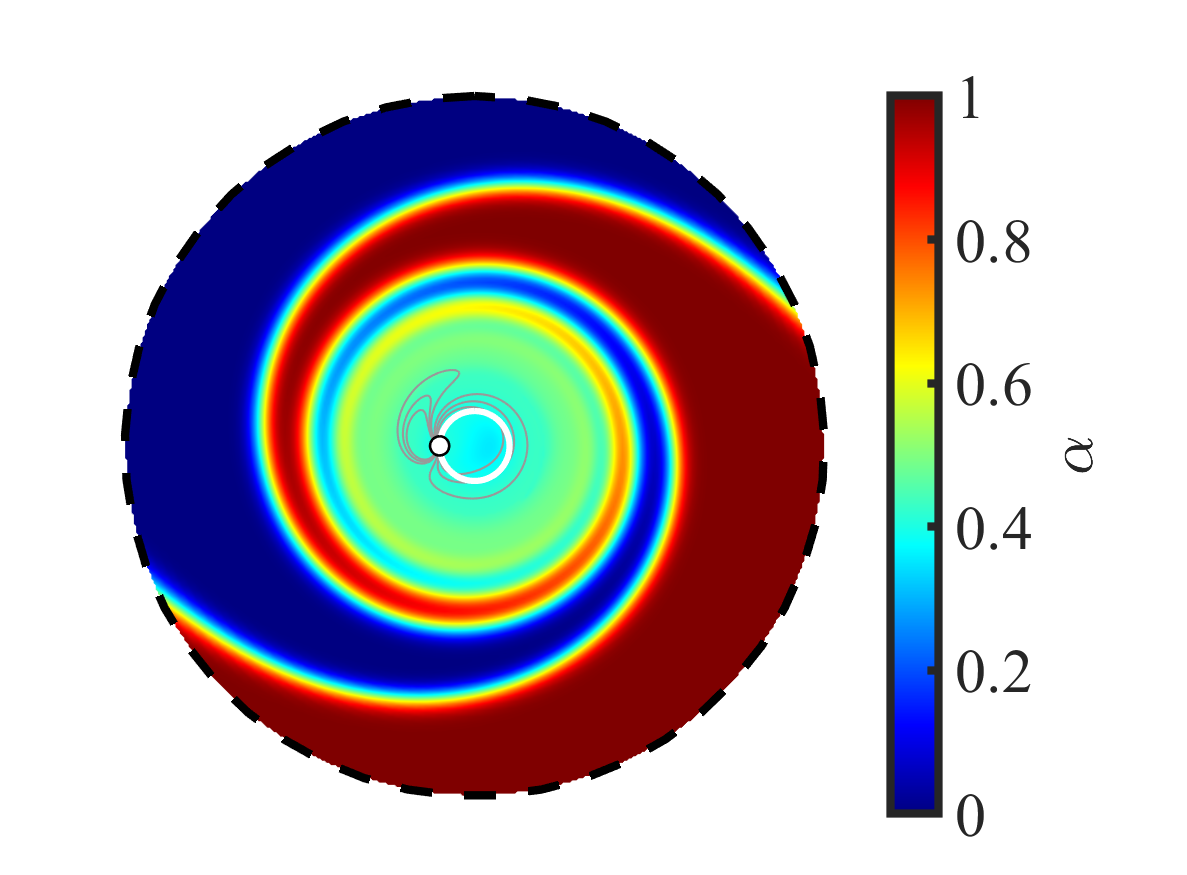}
		}~
		\hspace{-0.3cm}
		\subfloat[$R_{sd}=15$\label{fig:alpha_Re=5_H}]{
			\includegraphics[trim=2cm 0.5cm 1.5cm 1.25cm, clip=true,height=\subfigthreeB\textwidth]{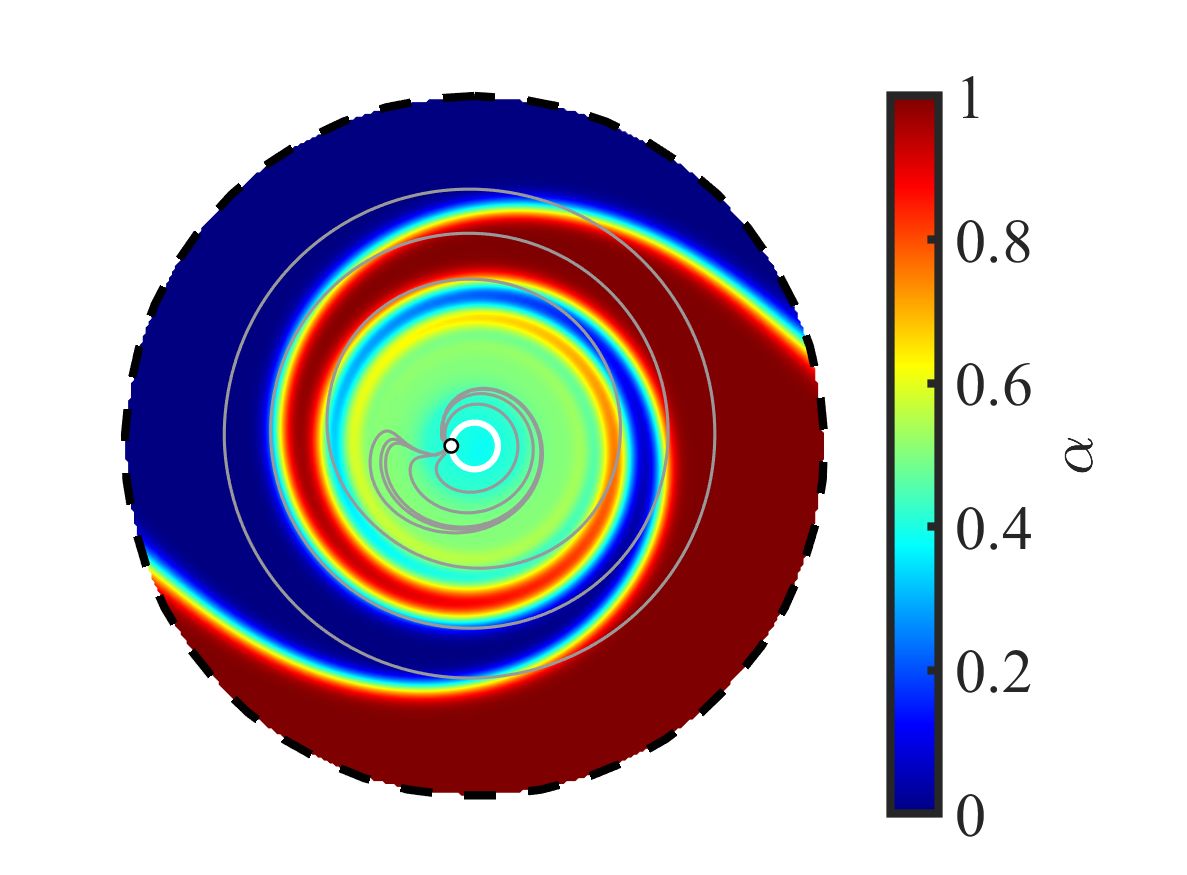}
		}
		\\
		\subfloat[$R_{sd}=4$\label{fig:vorticity_Re=5_A}]{
			\includegraphics[trim=2cm 0.5cm 6cm 1.25cm, clip=true,height=\subfigthreeB\textwidth]{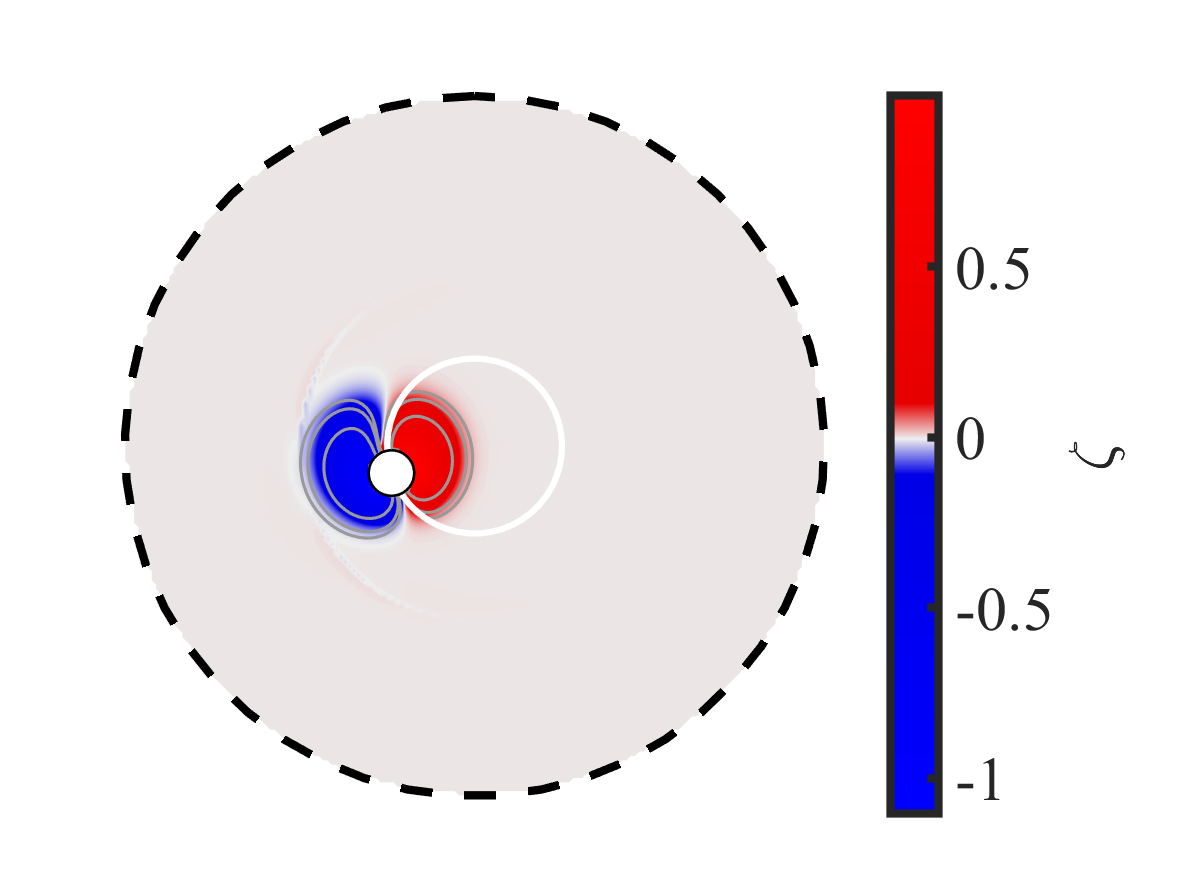}
		}~
		\hspace{-0.3cm}
		\subfloat[$R_{sd}=4$\label{fig:vorticity_Re=5_B}]{
			\includegraphics[trim=2cm 0.5cm 6cm 1.25cm, clip=true,height=\subfigthreeB\textwidth]{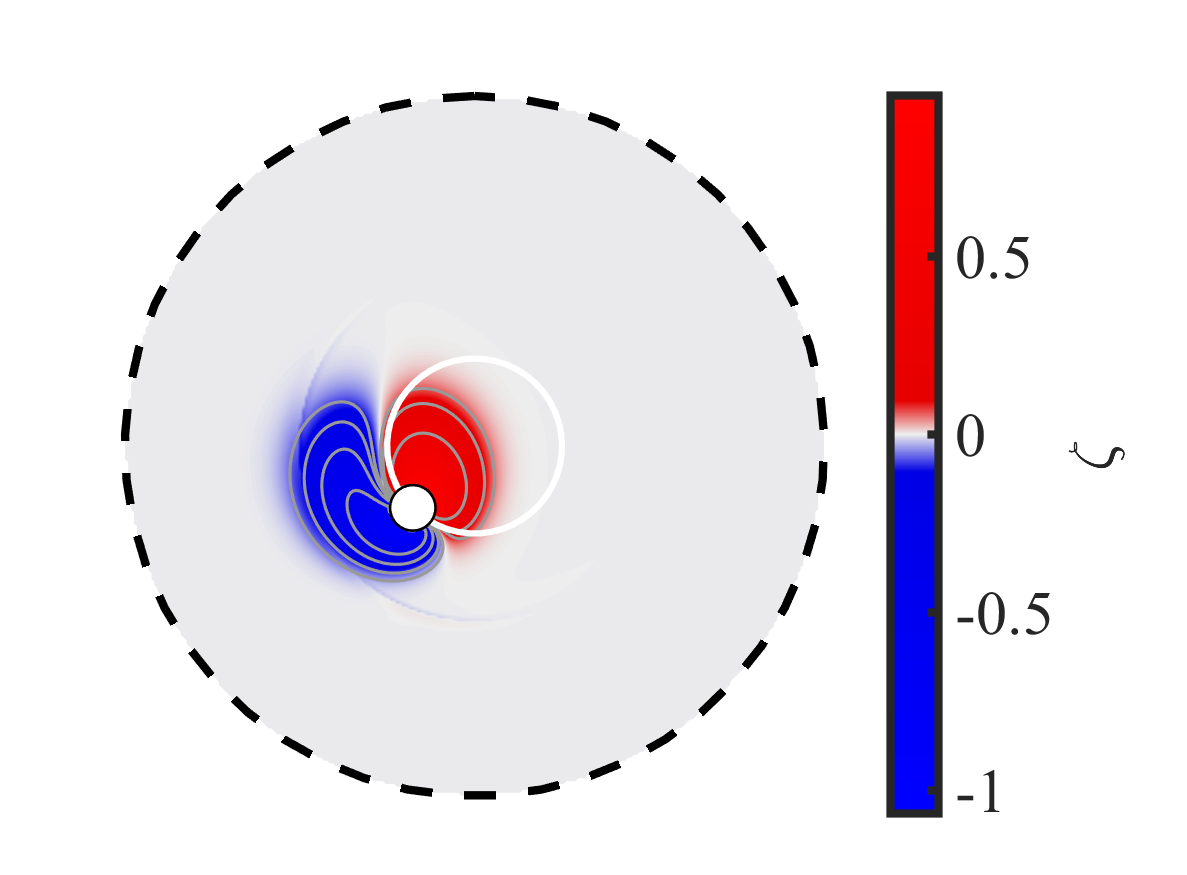}
		}~
		\hspace{-0.3cm}
		\subfloat[$R_{sd}=4$\label{fig:vorticity_Re=5_C}]{
			\includegraphics[trim=2cm 0.5cm 6cm 1.25cm, clip=true,height=\subfigthreeB\textwidth]{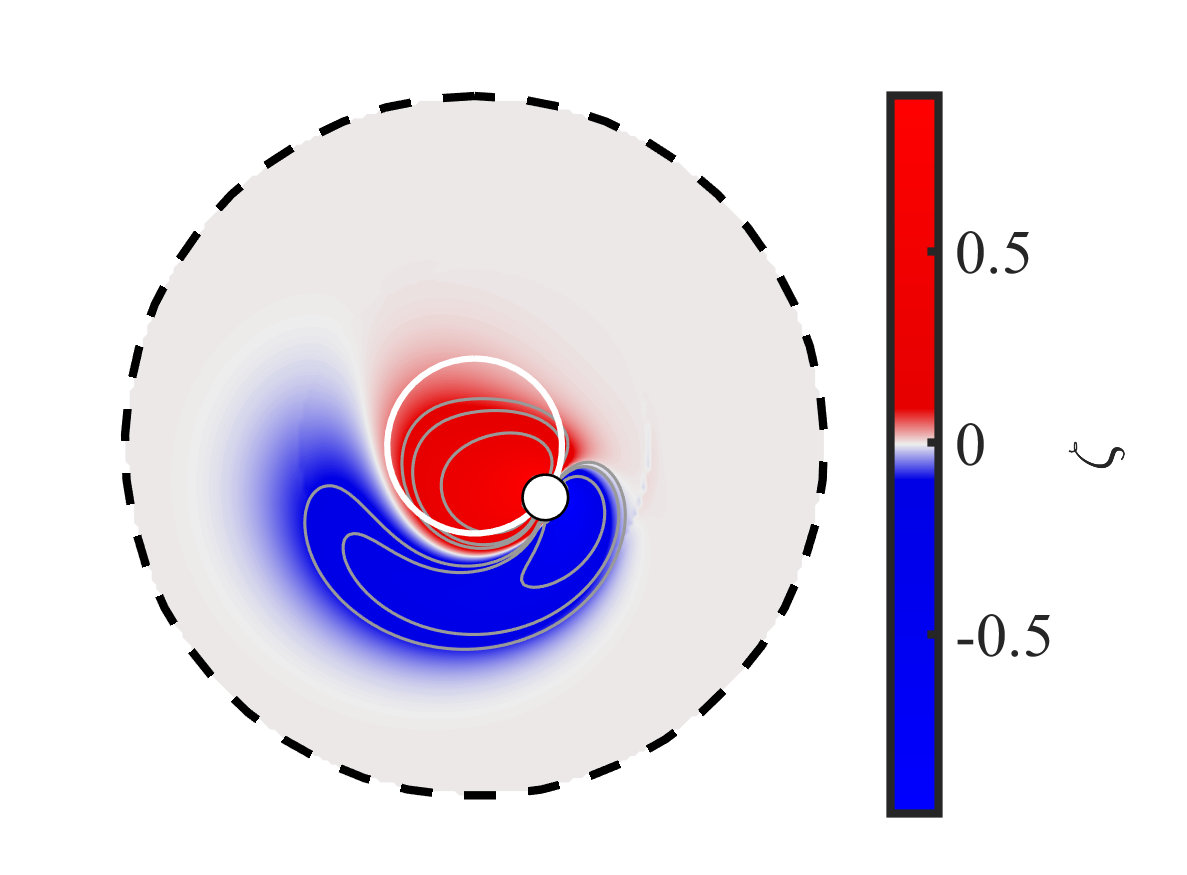}
		}~
		\hspace{-0.3cm}
		\subfloat[$R_{sd}=4$\label{fig:vorticity_Re=5_D}]{
			\includegraphics[trim=2cm 0.5cm 6cm 1.25cm, clip=true,height=\subfigthreeB\textwidth]{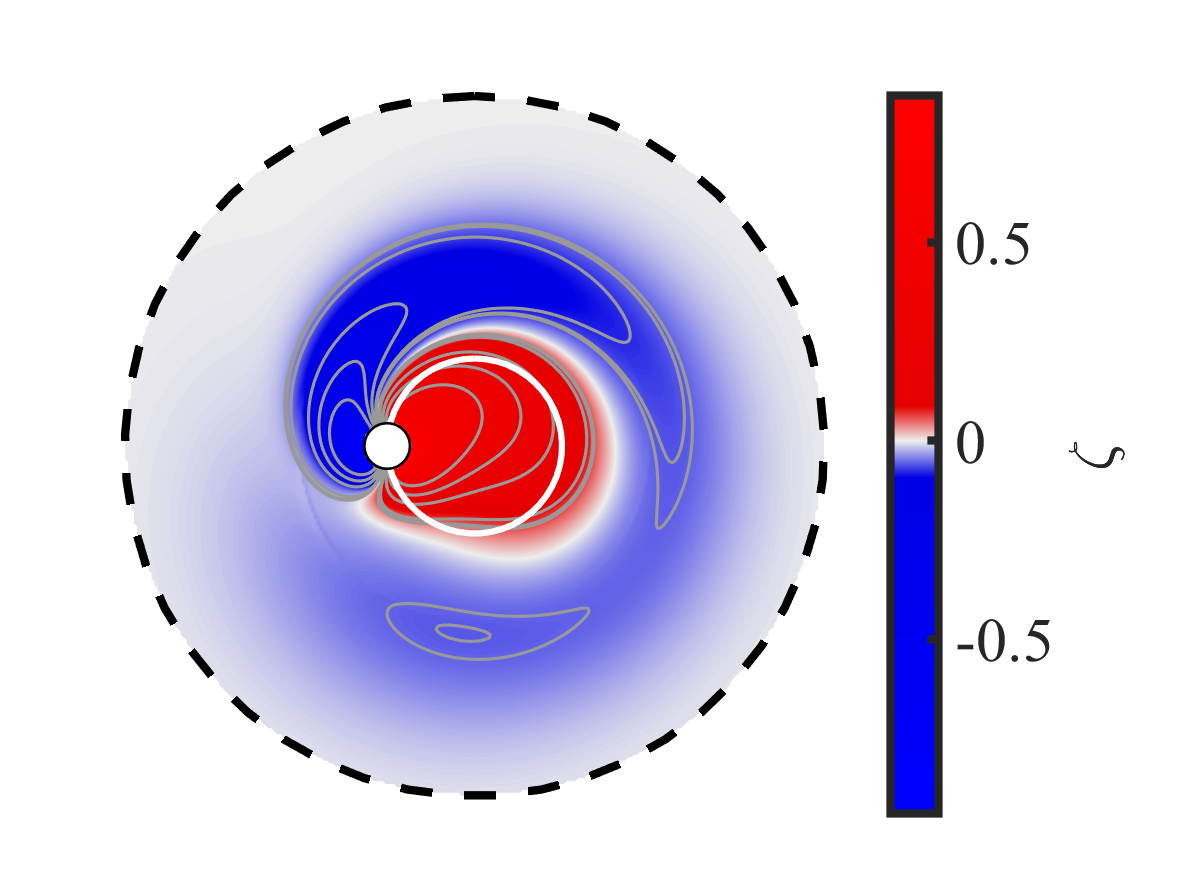}
		}\\
		\vspace{-0.3cm}
		\subfloat[$R_{sd}=4$\label{fig:vorticity_Re=5_E}]{
			\includegraphics[trim=2cm 0.5cm 6cm 1.25cm, clip=true,height=\subfigthreeB\textwidth]{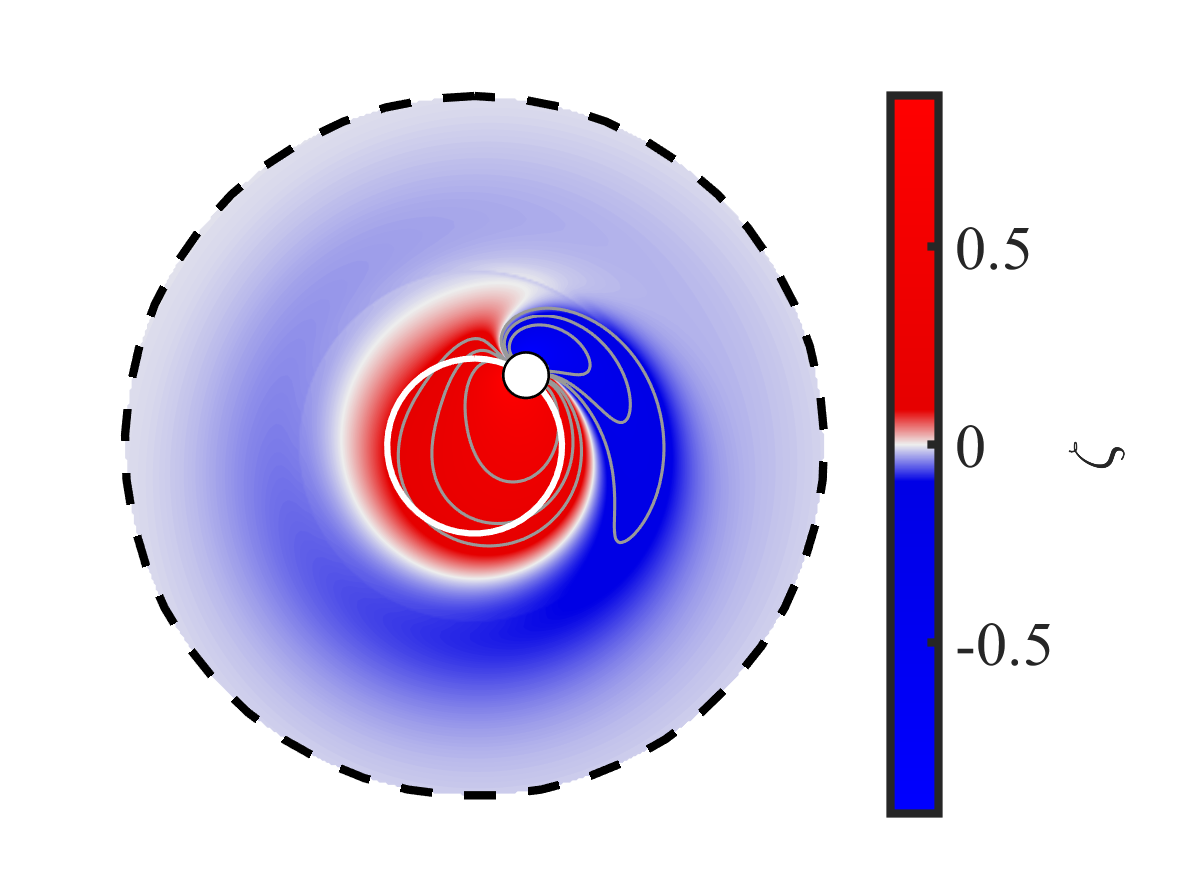}
		}~
		\hspace{-0.3cm}
		\subfloat[$R_{sd}=6$\label{fig:vorticity_Re=5_F}]{
			\includegraphics[trim=2cm 0.5cm 6cm 1.25cm, clip=true,height=\subfigthreeB\textwidth]{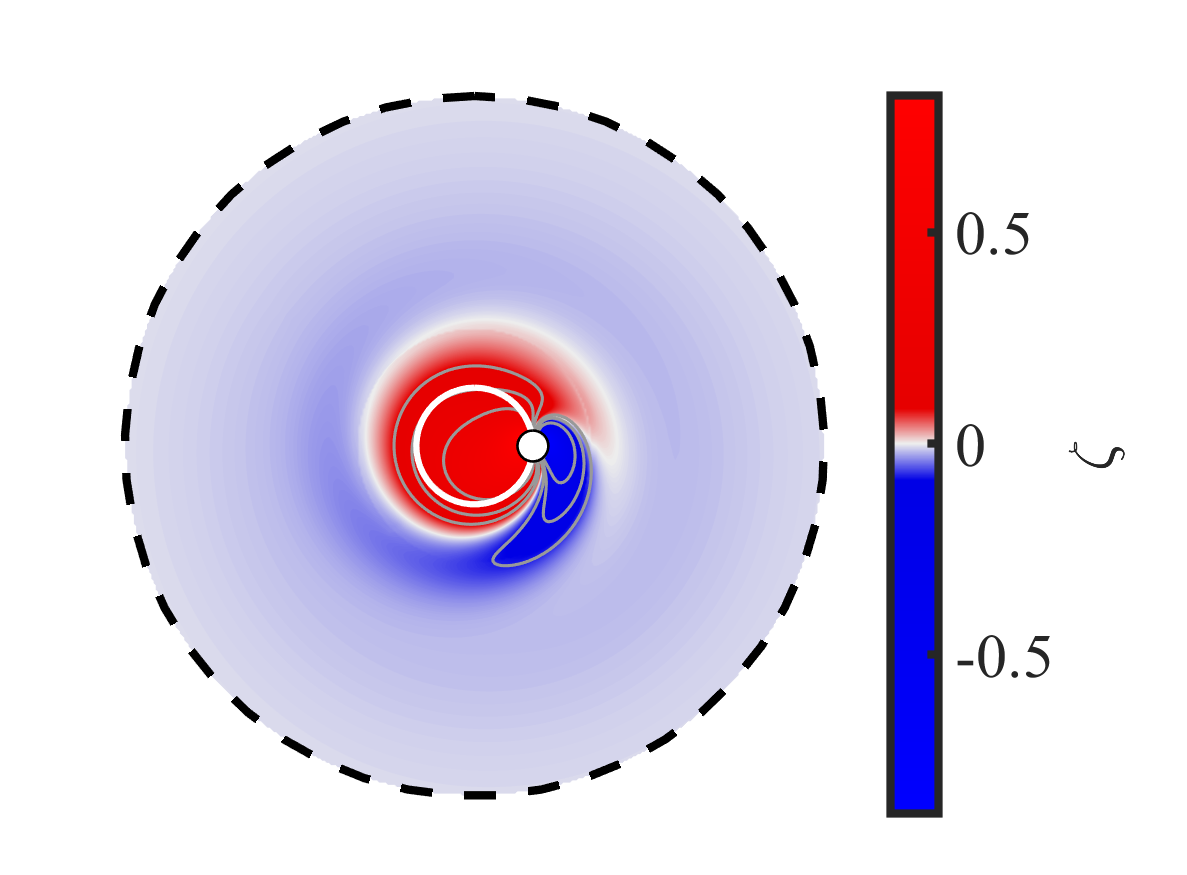}
		}~
		\hspace{-0.3cm}
		\subfloat[$R_{sd}=6$\label{fig:vorticity_Re=5_G}]{
			\includegraphics[trim=2cm 0.5cm 6cm 1.25cm, clip=true,height=\subfigthreeB\textwidth]{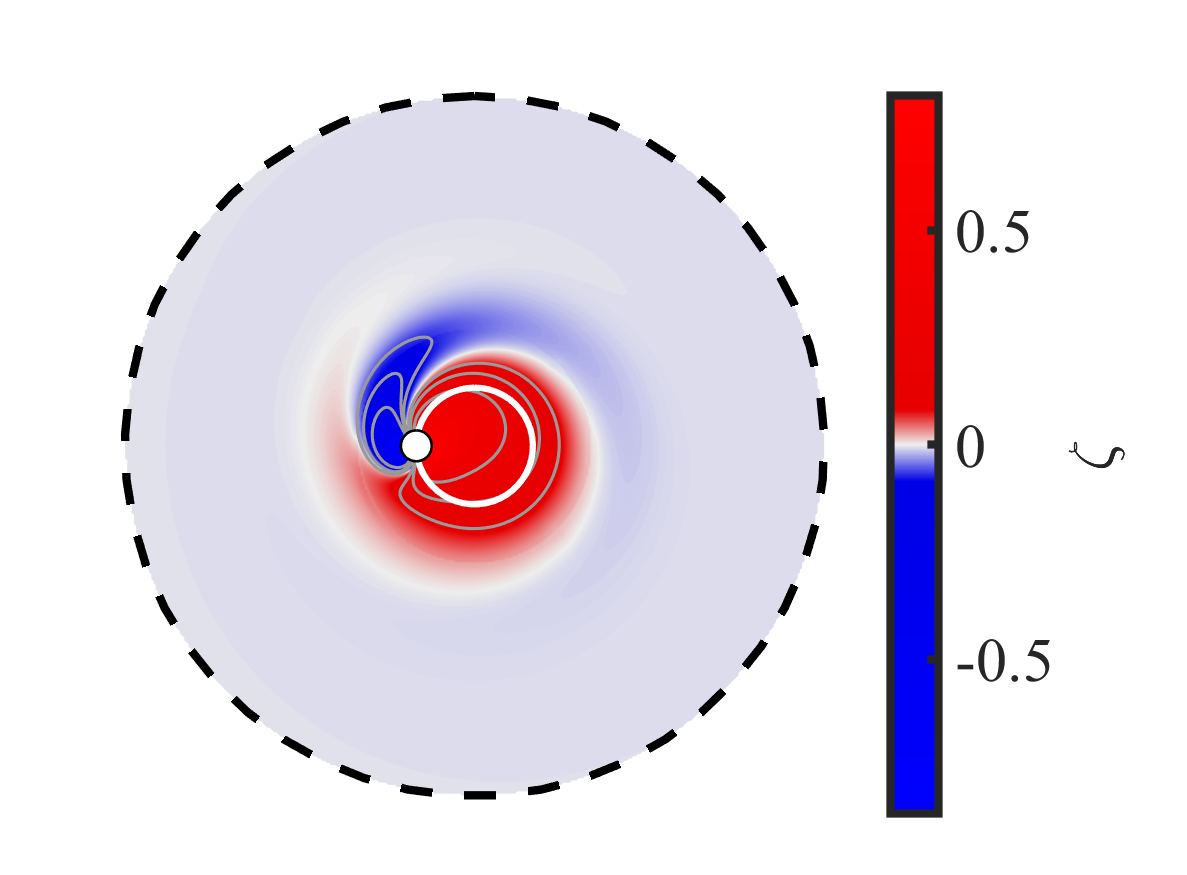}
		}~
		\hspace{-0.3cm}
		\subfloat[$R_{sd}=15$\label{fig:vorticity_Re=5_H}]{
			\includegraphics[trim=2cm 0.5cm 6cm 1.25cm, clip=true,height=\subfigthreeB\textwidth]{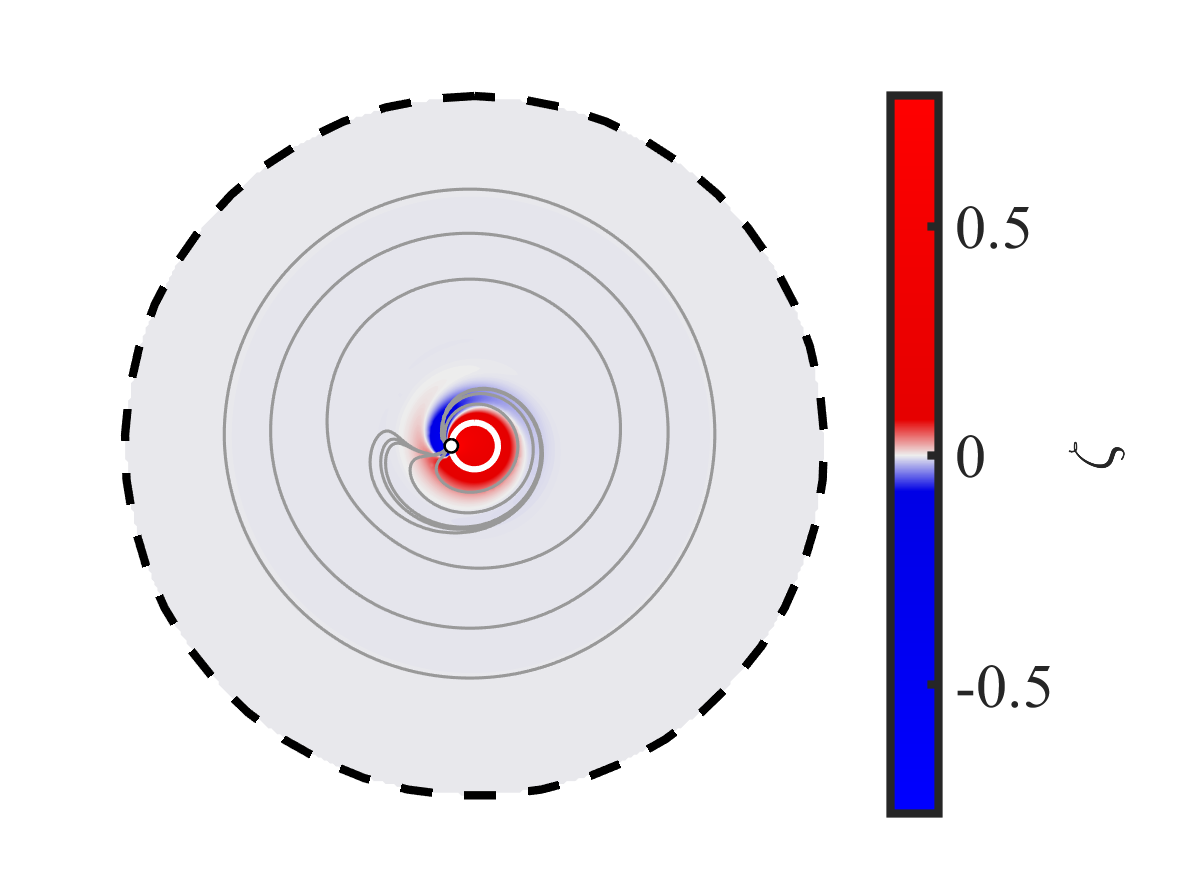}
		}
		\caption{{\color{black}Snapshots of the dye concentration field (a-h) and vorticity field (i-p) at times $T= 0.05, 0.125, 0.4, 1, 1.65, 7.5, 45, \text{and}~90$. In the bottom two rows, the red, blue, and grey colors denote counterclockwise (CCW), clockwise (CW), and near-zero vorticity values, respectively. The white and grey lines indicate the stirrer’s path and the streamlines, respectively.} The radius of the field of view is provided in the caption. $Re=5${\color{black}}.
		}
		\label{fig:alpha_Re=5}
	\end{figure}

	Figure \ref{fig:alpha_Re=5} illustrates the evolution of dye concentration {\color{black}and vorticity fields} for $Re = O(1)$. To facilitate comparison, the same time instances are shown in Figures \ref{fig:alpha_Re=0.05} and \ref{fig:alpha_Re=5}. The overall mixing dynamics are similar to those observed at $Re \ll 1$. The key difference is the enhanced stretching and striation of the interface at early times (see Figures \ref{fig:alpha_Re=5_D} \& \ref{fig:alpha_Re=5_E}). During the first period, more pronounced striation is particularly evident behind the stirrer (compare Figures \ref{fig:alpha_Re=0.05_D} and \ref{fig:alpha_Re=5_D}). Additionally, the stirrer briefly re-enters the dyed region after the first period, further stretching the interface (see Figure \ref{fig:alpha_Re=5_E}). After several periods, the dye concentration assumes a self-similar spiral pattern, with a well-mixed region forming at the center. Mixing then progresses through enhanced diffusion across streamlines and gradual interface stretching, \ida{following the same overall mechanism observed at $Re\ll 1$}.	
	
	Increasing $Re$ reduces the advection timescale ($\hat{t}_a$) relative to the momentum diffusion timescale ($\hat{t}_{\Df}$) \ida{since $\hat{t}_a/\hat{t}_{\Df} \propto 1/Re$}. As a result, the growth of the attached vortices during the first period is dominated by advection (see Figure \ref{fig:vorticity_Re=5_B}). The fore-and-aft symmetry observed at low $Re$ does not persist here (compare Figures \ref{fig:vorticity_Re=0.05_C} and \ref{fig:vorticity_Re=5_C}). The attached vortices extend downstream of the stirrer (Figures \ref{fig:vorticity_Re=5_C} and \ref{fig:vorticity_Re=5_D}), while the concentration and alignment of streamlines in this region indicate stronger local flow acceleration. This explains the enhanced striation behind the stirrer at $Re = O(1)$: dyed and dye-free fluid are rapidly drawn into this region as the stirrer moves across the interface (compare Figures \ref{fig:alpha_Re=0.05_C} and \ref{fig:alpha_Re=5_C}).
	
	The counterclockwise (CCW) vortex remains largely confined within the stirrer’s path throughout the process, while the clockwise (CW) vortex extends outward, forming on the exterior side of the stirrer’s path. {\color{black}A relatively weak CW vortex is shed after approximately one period, but it dissipates quickly and has no significant effect on the deformation of the dye interface (see Figures \ref{fig:alpha_Re=5_D} and \ref{fig:vorticity_Re=5_D}).} {\color{black}Here, the streamlines farther from the stirrer remain nearly circular and concentric with its path, indicating negligible radial flow (see Figures \ref{fig:alpha_Re=5_H} and \ref{fig:vorticity_Re=5_H})}. This \ida{explains} the persistence of the spiral dye pattern observed in Figure \ref{fig:alpha_Re=5} over long timescales.

	\begin{figure}
		\centering
		\subfloat{
			\includegraphics[trim=0cm 0cm 0cm 0cm, clip=true,height=.35\textwidth]{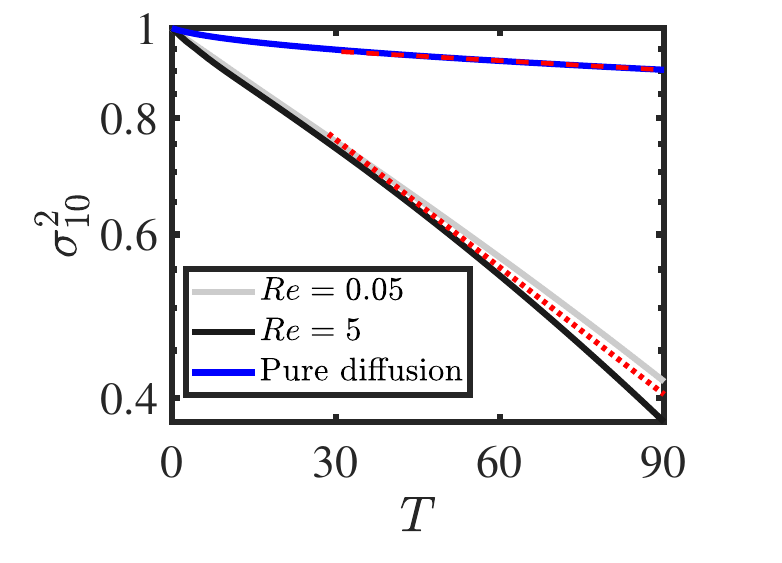}
		}
		\caption{Time evolution of normalized variance of dye concentration at $Re=0.05$ and 5. The solid blue line corresponds to the pure diffusion case. The red dashed and dotted lines depict the exponential fits for pure diffusion and $Re=0.05-5$ cases, respectively.}
		\label{fig:variance_regimeI}
	\end{figure}

To compare mixing rates, Figure~\ref{fig:variance_regimeI} illustrates the evolution of the normalized variance of dye concentration, $\sigma_{10}^2$, as a function of time for both cases discussed above. The blue line represents the purely diffusive case for reference. While both cases show significant improvement over the purely diffusive case, their decay rates are quite similar and closely follow an exponential fit, \ida{consistent with mixing remaining dominated by diffusion despite the enhanced transport produced by stirring.}

The dashed lines in the figure represent exponential fits to the normalized variance,
\begin{align*} 
\sigma_{10}^2 \approx \sigma_0 \exp\left(-\lambda T \right).
\end{align*}

Here, $\lambda$ is the decay constant. The parameter $\sigma_{0}$ approximates the normalized variance when the diffusion-dominated stage begins. Following \cite{christov2009enhancement}, two enhancement factors can be defined: $\eta_{\lambda} = \lambda / \lambda_p$ and $\eta_{\sigma} = \sigma_{0} / \sigma_{0p}$, where $\lambda_p$ and $\sigma_{0p}$ are the decay constant and intercept of the exponential fit for the purely diffusive case, respectively. The results yield $\eta_{\lambda} \approx 14$ and $\eta_{\sigma} \approx 1$, confirming that \ida{mixing remains diffusion-dominated while stirring enhances the mixing rate by approximately an order of magnitude}.

\begin{figure}
\centering
\subfloat[ $Re=0.05$\label{fig:vortexCenters_Re=0.05a}]{
	\includegraphics[trim=0cm 0cm 0cm 0cm, clip=true,height=.25\textwidth]{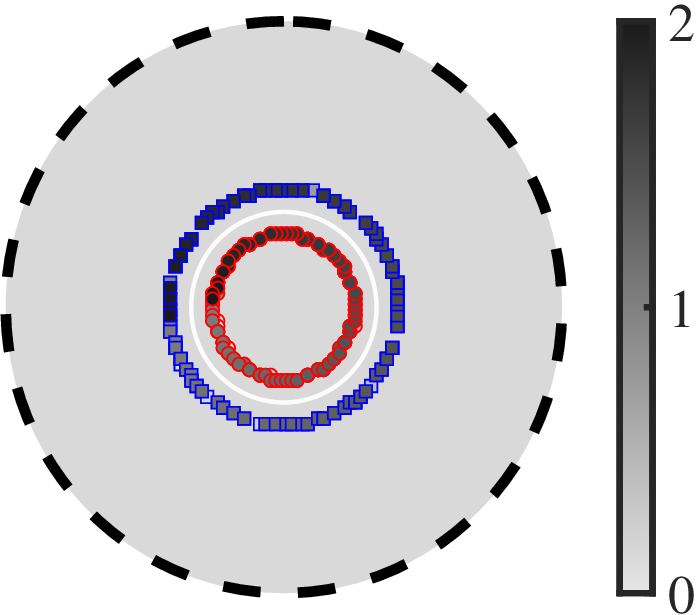}
}~
\subfloat[$Re=0.05$\label{fig:vortexCenters_Re=0.05b}]{
	\includegraphics[trim=0cm 0cm 0cm 0cm, clip=true,height=.25\textwidth]{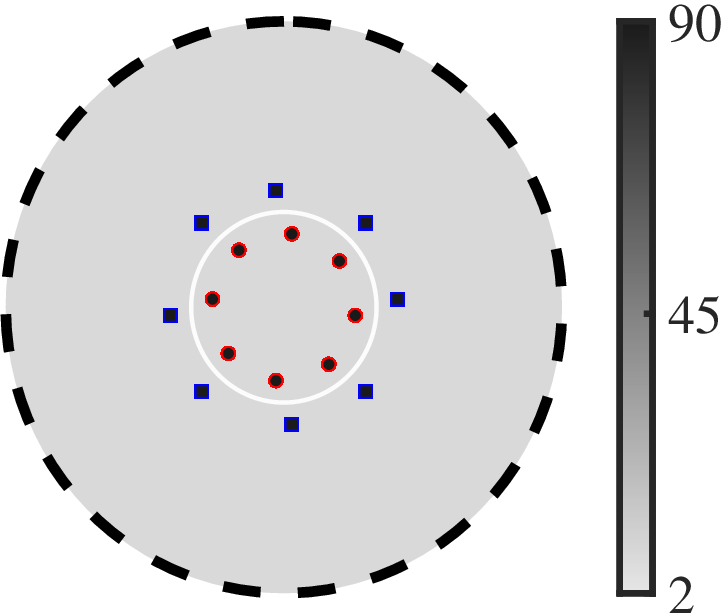}
} \\
\subfloat[ $Re=5$\label{fig:vortexCenters_Re=5a}]{
	\includegraphics[trim=0cm 0cm 0cm 0cm, clip=true,height=.25\textwidth]{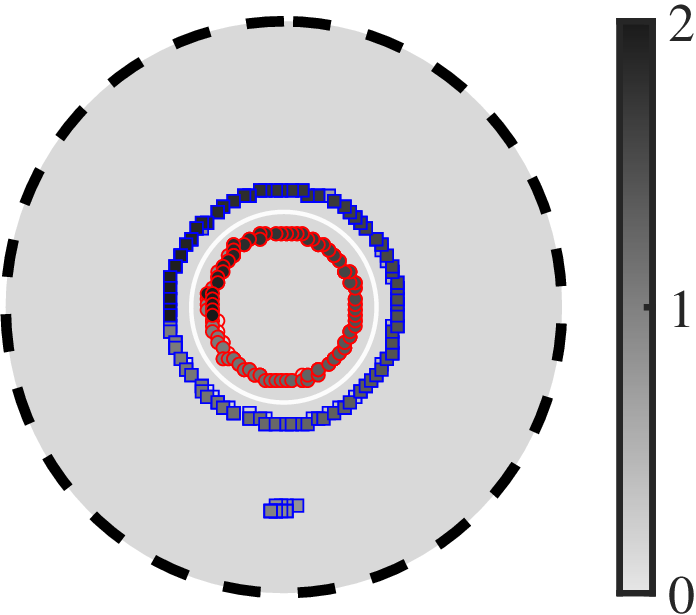}
}~
\subfloat[$Re=5$\label{fig:vortexCenters_Re=5b}]{
	\includegraphics[trim=0cm 0cm 0cm 0cm, clip=true,height=.25\textwidth]{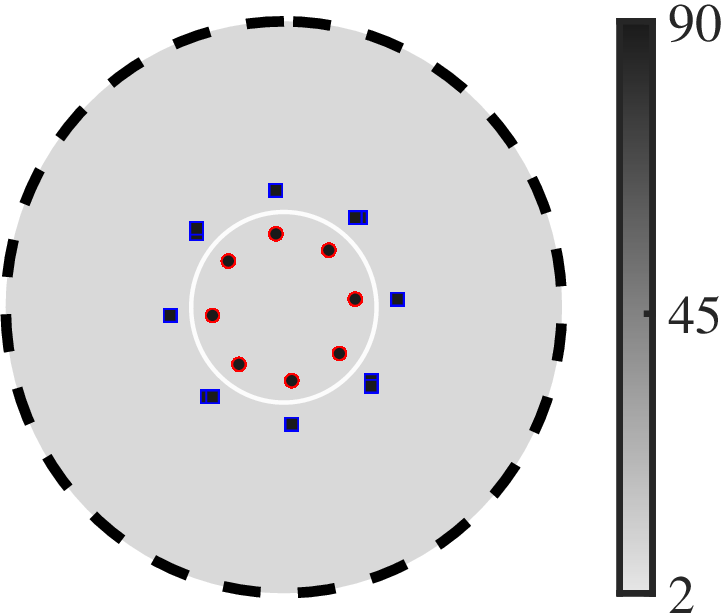}
} 

\caption{Time evolution of the vortex centers at different $Re$. The blue square and red circle mark center of CW and CCW vortices, respectively. Lighter shades correspond to earlier times. The white solid line represents the stirrer’s path and the black dashed lines denote the subdomain boundary. $R_{sd}=3$.}
\label{fig:vortexLocation_regimeI}

\end{figure}

We attribute the similarity in mixing dynamics and rate to the similarity \ida{of the underlying vortex dynamics}: in both cases, mixing is driven by two vortices attached to the stirrer. To further illustrate this, we present an approximation of the locus of the vortex centers in Figure \ref{fig:vortexLocation_regimeI}. These centers, identified as the local minima and maxima of the vorticity field, correspond to the CCW and CW vortices, respectively. For simplicity and brevity, we refer to this collection of points as the \emph{vortex centers} hereafter.

Figure \ref{fig:vortexLocation_regimeI} shows that in both cases ($Re\ll1$ and $Re=O(1)$) two attached vortices form at the onset of stirring (see Figures \ref{fig:vortexCenters_Re=0.05a} and \ref{fig:vortexCenters_Re=5a}). {\color{black}During the first period at $Re=O(1)$, a CW vortex is shed outside the stirrer's path (see Figures \ref{fig:vorticity_Re=5_D} and \ref{fig:vortexCenters_Re=5a}), but it dissipates quickly, as vortex centers show (Figure \ref{fig:vortexCenters_Re=5b}).
Comparing $Re\ll1$ and $Re=O(1)$, although the size and shape of vortices differ between the two cases, their locations relative to the stirrer’s path remain remarkably similar (Figures \ref{fig:vortexCenters_Re=0.05b} and \ref{fig:vortexCenters_Re=5b})}. We hypothesize that \ida{the similarity in vortex organization, rather than the detailed vortex geometry, is the primary reason for the similar mixing dynamics observed at $Re\lesssim O(1)$}.

\subsubsection{Emergence of advective mixing}
\label{sec:II}
When $10 \lesssim Re$, \ida{mixing is no longer confined to the central region. New advective transport mechanisms emerge while the flow remains laminar.}

\begin{figure}
\centering
\subfloat[$R_{sd}=3$\label{fig:alpha_Re=10_A}]{
	\includegraphics[trim=2cm 0.5cm 5.5cm 1.25cm, clip=true,height=\subfigthreeB\textwidth]{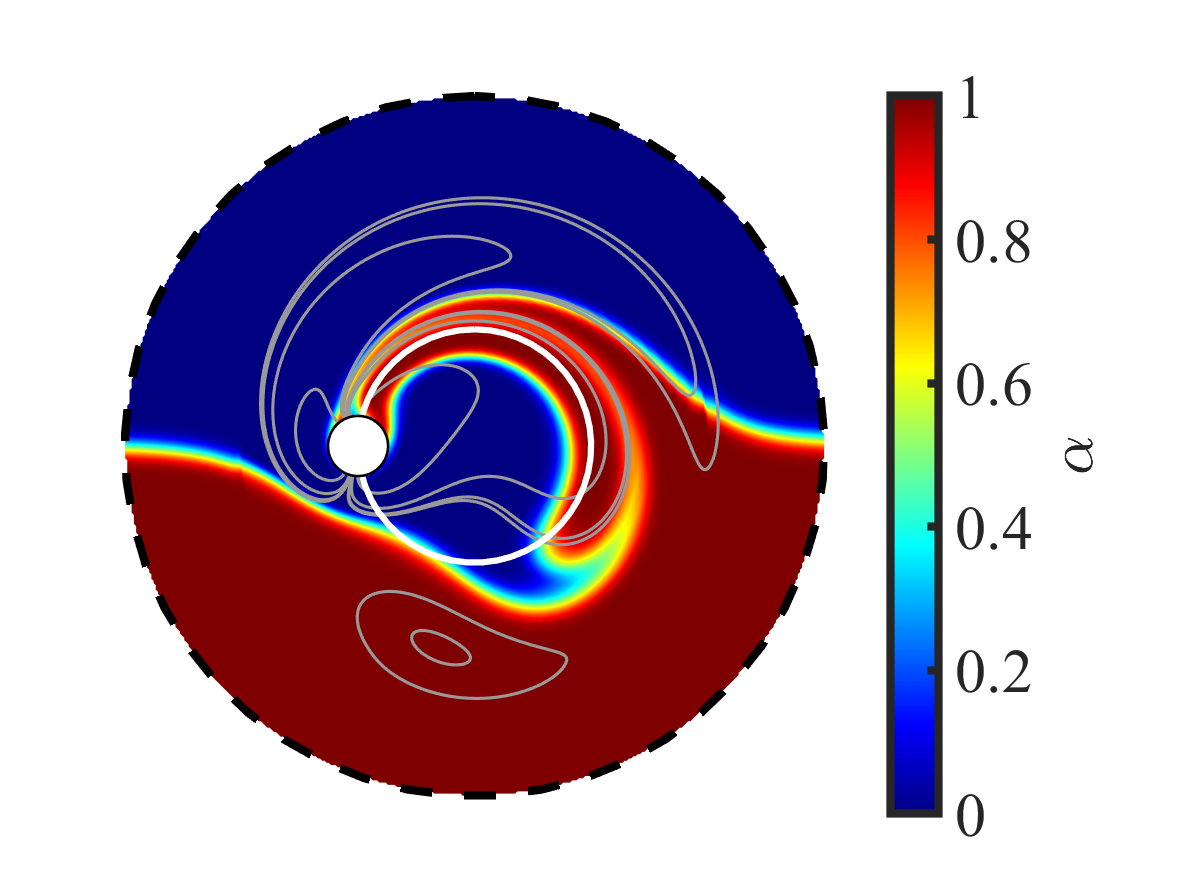}
}~
\hspace{-0.3cm}
\subfloat[$R_{sd}=3$\label{fig:alpha_Re=10_B}]{
	\includegraphics[trim=2cm 0.5cm 5.5cm 1.25cm, clip=true,height=\subfigthreeB\textwidth]{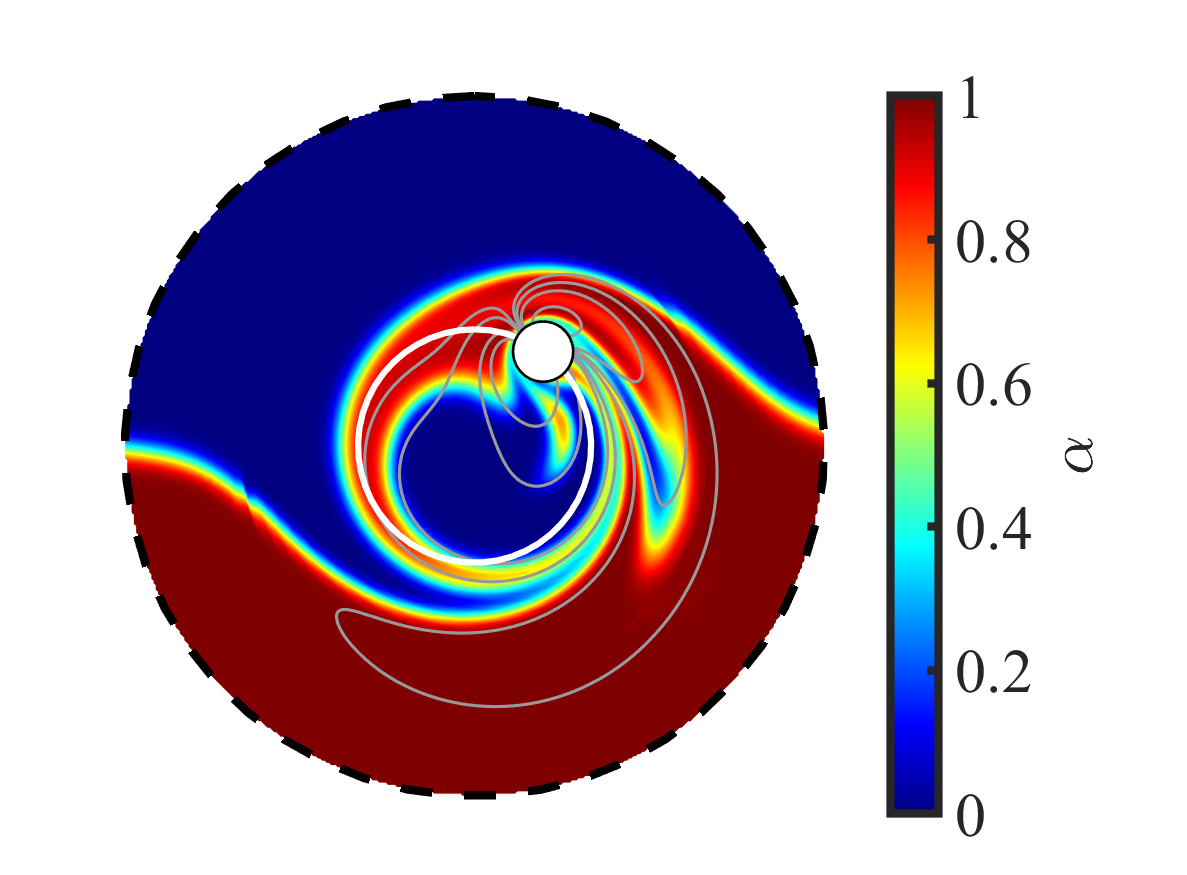}
}~
\hspace{-0.3cm}
\subfloat[$R_{sd}=6$\label{fig:alpha_Re=10_C}]{
	\includegraphics[trim=2cm 0.5cm 5.5cm 1.25cm, clip=true,height=\subfigthreeB\textwidth]{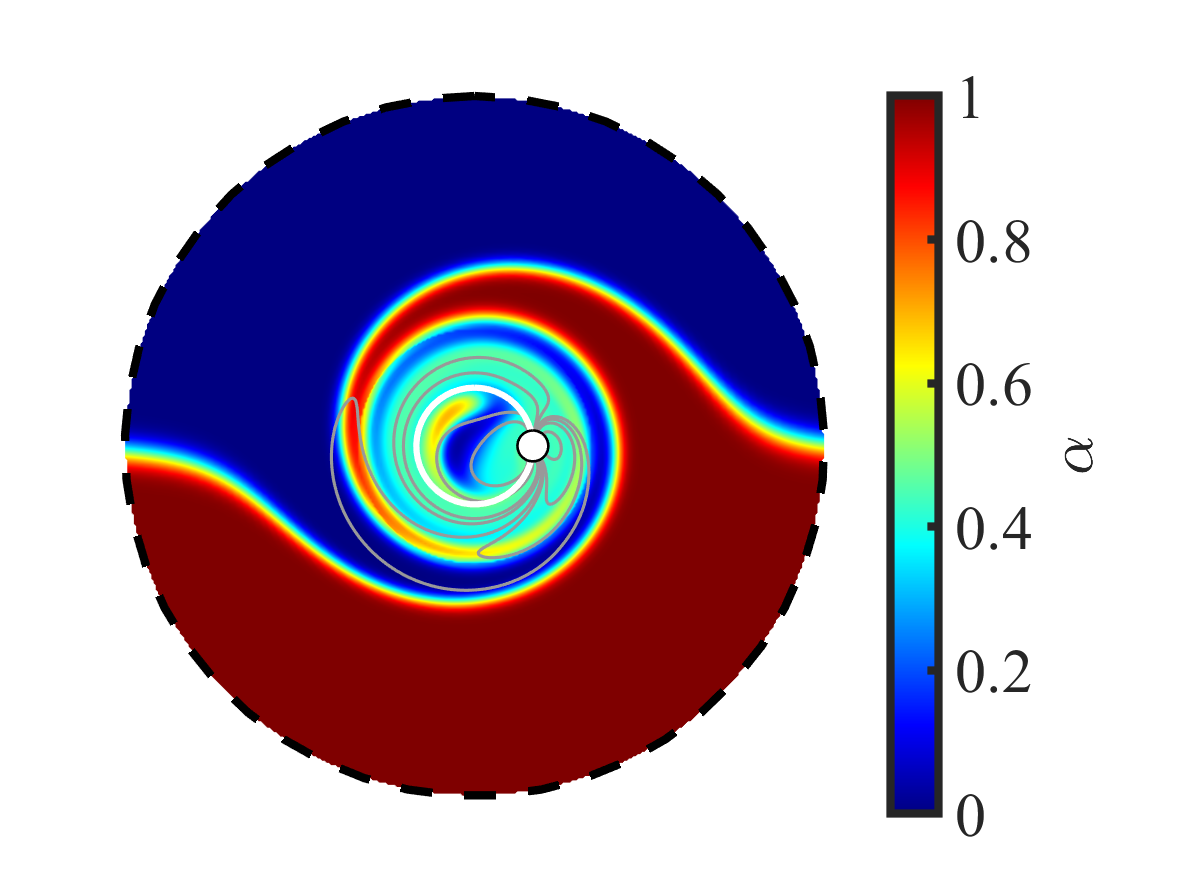}
}~
\hspace{-0.3cm}
\subfloat[$R_{sd}=8$\label{fig:alpha_Re=10_D}]{
	\includegraphics[trim=2cm 0.5cm 1.5cm 1.25cm, clip=true,height=\subfigthreeB\textwidth]{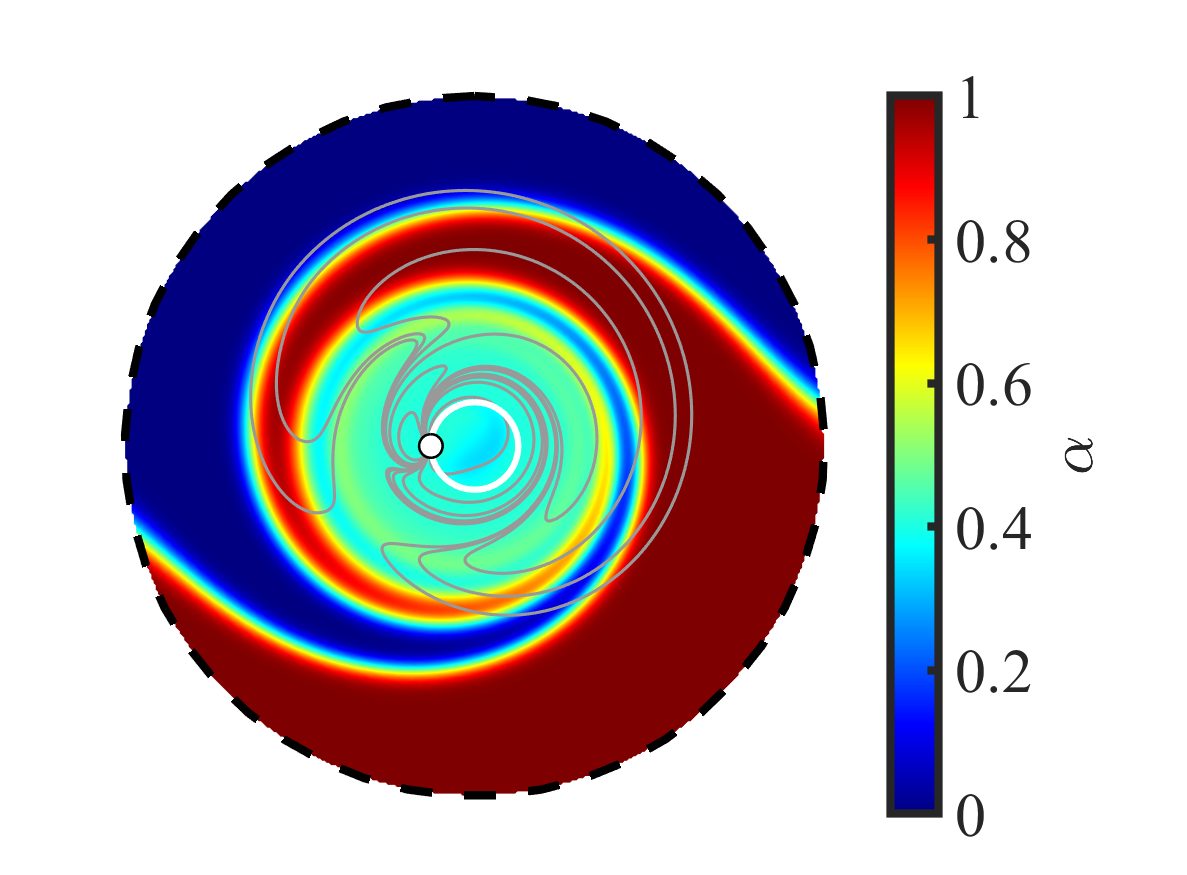}
}
\\
\vspace{-0.3cm}
\subfloat[$R_{sd}=8$\label{fig:alpha_Re=10_E}]{
	\includegraphics[trim=2cm 0.5cm 5.5cm 1.25cm, clip=true,height=\subfigthreeB\textwidth]{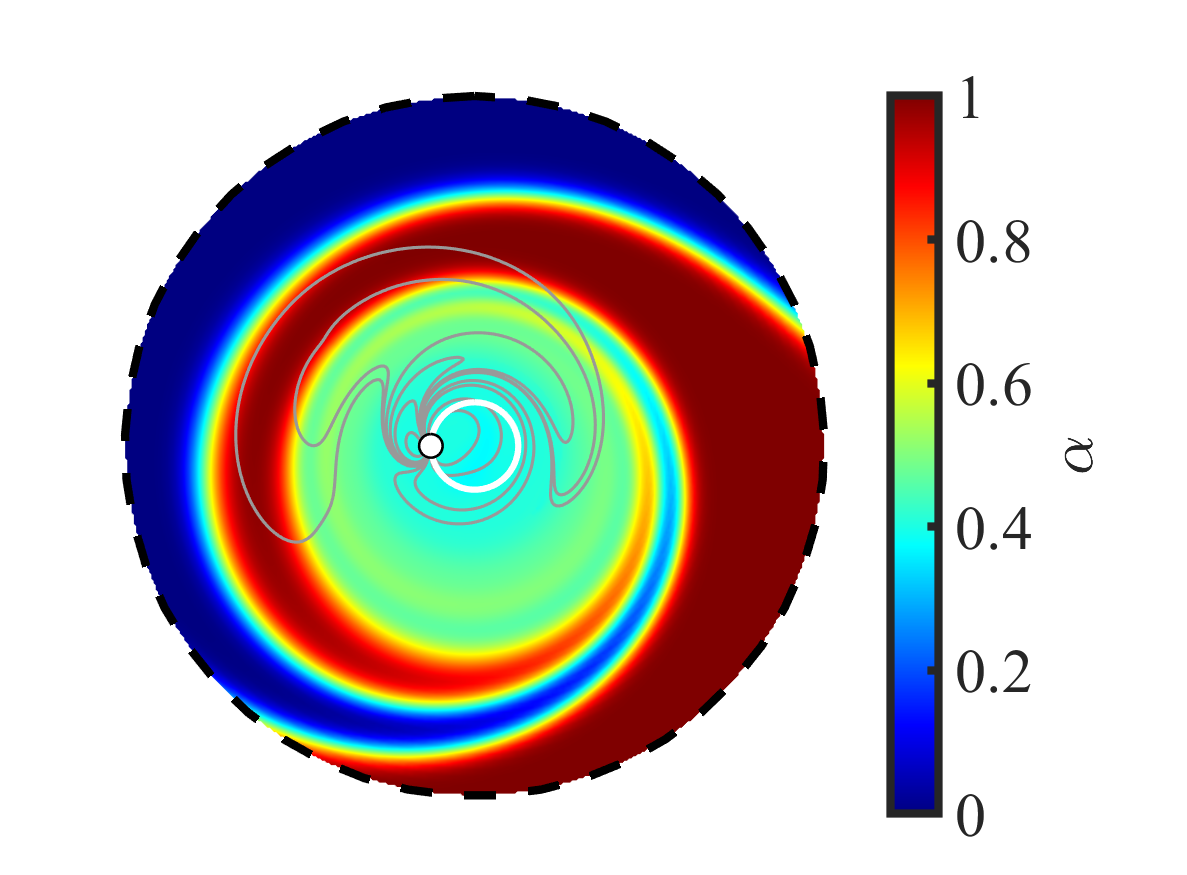}
}~
\hspace{-0.3cm}
\subfloat[$R_{sd}=8$\label{fig:alpha_Re=10_F}]{
	\includegraphics[trim=2cm 0.5cm 5.5cm 1.25cm, clip=true,height=\subfigthreeB\textwidth]{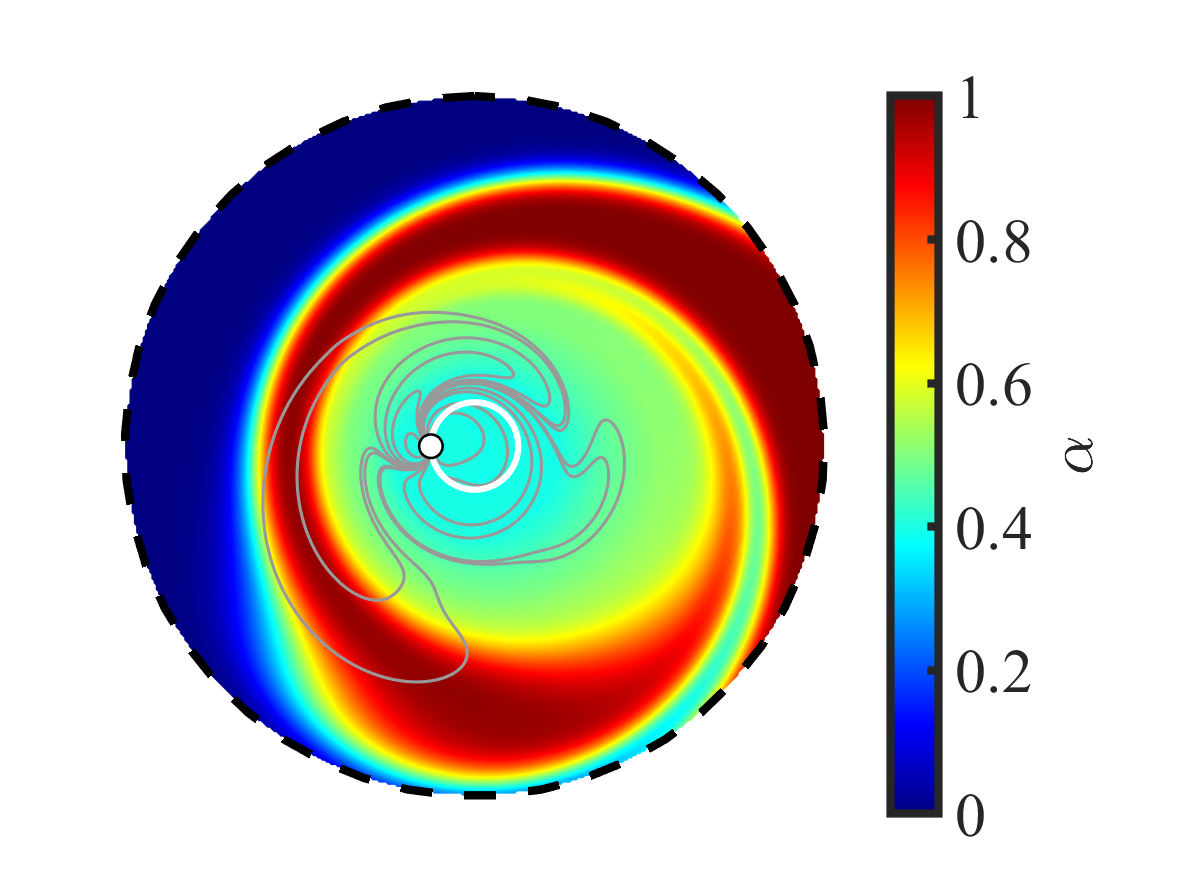}
}~
\hspace{-0.3cm}
\subfloat[$R_{sd}=12$\label{fig:alpha_Re=10_G}]{
	\includegraphics[trim=2cm 0.5cm 5.5cm 1.25cm, clip=true,height=\subfigthreeB\textwidth]{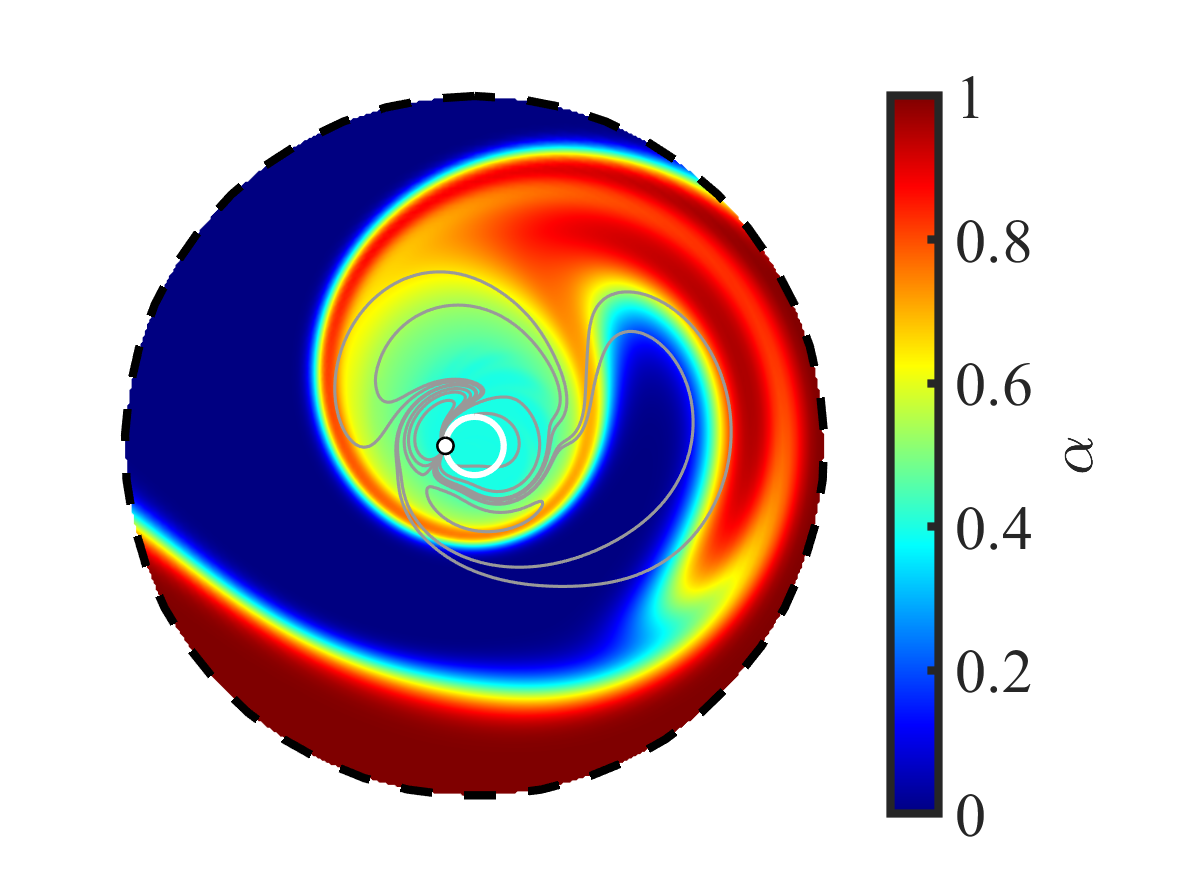}
}~
\hspace{-0.3cm}
\subfloat[$R_{sd}=18$\label{fig:alpha_Re=10_H}]{
	\includegraphics[trim=2cm 0.5cm 1.5cm 1.25cm, clip=true,height=\subfigthreeB\textwidth]{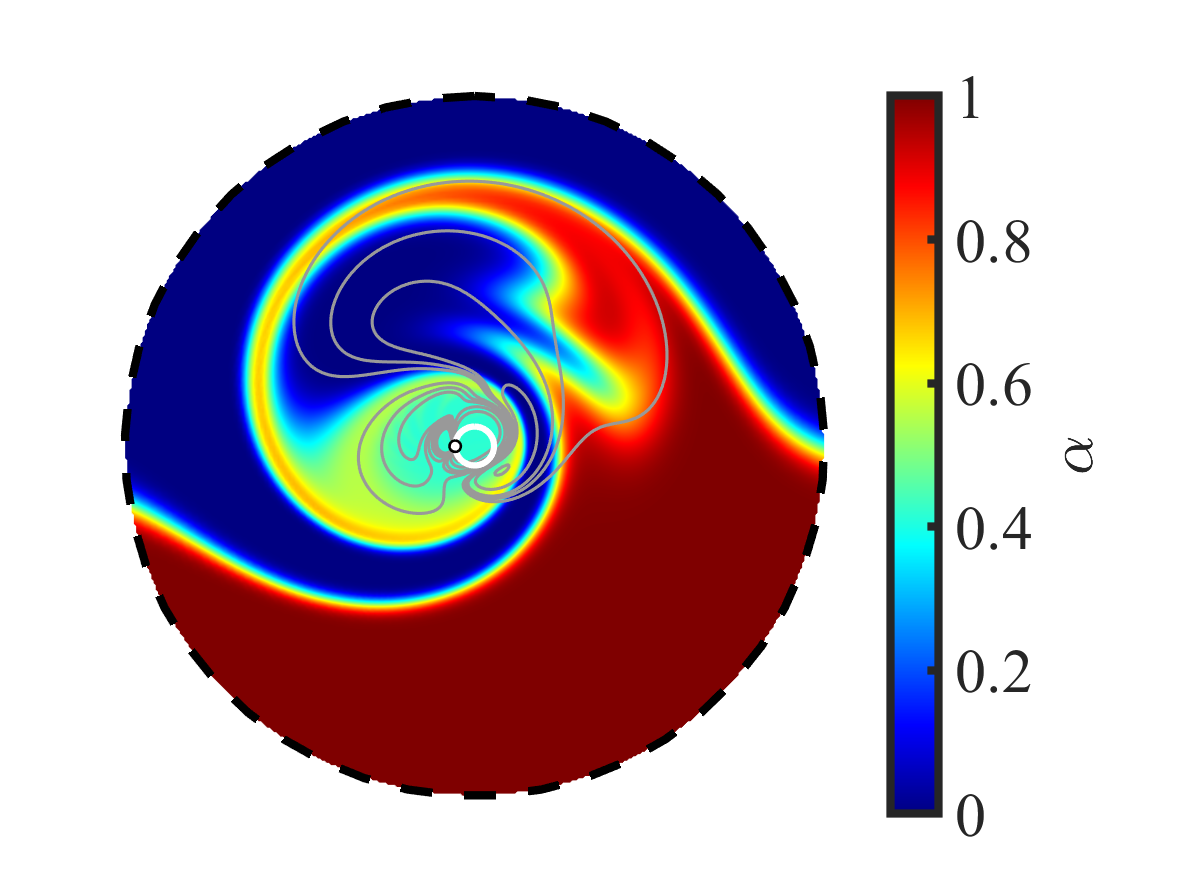}
}\\
\subfloat[$R_{sd}=3$\label{fig:vorticity_Re=10_A}]{
	\includegraphics[trim=2cm 0.5cm 5.5cm 1.25cm, clip=true,height=\subfigthreeB\textwidth]{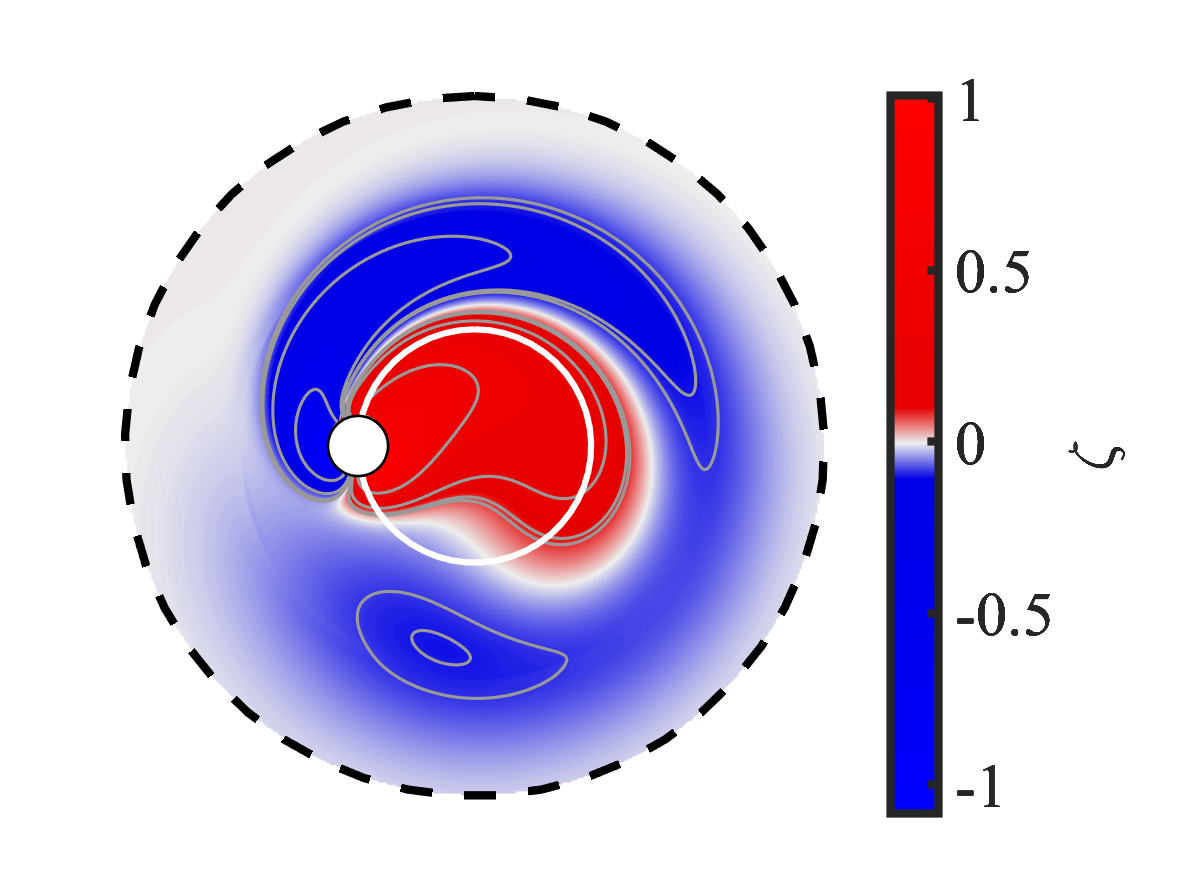}
}~
\hspace{-0.3cm}
\subfloat[$R_{sd}=3$\label{fig:vorticity_Re=10_B}]{
	\includegraphics[trim=2cm 0.5cm 5.5cm 1.25cm, clip=true,height=\subfigthreeB\textwidth]{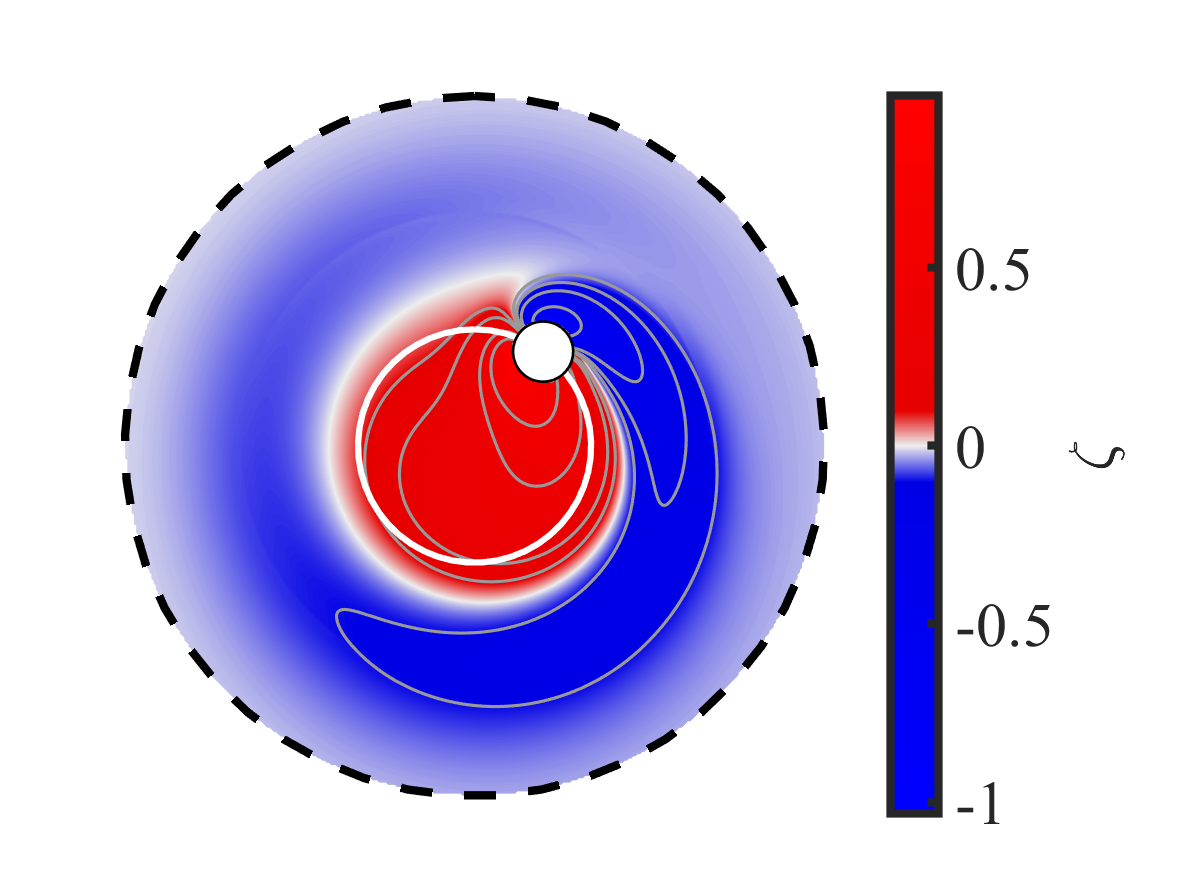}
}~
\hspace{-0.3cm}
\subfloat[$R_{sd}=6$\label{fig:vorticity_Re=10_C}]{
	\includegraphics[trim=2cm 0.5cm 5.5cm 1.25cm, clip=true,height=\subfigthreeB\textwidth]{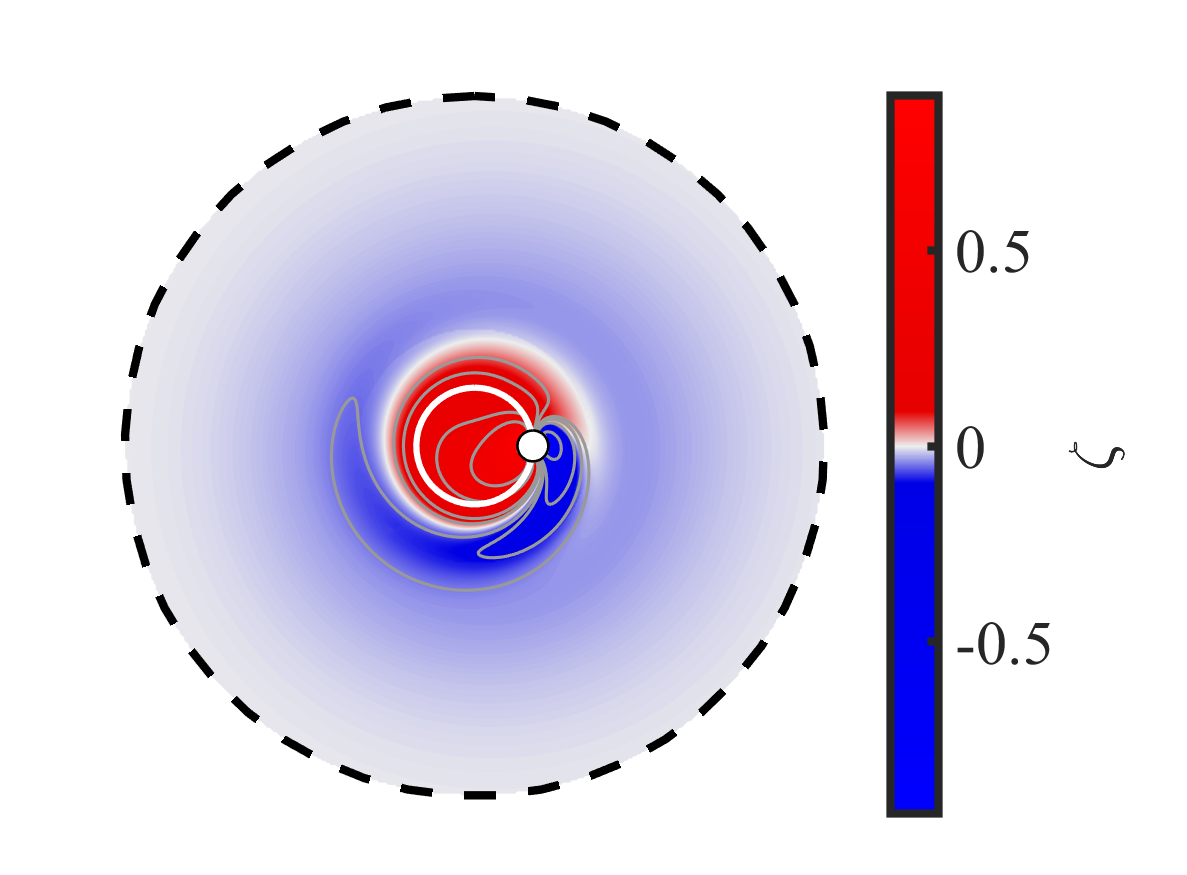}
}~
\hspace{-0.3cm}
\subfloat[$R_{sd}=8$\label{fig:vorticity_Re=10_D}]{
	\includegraphics[trim=2cm 0.5cm 5.5cm 1.25cm, clip=true,height=\subfigthreeB\textwidth]{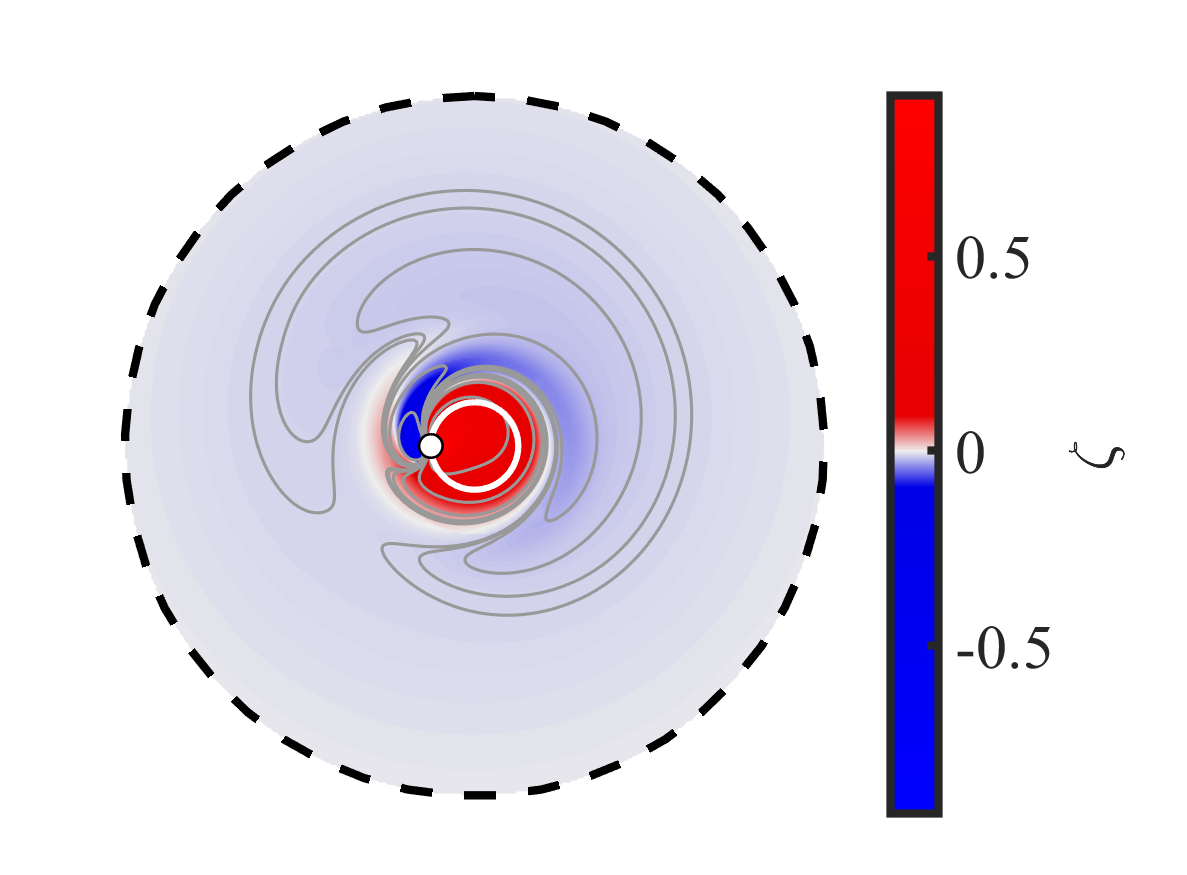}
}
\\
\vspace{-0.3cm}
\subfloat[$R_{sd}=8$\label{fig:vorticity_Re=10_E}]{
	\includegraphics[trim=2cm 0.5cm 5.5cm 1.25cm, clip=true,height=\subfigthreeB\textwidth]{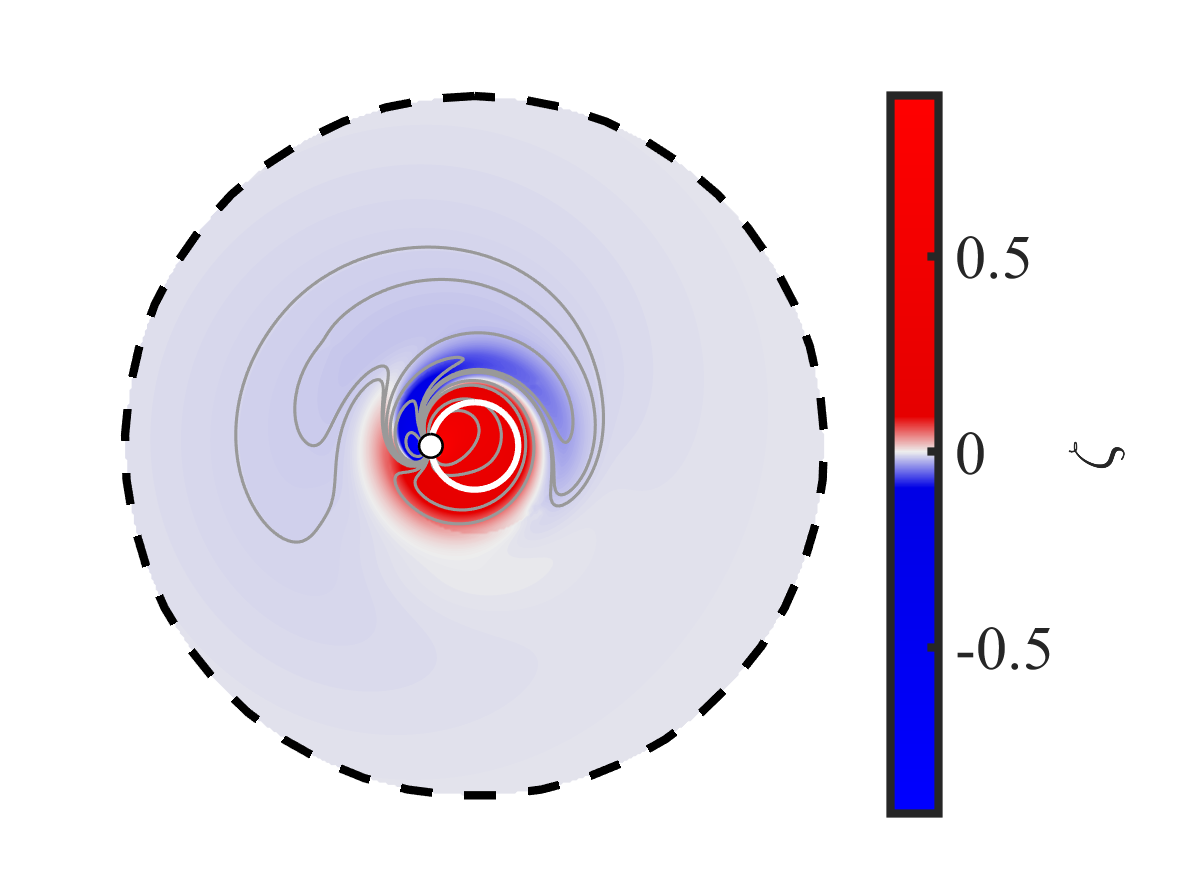}
}~
\hspace{-0.3cm}
\subfloat[$R_{sd}=8$\label{fig:vorticity_Re=10_F}]{
	\includegraphics[trim=2cm 0.5cm 5.5cm 1.25cm, clip=true,height=\subfigthreeB\textwidth]{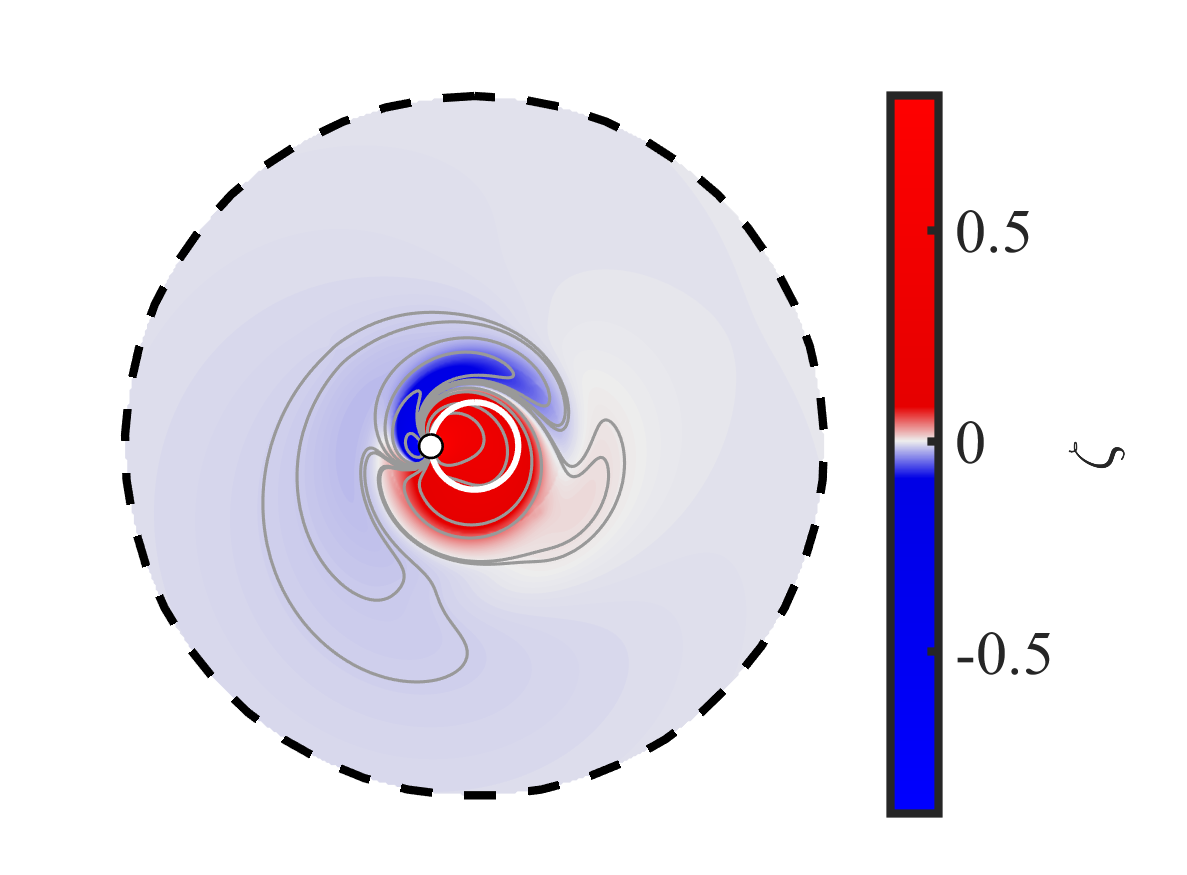}
}~
\hspace{-0.3cm}
\subfloat[$R_{sd}=12$\label{fig:vorticity_Re=10_G}]{
	\includegraphics[trim=2cm 0.5cm 5.5cm 1.25cm, clip=true,height=\subfigthreeB\textwidth]{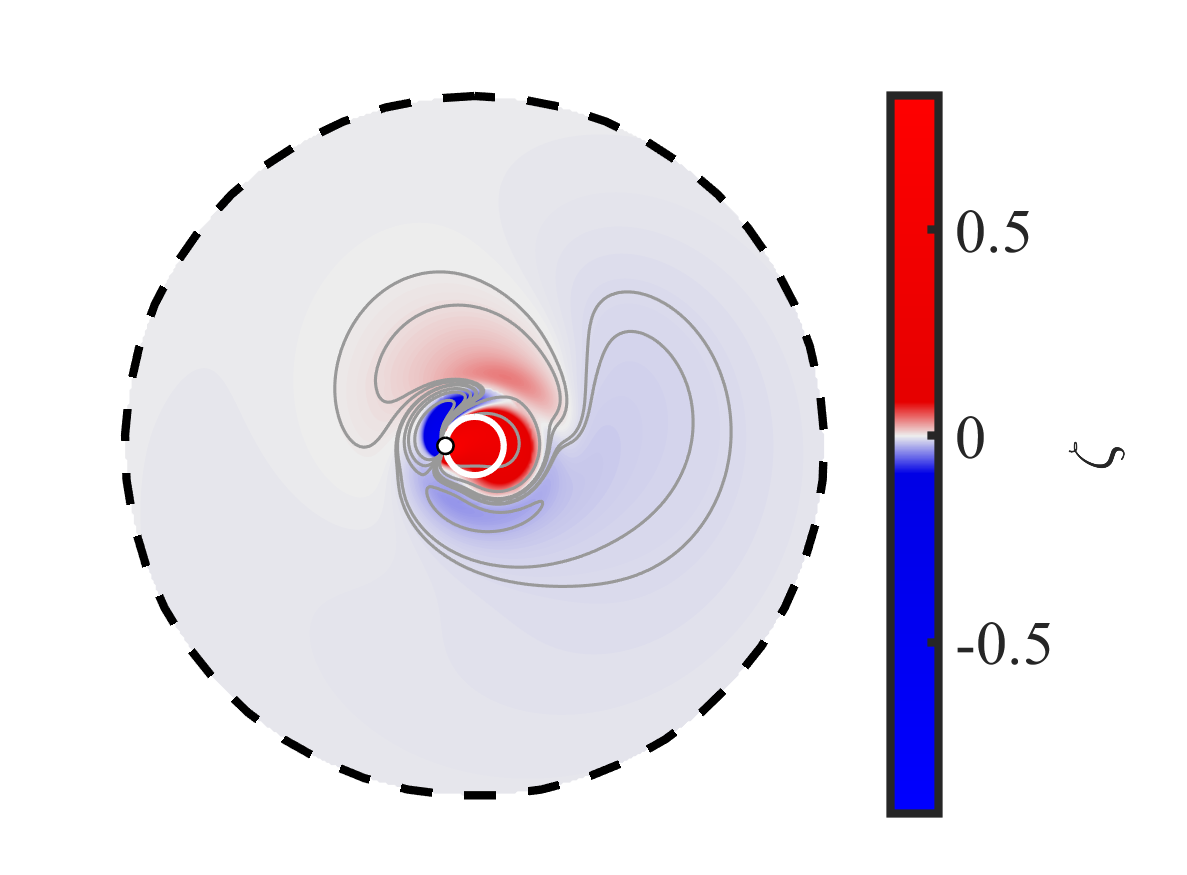}
}~
\hspace{-0.3cm}
\subfloat[$R_{sd}=18$\label{fig:vorticity_Re=10_H}]{
	\includegraphics[trim=2cm 0.5cm 5.5cm 1.25cm, clip=true,height=\subfigthreeB\textwidth]{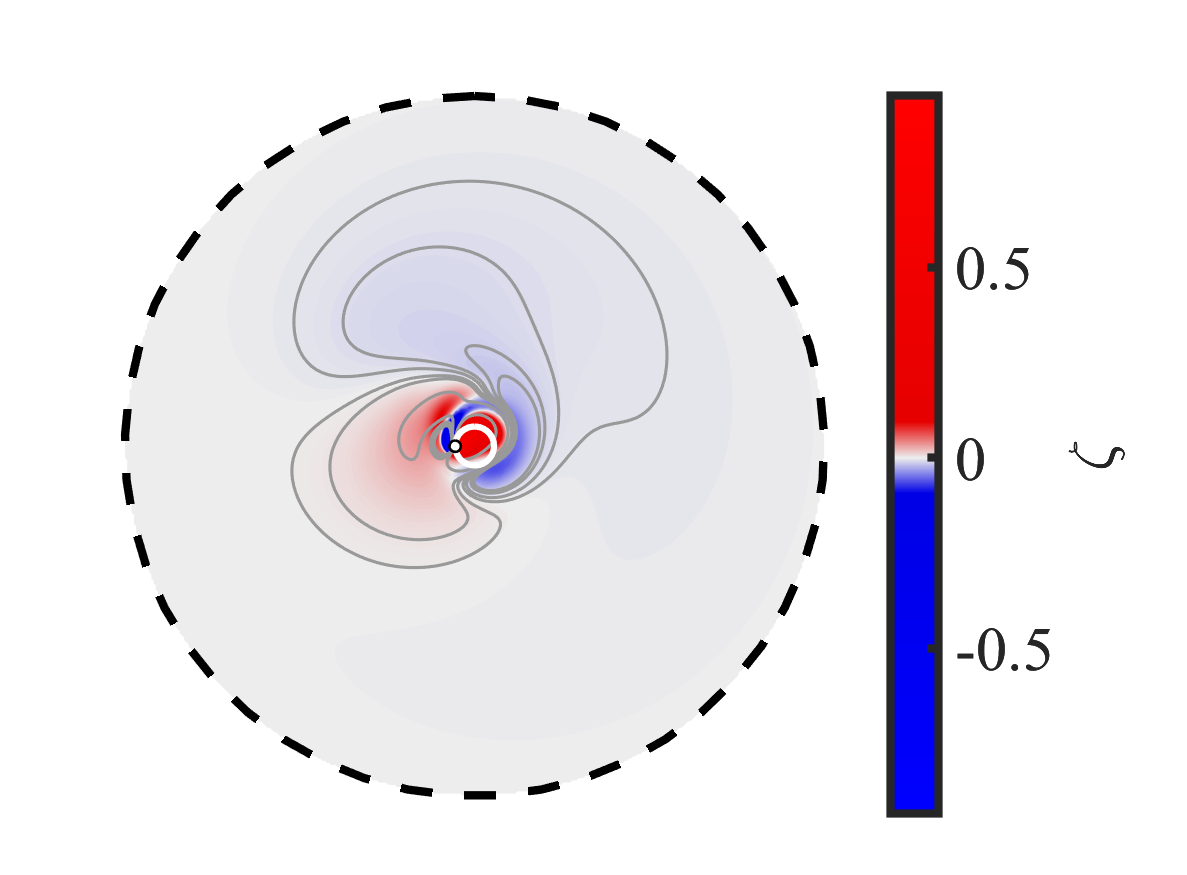}
}
\caption{{\color{black}Snapshots of the dye concentration field (a-h) and vorticity field (i-p) at times $T=1, 1.65, 7.5, 30, 45, 60, 90$, and 120. In the bottom two rows, the red, blue, and grey colors denote counterclockwise (CCW), clockwise (CW), and near-zero vorticity values, respectively. The white and grey lines indicate the stirrer’s path and the streamlines, respectively.} The radius of the field of view is provided in the caption. $Re=10${\color{black}.}
}
\label{fig:alpha_Re=10}
\end{figure}

Figure \ref{fig:alpha_Re=10} illustrates the evolution of dye concentration and \ida{the corresponding vorticity field} at $Re = 10$. The early stages of mixing resemble those at $Re = \idaa{o(1)}$. As expected, more pronounced interface deformation is observed at $Re = 10$ (Figures \ref{fig:alpha_Re=10_A} \& \ref{fig:alpha_Re=10_B}), due to the shorter advective timescale and the increased fluid acceleration behind the stirrer. This enhanced striation leads to the formation of a spiral dye pattern. While the spiral pattern persists, mixing remains dominated by cross-streamline diffusion (Figures \ref{fig:alpha_Re=10_C} \& \ref{fig:alpha_Re=10_D}). \ida{Unlike the lower-$Re$ cases, however, this spiral evolution is only transient.} The dye concentration begins to deviate from the self-similar spiral pattern at later times (see $45 \le T$ in Figure \ref{fig:alpha_Re=10}). The dye interface undergoes extensive stretching beyond the central region, while the dye-free region is advected primarily azimuthally (Figures \ref{fig:alpha_Re=10_F} - \ref{fig:alpha_Re=10_H}).

The early evolution of the vorticity field is qualitatively similar to that observed at $Re = O(1)$ (compare with $T \leq 45$ in Figure~\ref{fig:alpha_Re=5}). However, unlike in the lower-$Re$ cases, the counterclockwise (CCW) vorticity region extends beyond the central region at later times (see Figure~\ref{fig:vorticity_Re=10_F}). As the stirrer moves periodically across this extended CCW vorticity region, a new vortex pair forms \ida{outside the stirrer's path} (Figure~\ref{fig:vorticity_Re=10_G}).

We attribute the formation of this new vortex pair to two factors. First, analogous to the separation zone behind a cylinder in a free stream, the attached vortices grow in both length and width with increasing $Re$, \ida{enhancing} their interaction with the stirrer and the vortex tails.  {\color{black}Second, as $Re$ increases, the advective timescale decreases relative to the timescale of viscous dissipation, following $\hat{t}_a/\hat{t}_{\Df} \propto 1/Re$.} Consequently, the strengthening influence of stirring outweighs vorticity dissipation due to viscosity, facilitating the formation of a new vortex pair outside the stirrer’s path. This pair grows over time, advecting primarily in the azimuthal direction.

The radial extension of these \textit{satellite vortices} from the central region explains the extensive, albeit slow, deformation of the dye interface at long timescales, \ida{thereby marking the emergence of advective mixing beyond the central region} (see $T \geq 60$ in \ida{Figures~\ref{fig:alpha_Re=10_G} \& \ref{fig:alpha_Re=10_H}}).

\begin{figure}
	\centering
	\subfloat[ $R_{sd}=3$\label{fig:locationVortex_Re=10_A}]{
		\includegraphics[trim=0cm 0cm 0cm 0cm, clip=true,height=.235\textwidth]{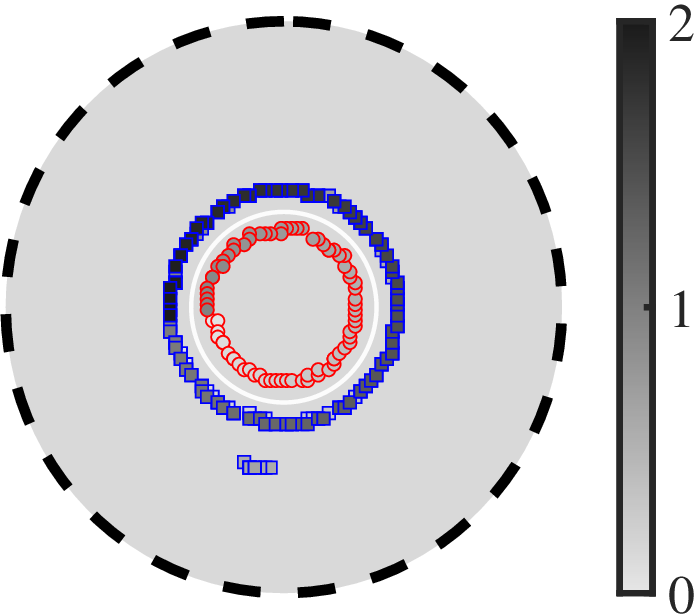}
	}~
	\subfloat[ $R_{sd}=4$\label{fig:locationVortex_Re=10_B}]{
		\includegraphics[trim=0cm 0cm 0cm 0cm, clip=true,height=.235\textwidth]{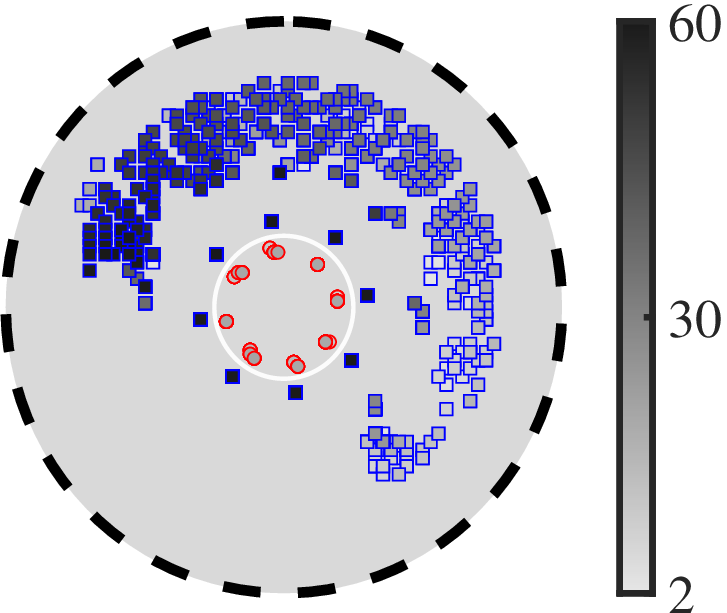}
	}~
	\subfloat[$R_{sd}=4$\label{fig:locationVortex_Re=10_C}]{
		\includegraphics[trim=0cm 0cm 0cm 0cm, clip=true,height=.235\textwidth]{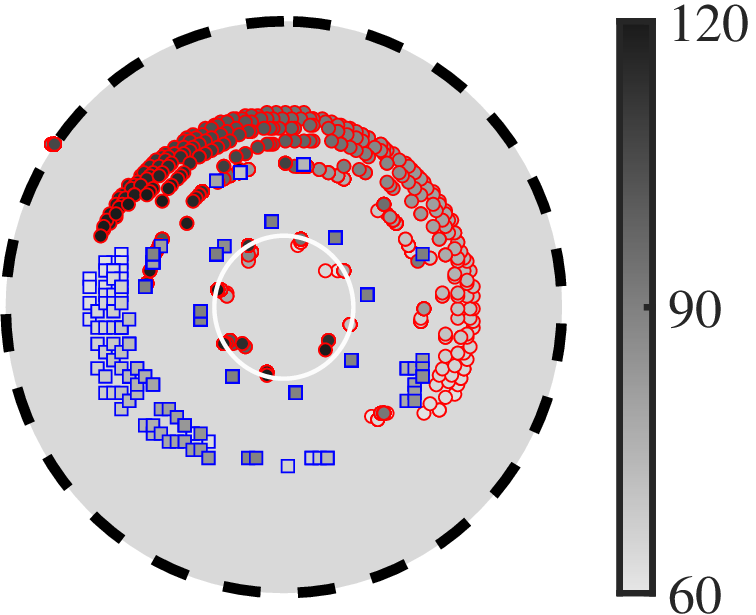}
	} 
	
	\caption{Time evolution of the vortex centers at different time intervals. The square and circle marks represent center of CW and CCW vortices, respectively. Lighter shades correspond to earlier times. The white solid line represents the stirrer’s path and the black dashed lines denote the subdomain boundary. The radius of the subdomain is presented in the captions. $Re=10$.}
	\label{fig:locationVortex_Re=10}
\end{figure}

The formation and evolution of these satellite vortices are illustrated more clearly in Figure \ref{fig:locationVortex_Re=10}. Blue square and red circular markers indicate the centers of clockwise (CW) and CCW vortices, respectively. The square markers near the stirrer’s path and the circular markers just inside it correspond to the CW and CCW vortices attached to the stirrer. The outermost square markers in Figure \ref{fig:locationVortex_Re=10_A} highlight the shedding of CW vortices shortly after the onset of stirring.
Figure~\ref{fig:locationVortex_Re=10_B} illustrates the  slow azimuthal drift of the shed CW vortex around the stirrer’s path. Similarly, Figure~\ref{fig:locationVortex_Re=10_C} shows that at $Re = 10$, CCW vortices persist outside the stirrer’s path around $60 \lesssim T$. Like the satellite CW vortices, these CCW vortices advect \ida{predominantly in the azimuthal direction} at a speed much slower than the stirrer, \ida{thereby} explaining the extensive yet slow deformation of the dye interface outside the central region during the later stages of mixing.

\begin{figure}
	\centering
	\subfloat[$R_{sd}=3$\label{fig:alpha_Re=20_A}]{
		\includegraphics[trim=2cm 0.5cm 5.5cm 1.25cm, clip=true,height=\subfigthreeB\textwidth]{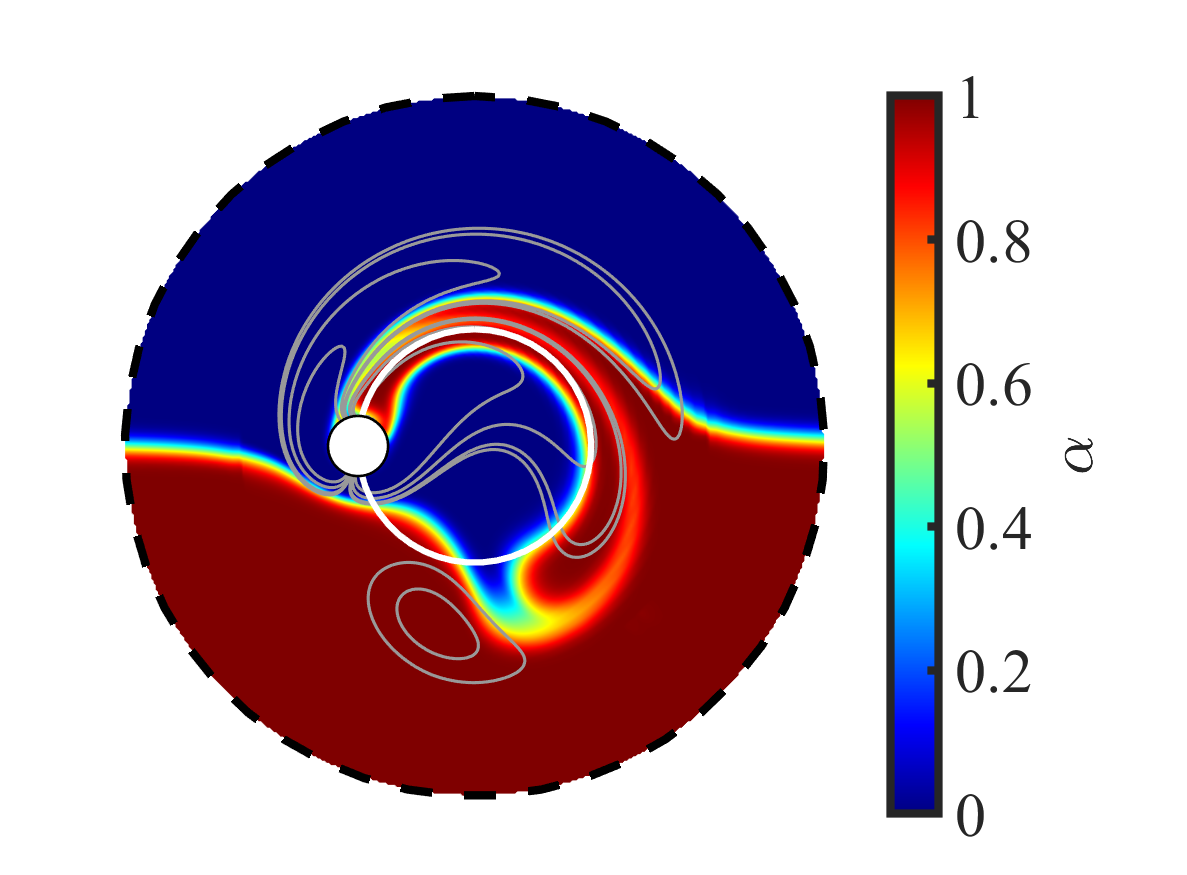}
	}~
	\hspace{-0.3cm}
	\subfloat[$R_{sd}=3$\label{fig:alpha_Re=20_B}]{
		\includegraphics[trim=2cm 0.5cm 5.5cm 1.25cm, clip=true,height=\subfigthreeB\textwidth]{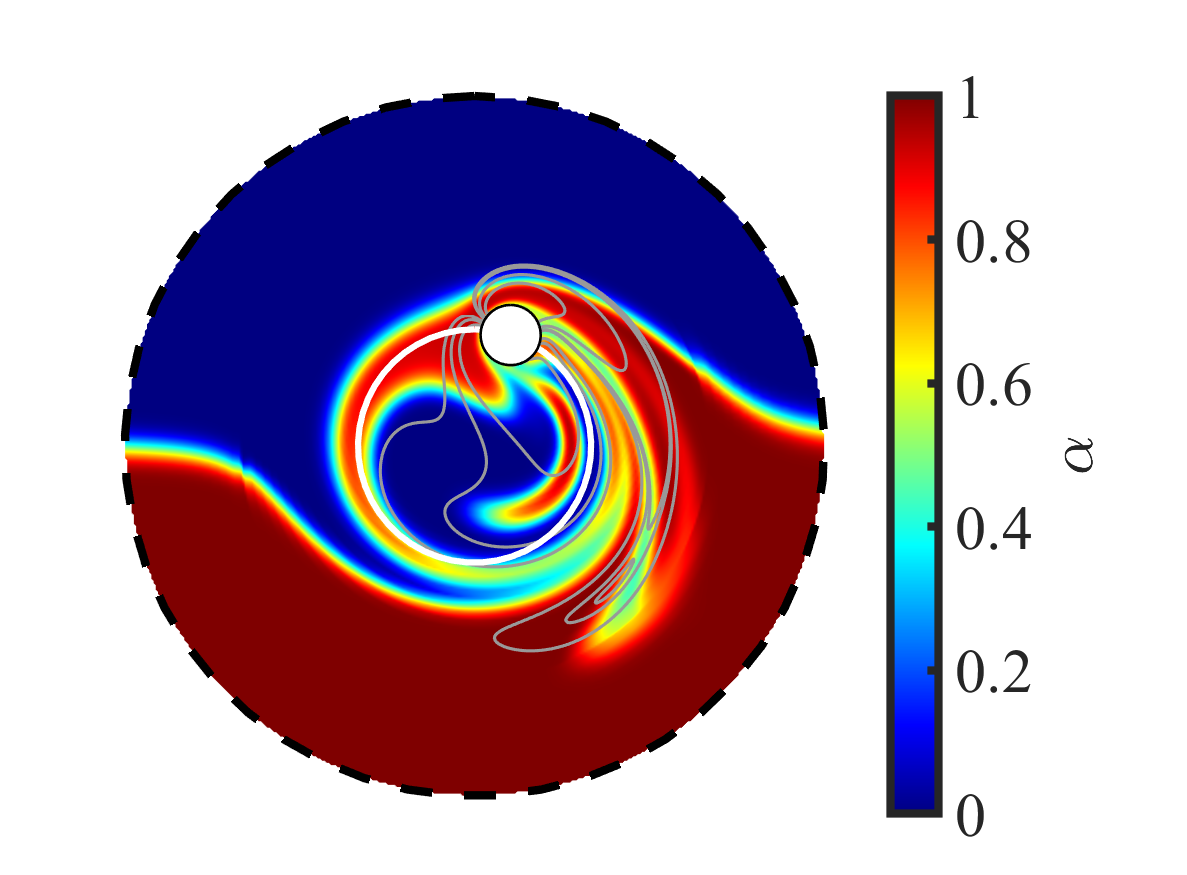}
	}~
	\hspace{-0.3cm}
	\subfloat[$R_{sd}=6$\label{fig:alpha_Re=20_C}]{
		\includegraphics[trim=2cm 0.5cm 5.5cm 1.25cm, clip=true,height=\subfigthreeB\textwidth]{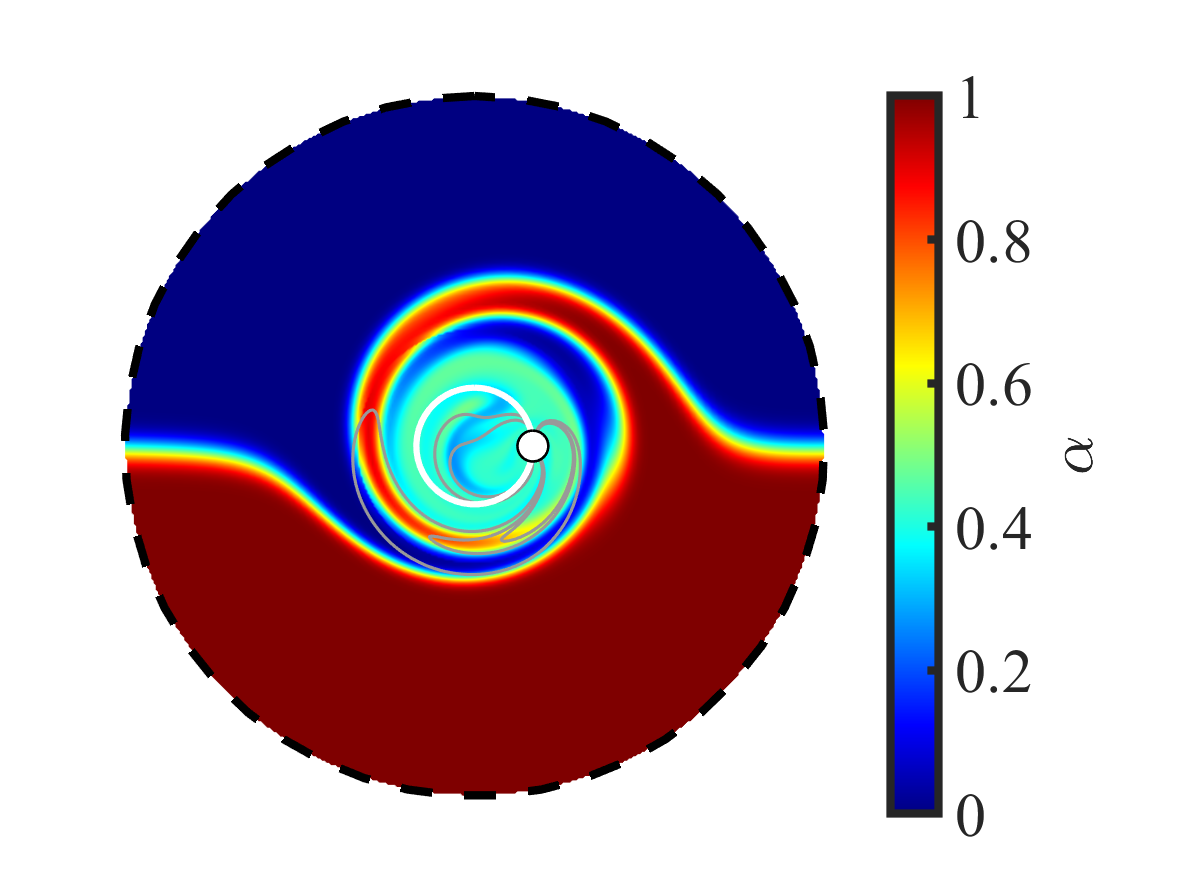}
	}~
	\hspace{-0.3cm}
	\subfloat[$R_{sd}=6$\label{fig:alpha_Re=20_D}]{
		\includegraphics[trim=2cm 0.5cm 1.5cm 1.25cm, clip=true,height=\subfigthreeB\textwidth]{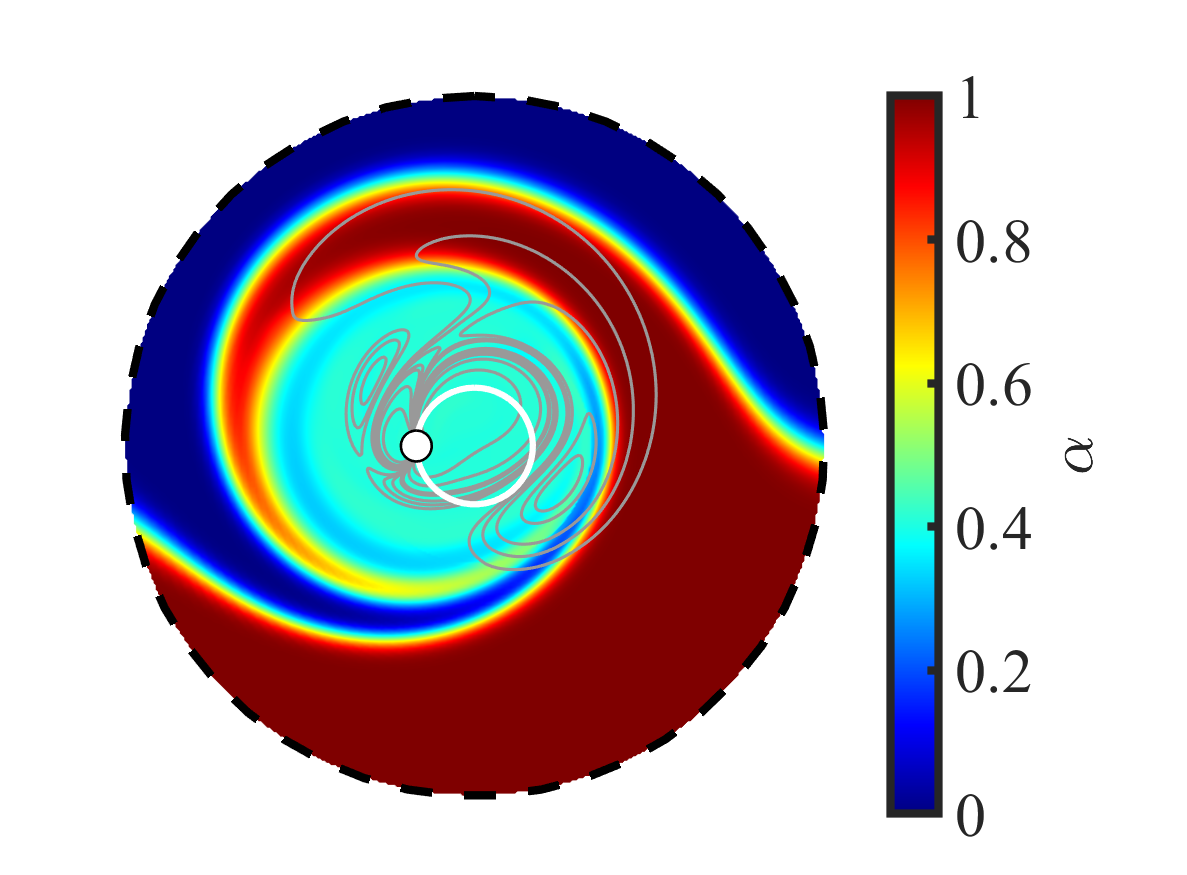}
	}\\
	\vspace{-0.3cm}
	\subfloat[$R_{sd}=9$\label{fig:alpha_Re=20_E}]{
		\includegraphics[trim=2cm 0.5cm 5.5cm 1.25cm, clip=true,height=\subfigthreeB\textwidth]{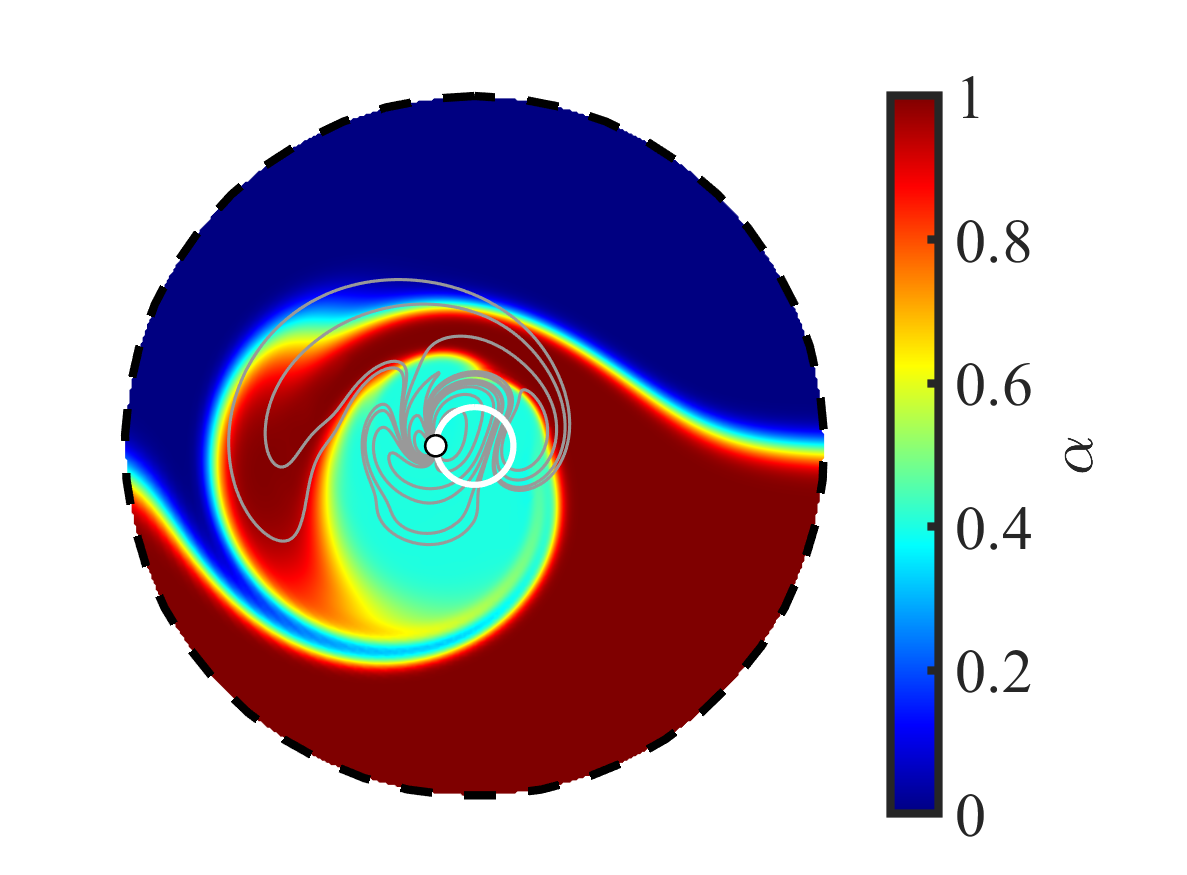}
	}~
	\hspace{-0.3cm}
	\subfloat[$R_{sd}=9$\label{fig:alpha_Re=20_F}]{
		\includegraphics[trim=2cm 0.5cm 5.5cm 1.25cm, clip=true,height=\subfigthreeB\textwidth]{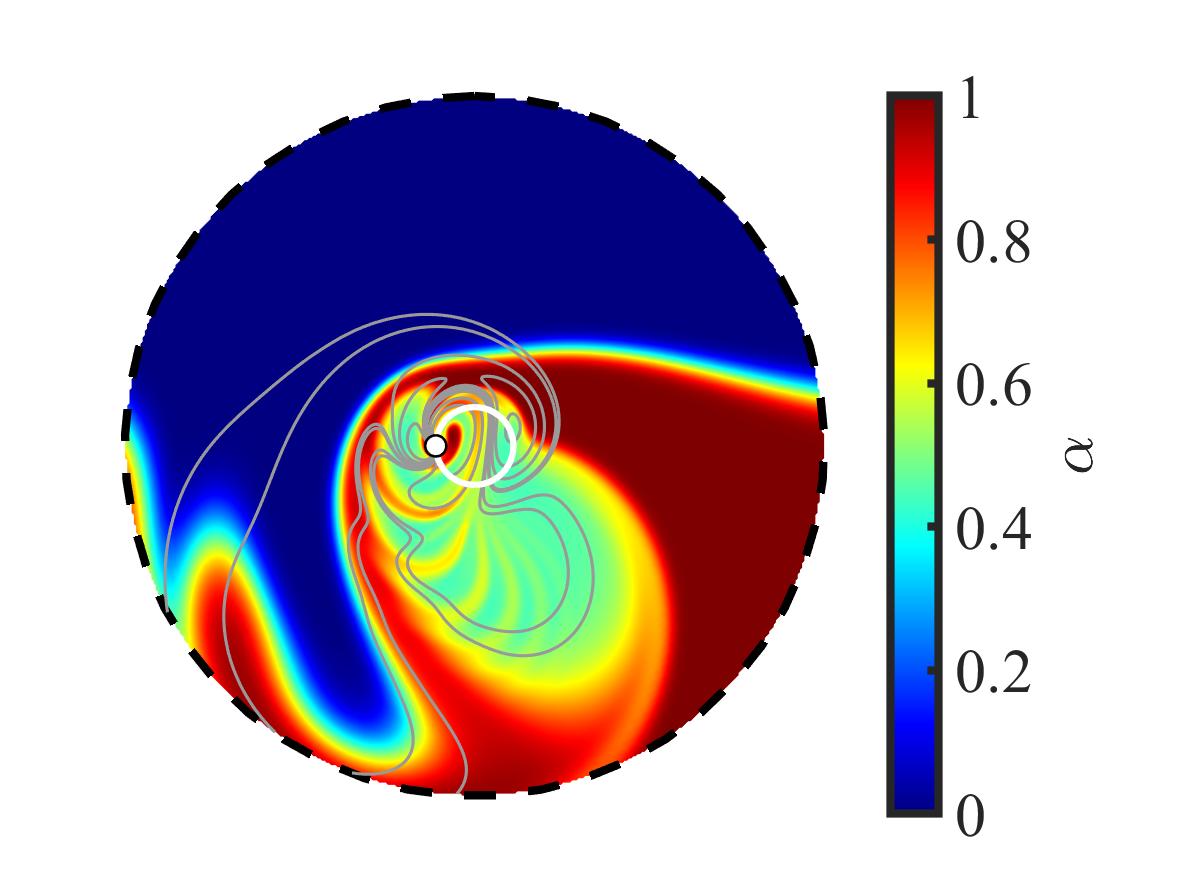}
	}~
	\hspace{-0.3cm}
	\subfloat[$R_{sd}=9$\label{fig:alpha_Re=20_G}]{
		\includegraphics[trim=2cm 0.5cm 5.5cm 1.25cm, clip=true,height=\subfigthreeB\textwidth]{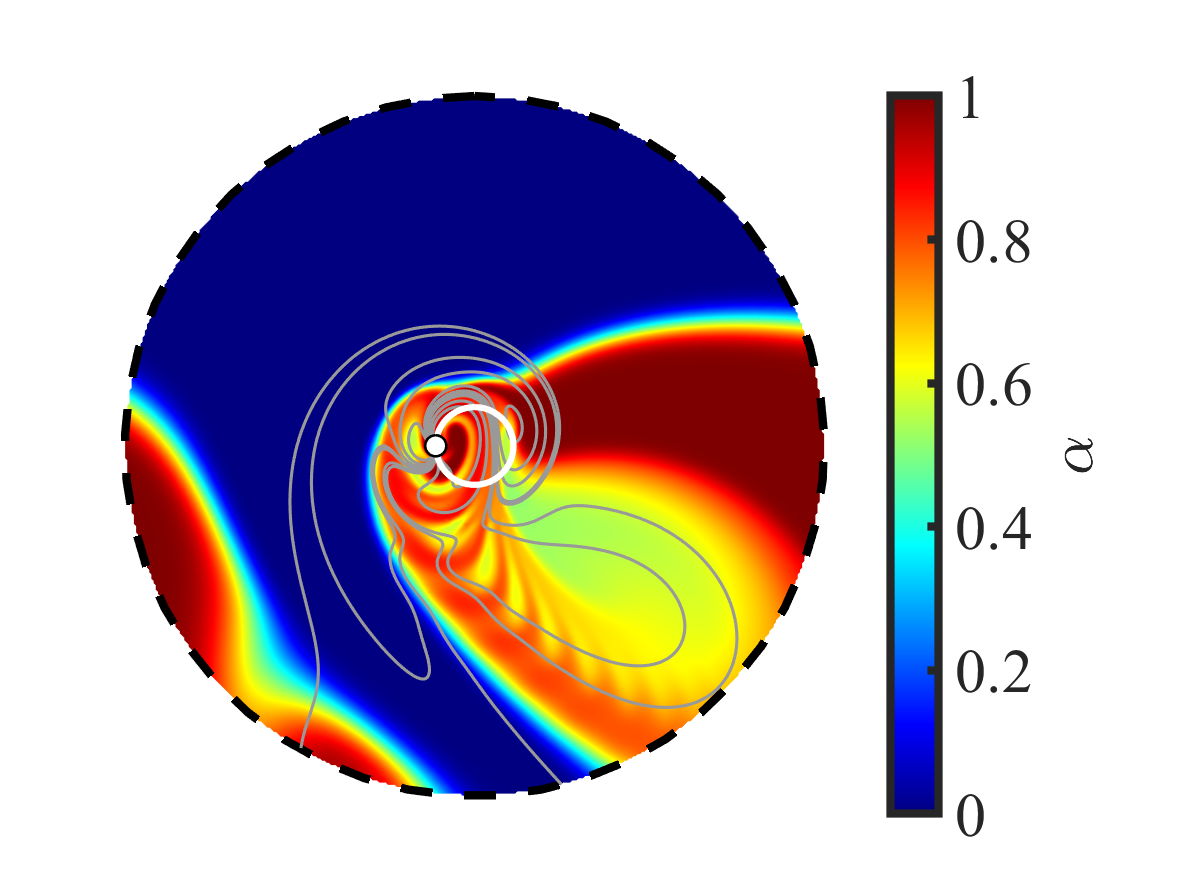}
	}~
	\hspace{-0.3cm}
	\subfloat[$R_{sd}=25$\label{fig:alpha_Re=20_H}]{
		\includegraphics[trim=2cm 0.5cm 1.5cm 1.25cm, clip=true,height=\subfigthreeB\textwidth]{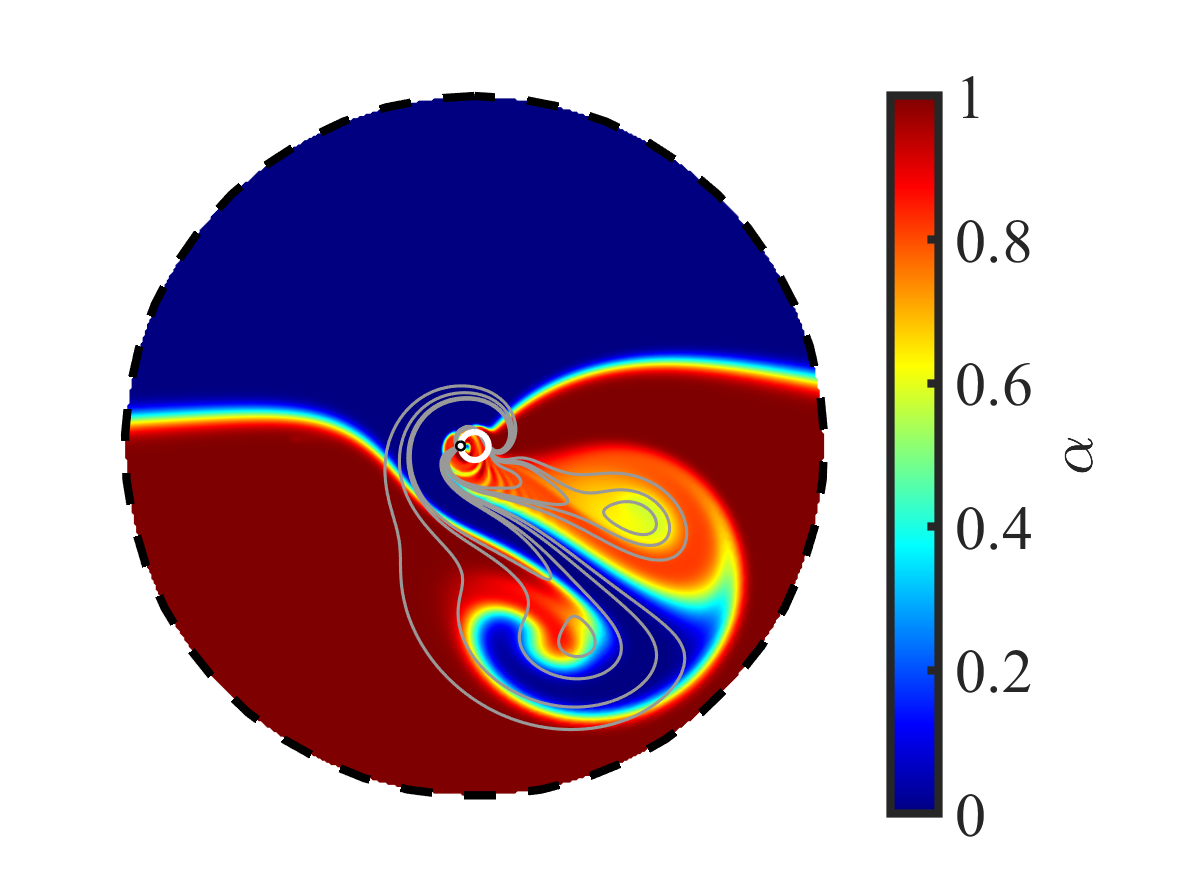}
	}
	\\
	\subfloat[$R_{sd}=3$\label{fig:vorticity_Re=20_A}]{
		\includegraphics[trim=2cm 0.5cm 5.5cm 1.25cm, clip=true,height=\subfigthreeB\textwidth]{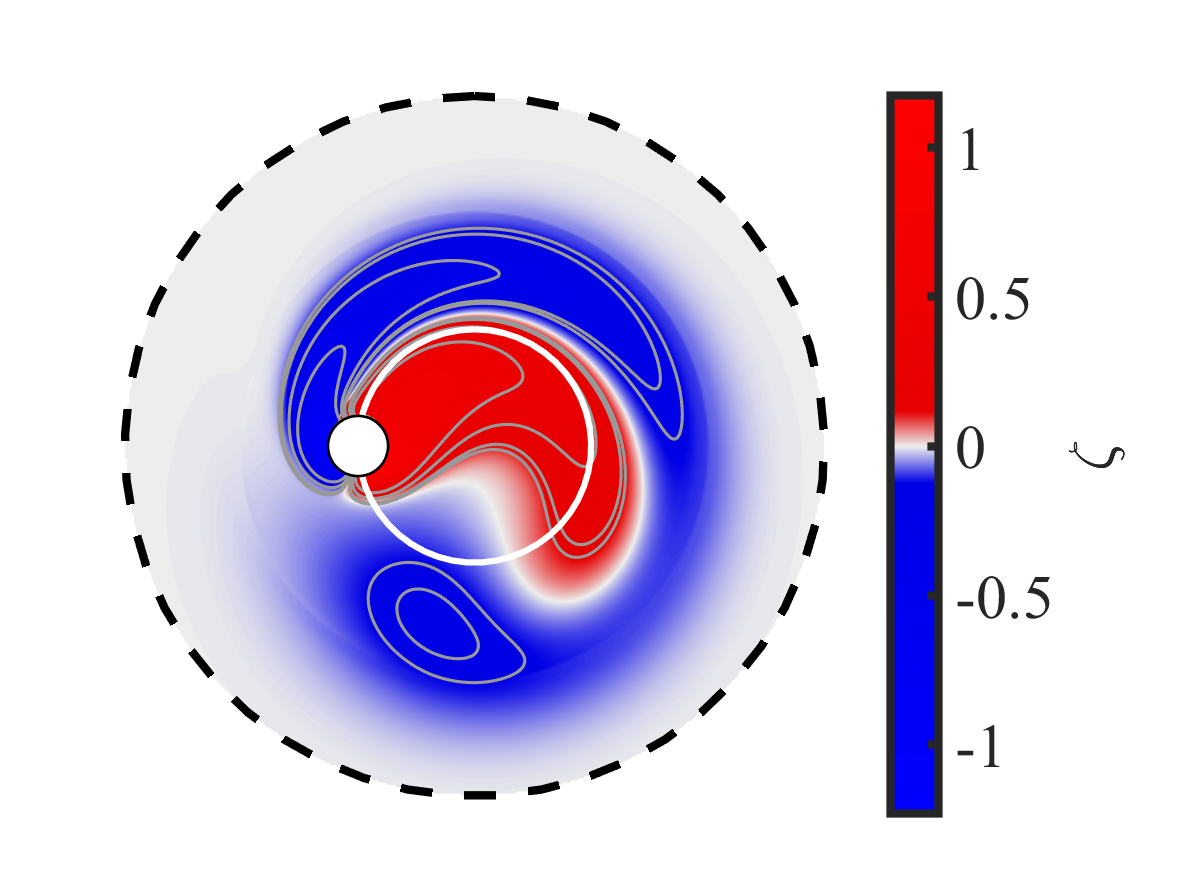}
	}~
	\hspace{-0.3cm}
	\subfloat[$R_{sd}=3$\label{fig:vorticity_Re=20_B}]{
		\includegraphics[trim=2cm 0.5cm 5.5cm 1.25cm, clip=true,height=\subfigthreeB\textwidth]{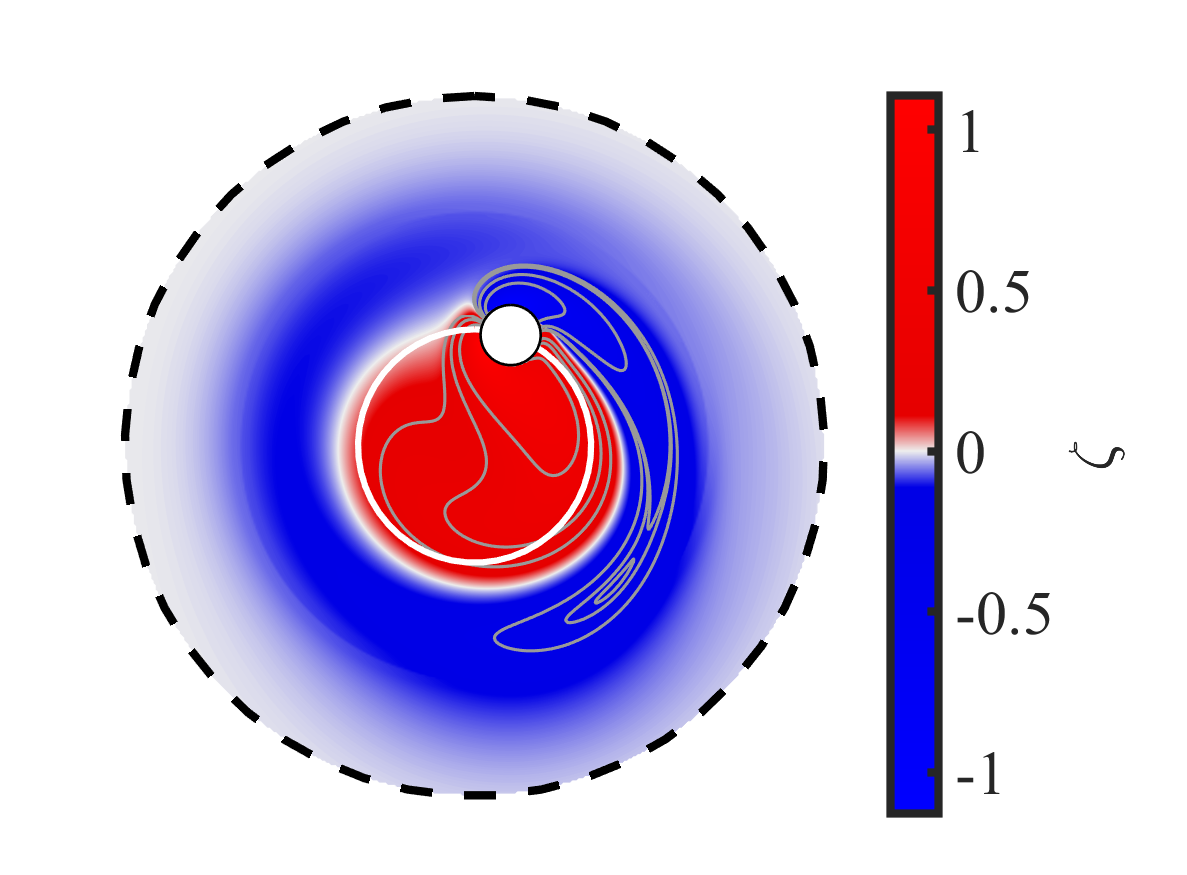}
	}~
	\hspace{-0.3cm}
	\subfloat[$R_{sd}=6$\label{fig:vorticity_Re=20_C}]{
		\includegraphics[trim=2cm 0.5cm 5.5cm 1.25cm, clip=true,height=\subfigthreeB\textwidth]{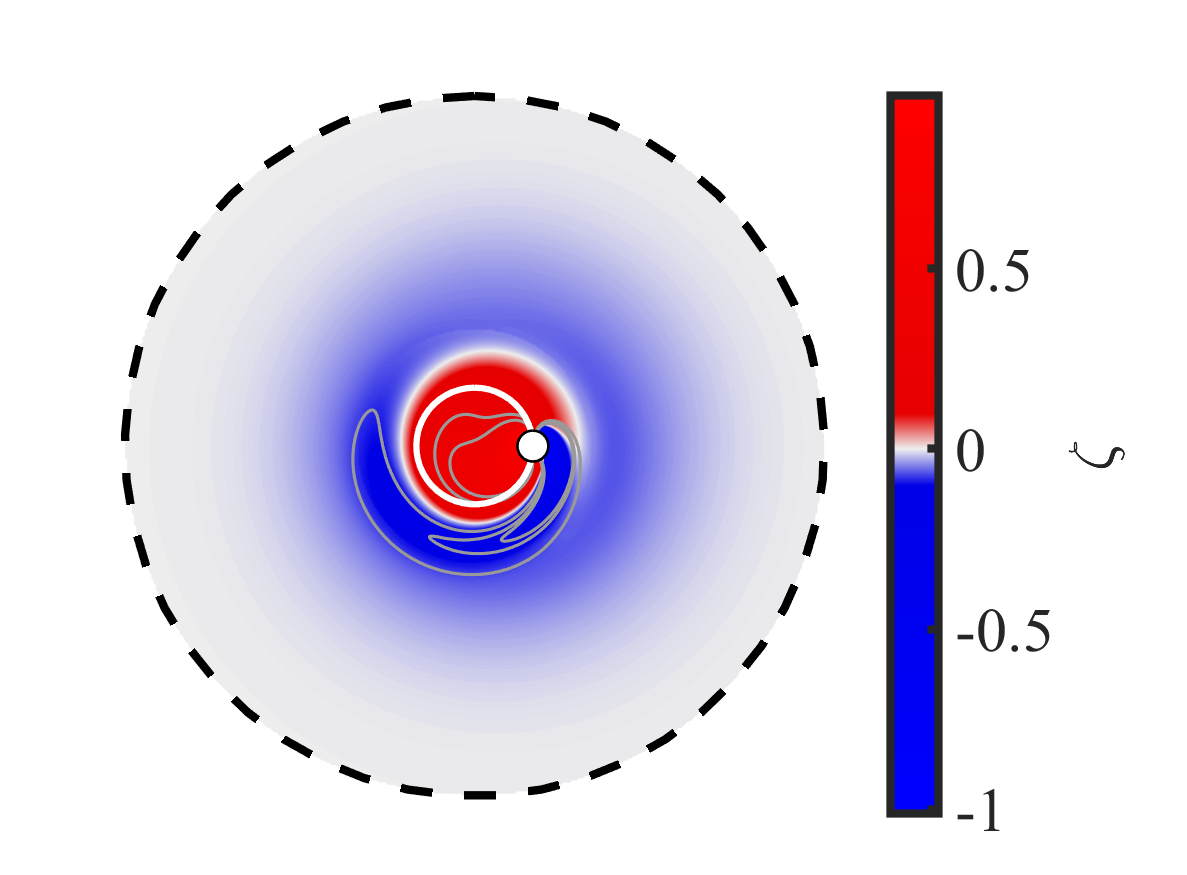}
	}~
	\hspace{-0.3cm}
	\subfloat[$R_{sd}=6$\label{fig:vorticity_Re=20_D}]{
		\includegraphics[trim=2cm 0.5cm 5.5cm 1.25cm, clip=true,height=\subfigthreeB\textwidth]{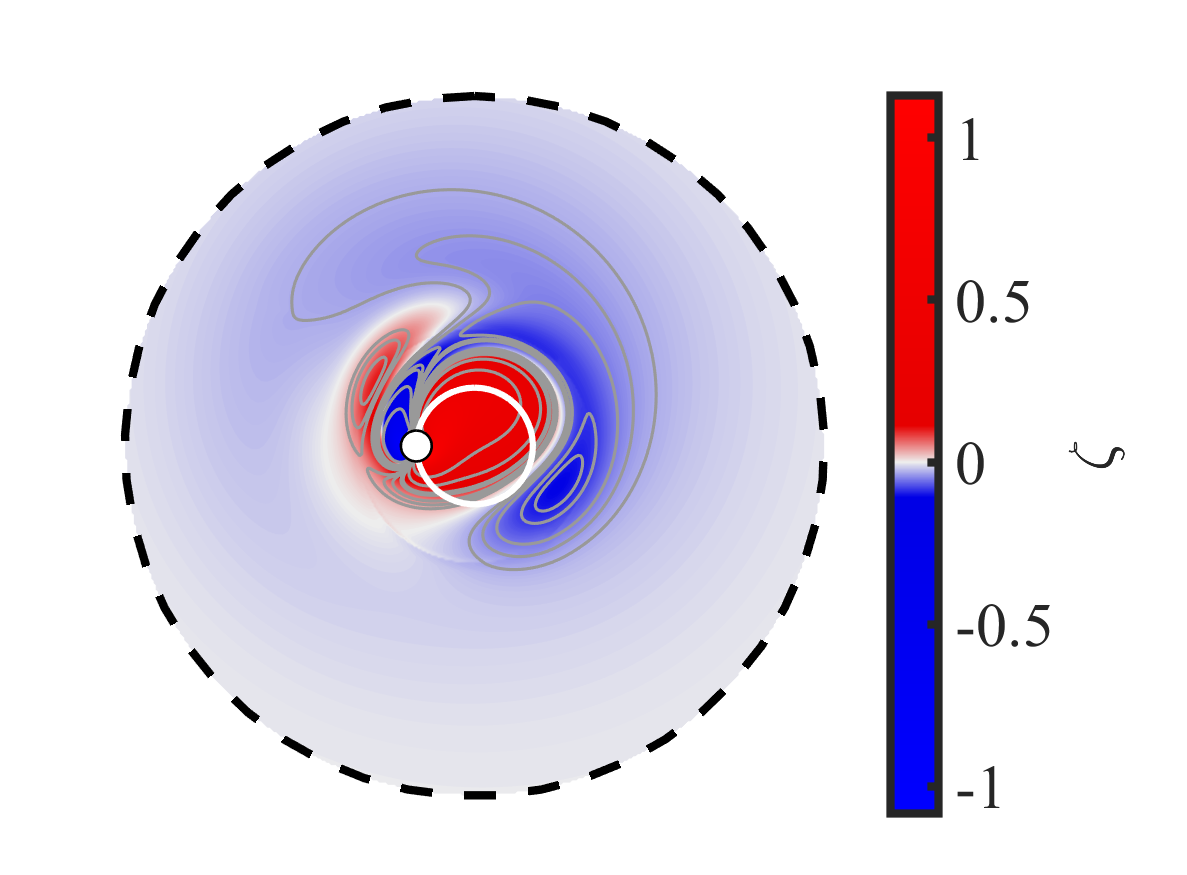}
	}\\
	\vspace{-0.3cm}
	\subfloat[$R_{sd}=9$\label{fig:vorticity_Re=20_E}]{
		\includegraphics[trim=2cm 0.5cm 5.5cm 1.25cm, clip=true,height=\subfigthreeB\textwidth]{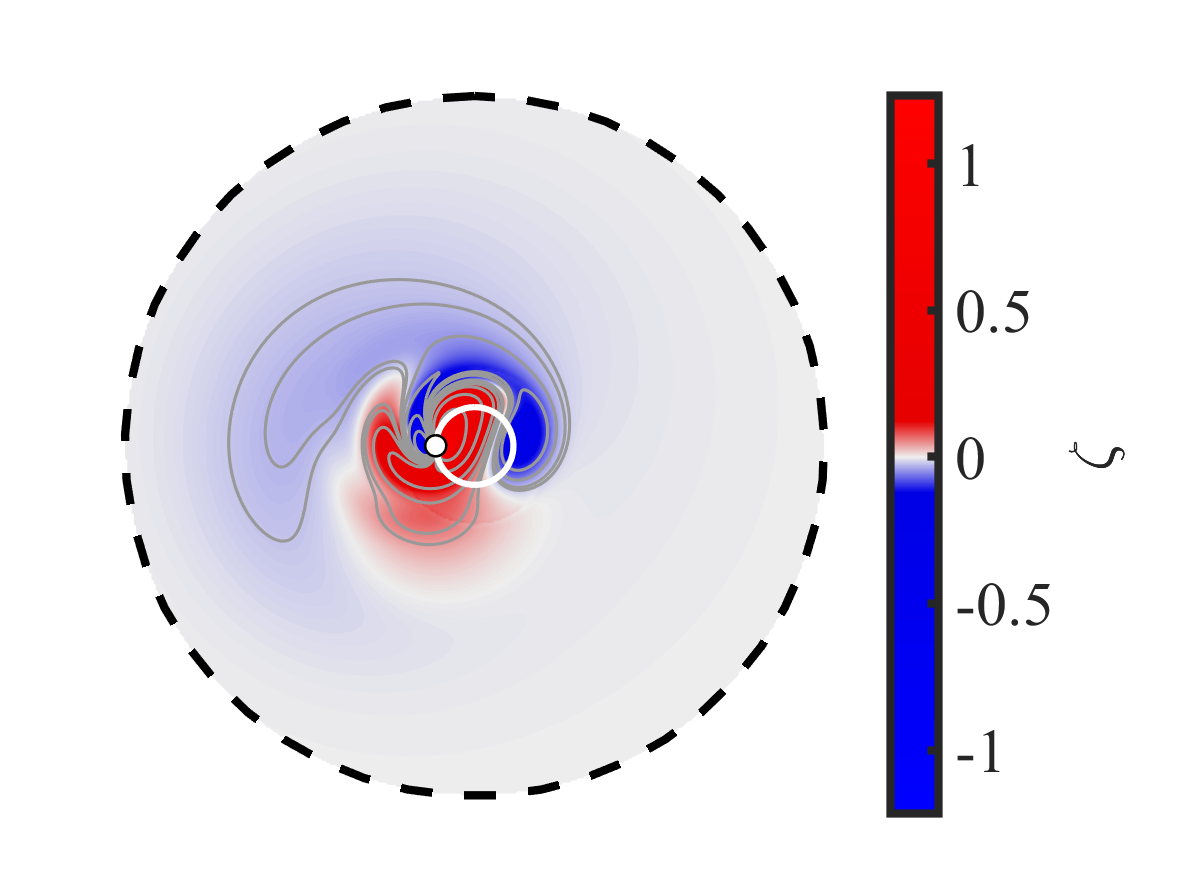}
	}~
	\hspace{-0.3cm}
	\subfloat[$R_{sd}=9$\label{fig:vorticity_Re=20_F}]{
		\includegraphics[trim=2cm 0.5cm 5.5cm 1.25cm, clip=true,height=\subfigthreeB\textwidth]{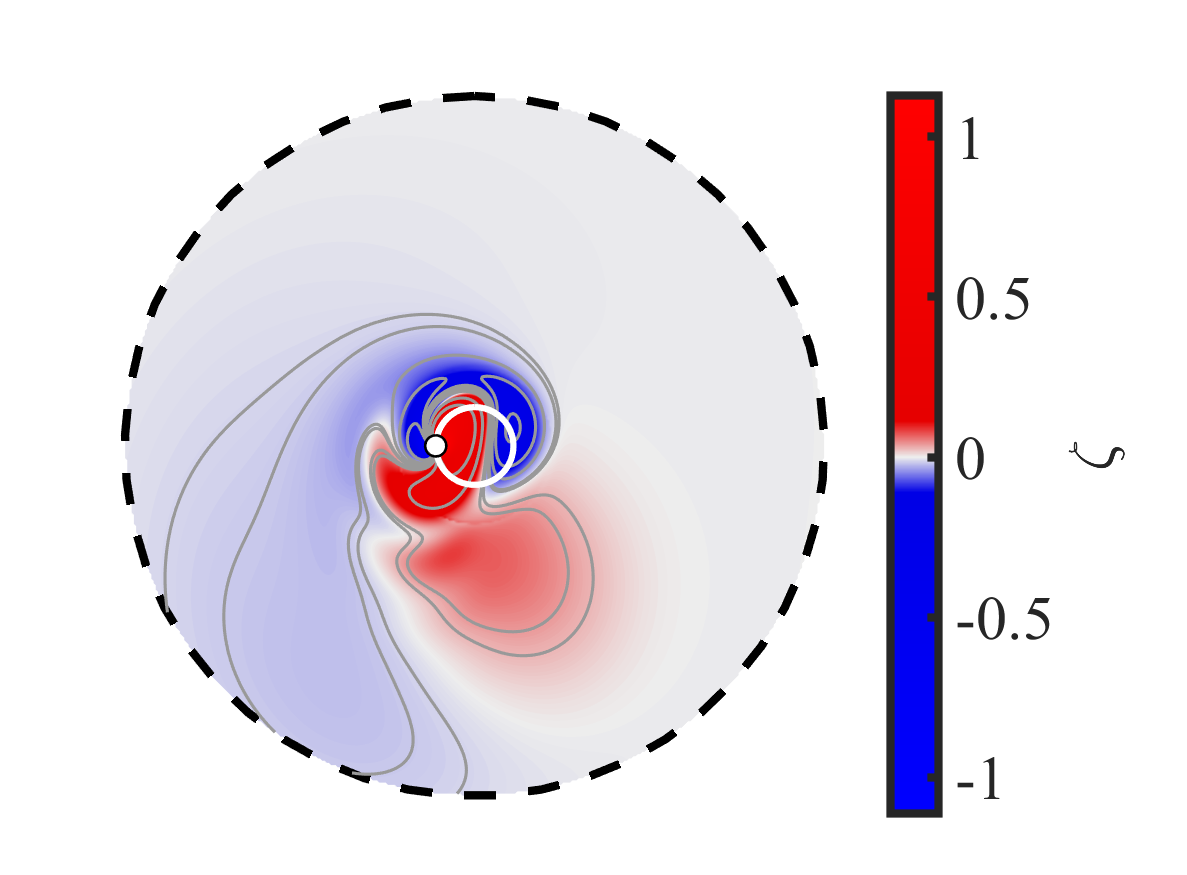}
	}~
	\hspace{-0.3cm}
	\subfloat[$R_{sd}=9$\label{fig:vorticity_Re=20_G}]{
		\includegraphics[trim=2cm 0.5cm 5.5cm 1.25cm, clip=true,height=\subfigthreeB\textwidth]{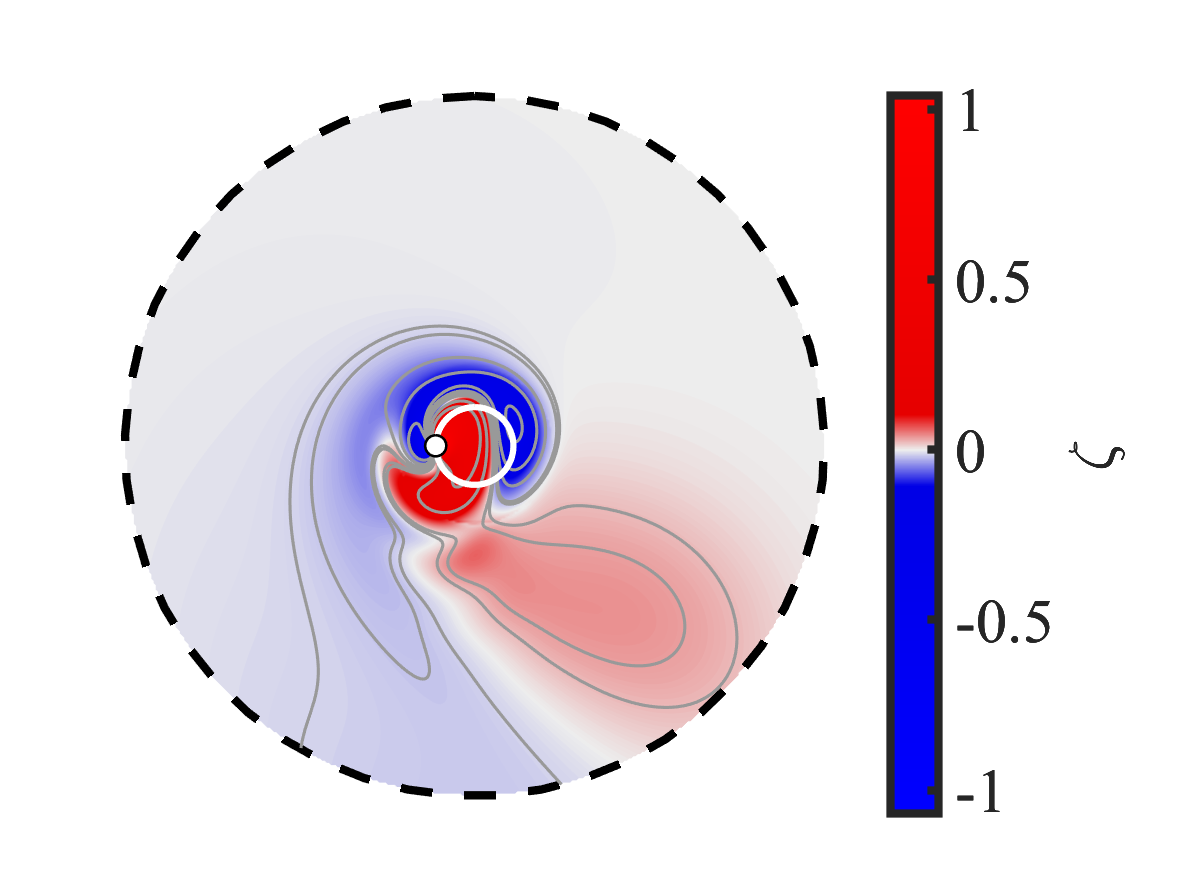}
	}~
	\hspace{-0.3cm}
	\subfloat[$R_{sd}=25$\label{fig:vorticity_Re=20_H}]{
		\includegraphics[trim=2cm 0.5cm 5.5cm 1.25cm, clip=true,height=\subfigthreeB\textwidth]{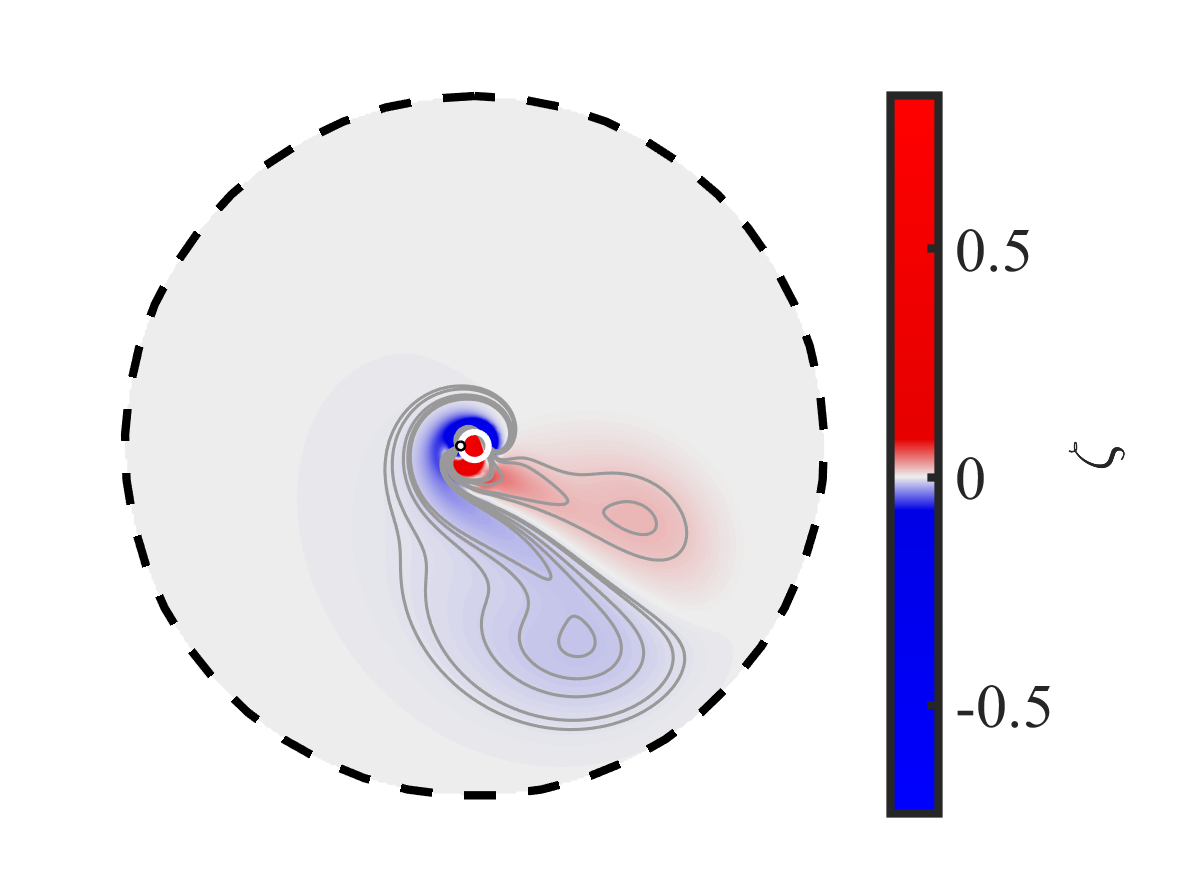}
	}
	\caption{{\color{black}Snapshots of the dye concentration field (a-h) and vorticity field (i-p) at times $T= 1, 1.65, 7.5, 20, 30, 45, 60, \text{and}~90$. In the bottom two rows, the red, blue, and grey colors denote counterclockwise (CCW), clockwise (CW), and near-zero vorticity values, respectively. The white and grey lines indicate the stirrer’s path and the streamlines, respectively.} The radius of the field of view is provided in the caption. $Re=20${\color{black}}.
	}
	\label{fig:alpha_Re=20}
\end{figure}

As $Re$ increases, advection intensifies, enhancing mixing both by strengthening the previously discussed mechanisms and introducing new ones. As an illustrative case, Figure~\ref{fig:alpha_Re=20} presents snapshots of dye concentration and vorticity fields at $Re = 20$. As expected, the initial stretching and deformation of the interface are more pronounced than at lower $Re$ (see Figures~\ref{fig:alpha_Re=20_A} \& \ref{fig:alpha_Re=20_B}). Although a well-mixed region forms at the center, the self-similar spiral-like evolution of the dye concentration is short-lived (see Figures~\ref{fig:alpha_Re=20_C} \& \ref{fig:alpha_Re=20_D}). Around $T \approx 30$, the well-mixed region is advected away from the central area (see Figure~\ref{fig:alpha_Re=20_E}). The interface is stretched across the stirrer’s path (see Figure~\ref{fig:alpha_Re=20_F}), leading to further stretching and folding. Unlike at $Re = 10$, where late-stage interface deformation was primarily azimuthal, \ida{the interface now undergoes substantial radial deformation beyond the stirrer’s immediate vicinity (Figure~\ref{fig:alpha_Re=20_F}).}

Examining the development of the vorticity field in Figure \ref{fig:alpha_Re=20}, we observe that as $Re$ increases, the CCW vorticity region extends beyond the central region at an earlier stage due to the reduced advection timescale relative to viscous dissipation (Figure \ref{fig:vorticity_Re=20_D}). Similarly, higher $Re$ leads to the earlier emergence of satellite vortices (Figure \ref{fig:vorticity_Re=20_E}), which \ida{now} primarily advect in the radial direction (Figures \ref{fig:vorticity_Re=20_F}-\ref{fig:vorticity_Re=20_H}). 

Moreover, the strength of the satellite vortices increases with $Re$. This is evident in Figures \ref{fig:vorticity_Re=20_E}-\ref{fig:vorticity_Re=20_G}, where the vorticity of the satellite CW vortex is comparable to that of the attached CW vortex. In fact, the satellite CW vortex extends into the stirrer’s path, stretching the dye interface and drawing it into the stirrer’s path (see the corresponding dye concentration field in Figures \ref{fig:alpha_Re=20_E}-\ref{fig:alpha_Re=20_G}).

\begin{figure}
	\centering
	\subfloat[ $R_{sd}=3$\label{fig:locationVortex_Re=20_A}]{
		\includegraphics[trim=0cm 0cm 0cm 0cm, clip=true,height=.25\textwidth]{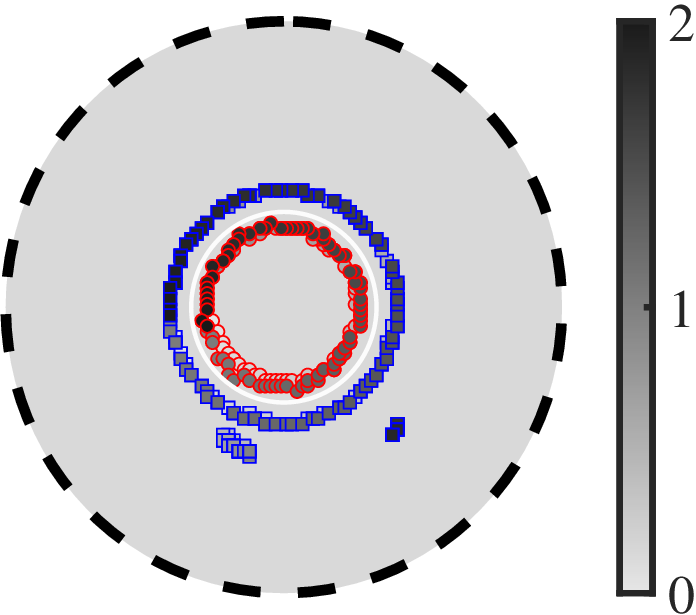}
	}~
	\subfloat[$R_{sd}=5$\label{fig:locationVortex_Re=20_B}]{
		\includegraphics[trim=0cm 0cm 0cm 0cm, clip=true,height=.25\textwidth]{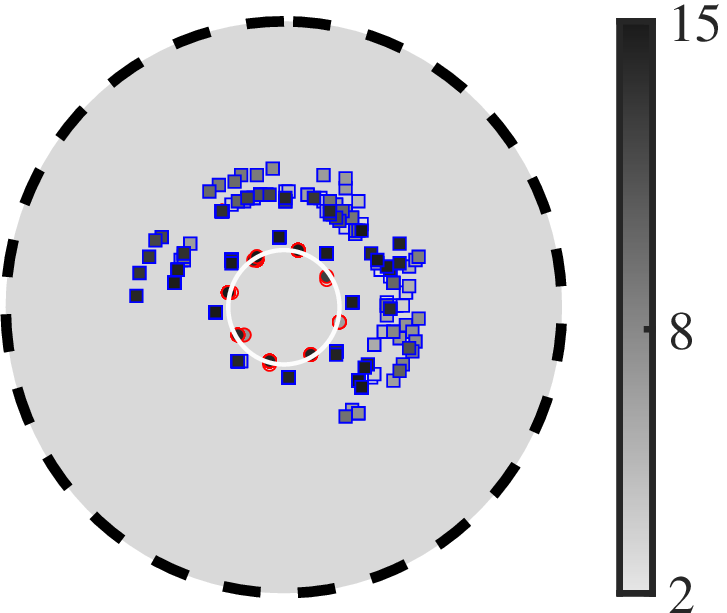}
	}~
	\subfloat[$R_{sd}=12$\label{fig:locationVortex_Re=20_C}]{
		\includegraphics[trim=0cm 0cm 0cm 0cm, clip=true,height=.25\textwidth]{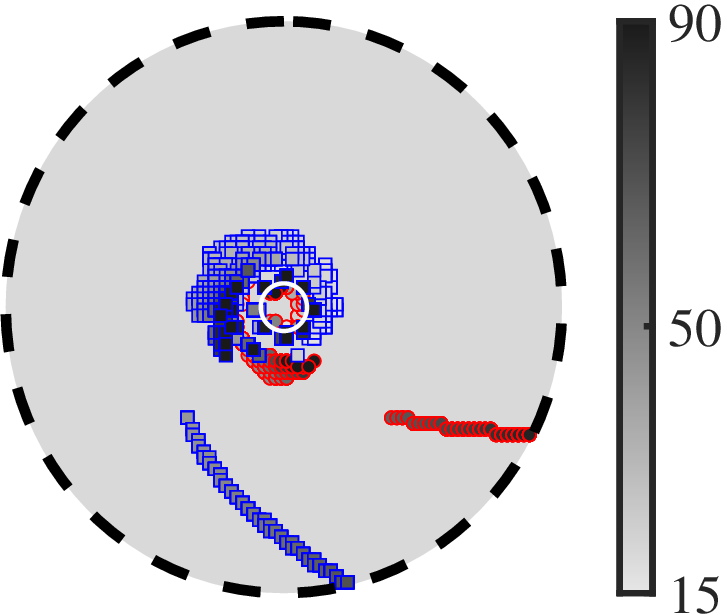}
	}
	\caption{Time evolution of the vortex centers at different time intervals. The square and circle marks represent center of CW and CCW vortices, respectively. Lighter shades correspond to earlier times. The white solid line represents the stirrer’s path and the black dashed lines denote the subdomain boundary. The radius of the subdomain is presented in the captions. $Re=20$.}
	\label{fig:locationVortex_Re=20}
	
\end{figure}

Figure \ref{fig:locationVortex_Re=20} shows the location of the vortex centers at $Re=20$. Stirring is immediately followed by the formation of the attached vortices (Figure \ref{fig:locationVortex_Re=20_A}). The satellite vortices initially grow and advect azimuthally, moving much more slowly than the stirrer (Figure \ref{fig:locationVortex_Re=20_B}). However, at later times, and in contrast to lower $Re$, these vortices begin to advect radially away from the central region, \ida{providing the transport mechanism responsible for the radial deformation of the dye interface} (Figure \ref{fig:locationVortex_Re=20_C}). 

\begin{figure}
	\centering
	\subfloat[$R_{sd}=3$\label{fig:alpha_Re=50_A}]{
		\includegraphics[trim=2cm 0.5cm 5.5cm 1.25cm, clip=true,height=\subfigthreeB\textwidth]{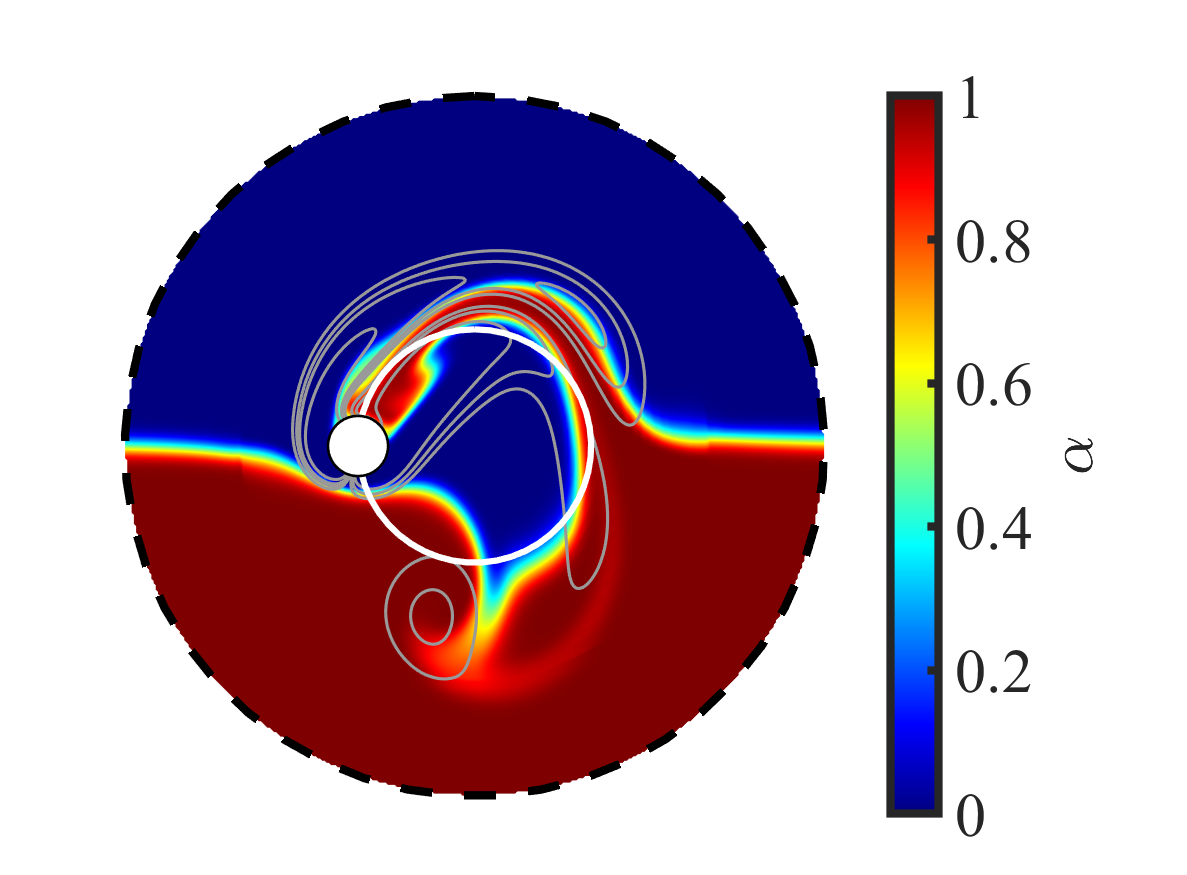}
	}~
	\hspace{-0.3cm}
	\subfloat[$R_{sd}=3$\label{fig:alpha_Re=50_B}]{
		\includegraphics[trim=2cm 0.5cm 5.5cm 1.25cm, clip=true,height=\subfigthreeB\textwidth]{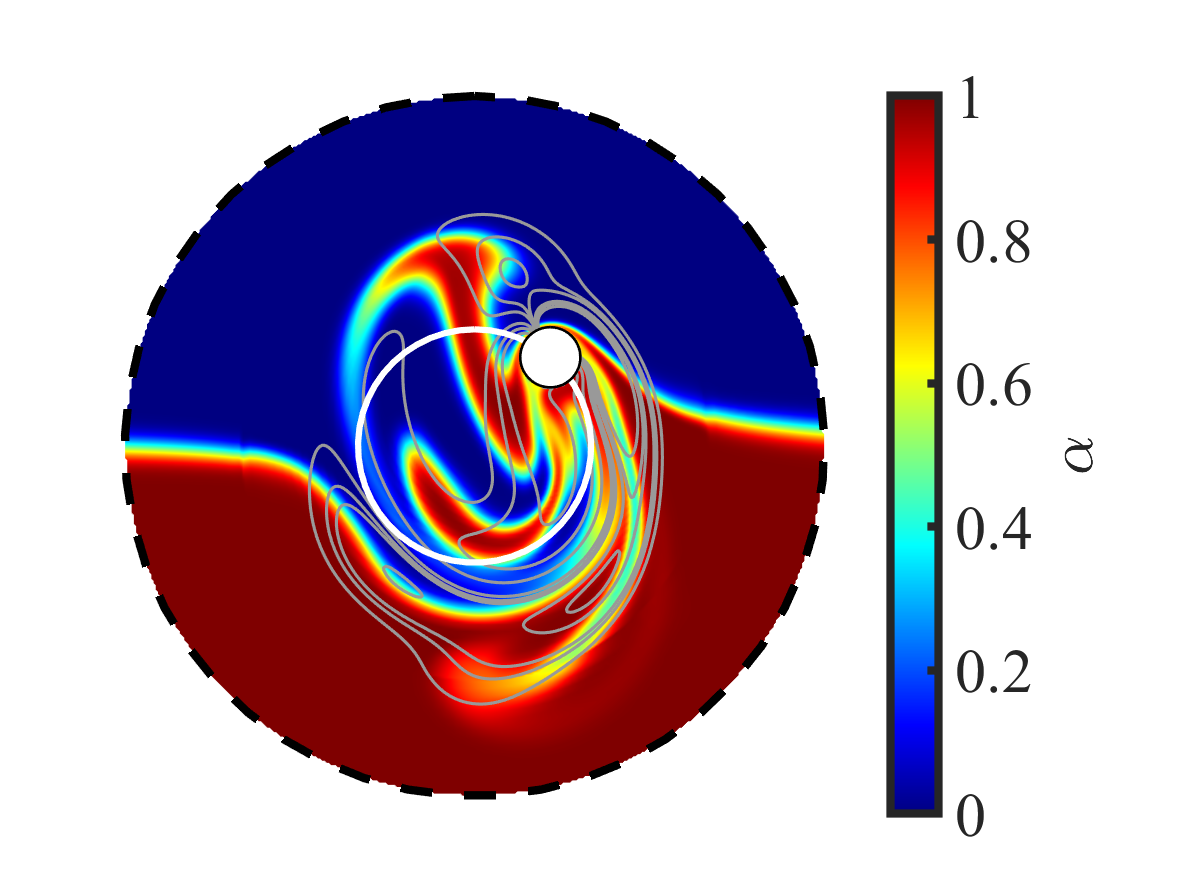}
	}~
	\hspace{-0.3cm}
	\subfloat[$R_{sd}=6$\label{fig:alpha_Re=50_C}]{
		\includegraphics[trim=2cm 0.5cm 5.5cm 1.25cm, clip=true,height=\subfigthreeB\textwidth]{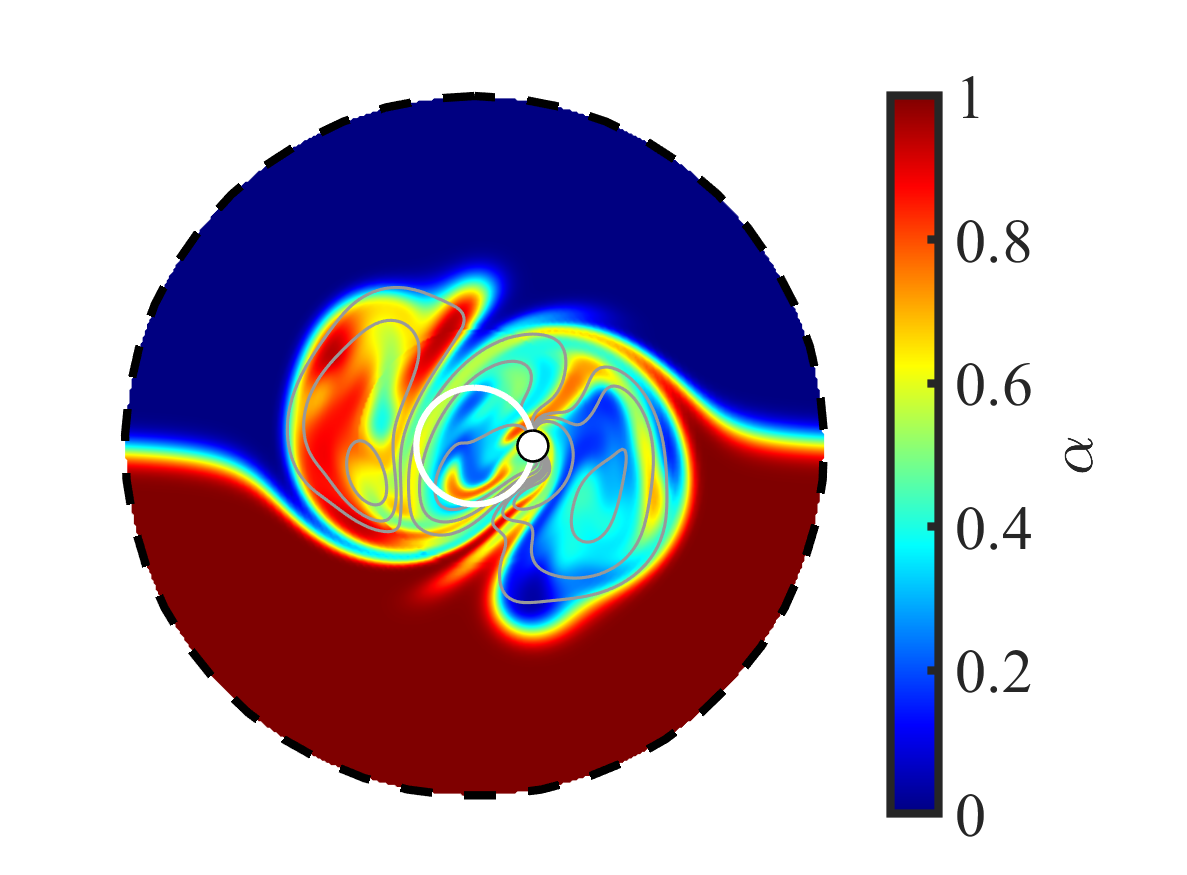}
	}~
	\hspace{-0.3cm}
	\subfloat[$R_{sd}=6$\label{fig:alpha_Re=50_D}]{
		\includegraphics[trim=2cm 0.5cm 1.5cm 1.25cm, clip=true,height=\subfigthreeB\textwidth]{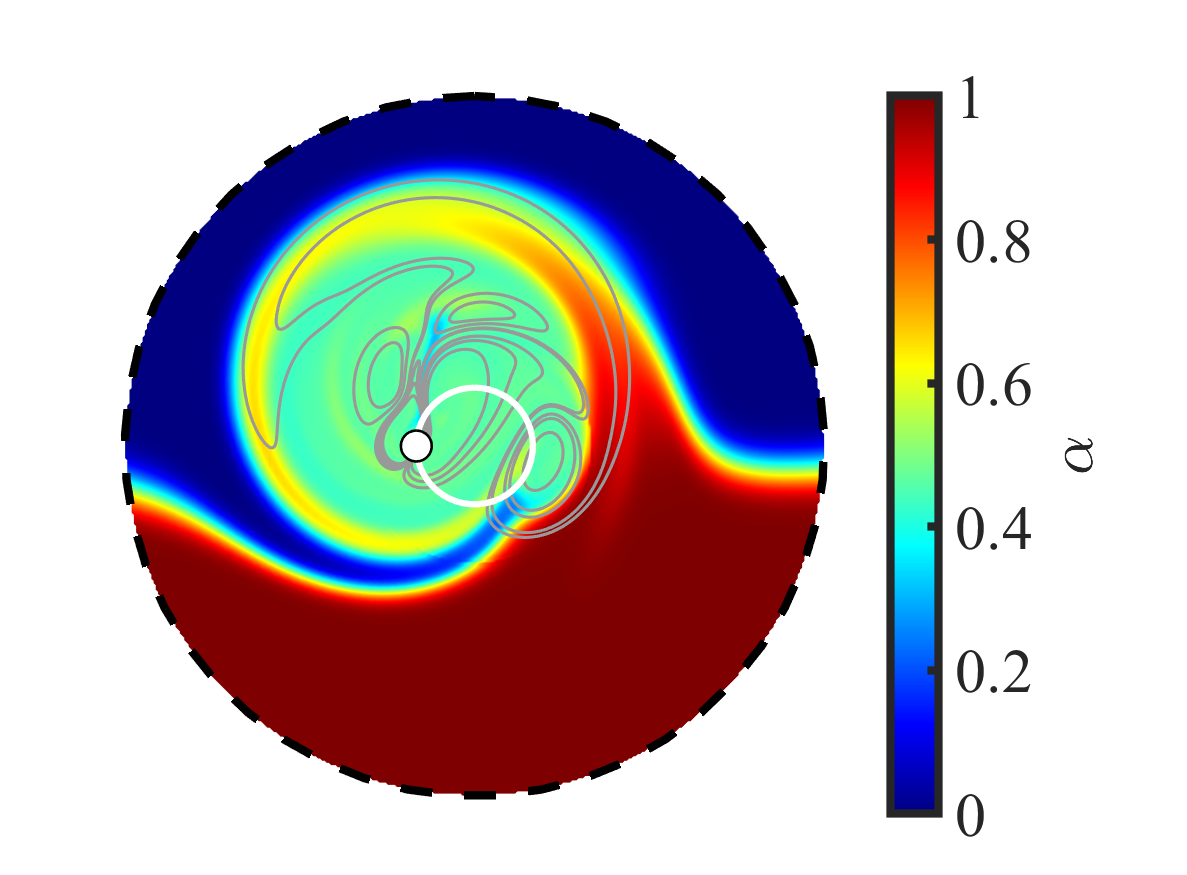}
	}\\
	\vspace{-0.3cm}
	\subfloat[$R_{sd}=9$\label{fig:alpha_Re=50_E}]{
		\includegraphics[trim=2cm 0.5cm 5.5cm 1.25cm, clip=true,height=\subfigthreeB\textwidth]{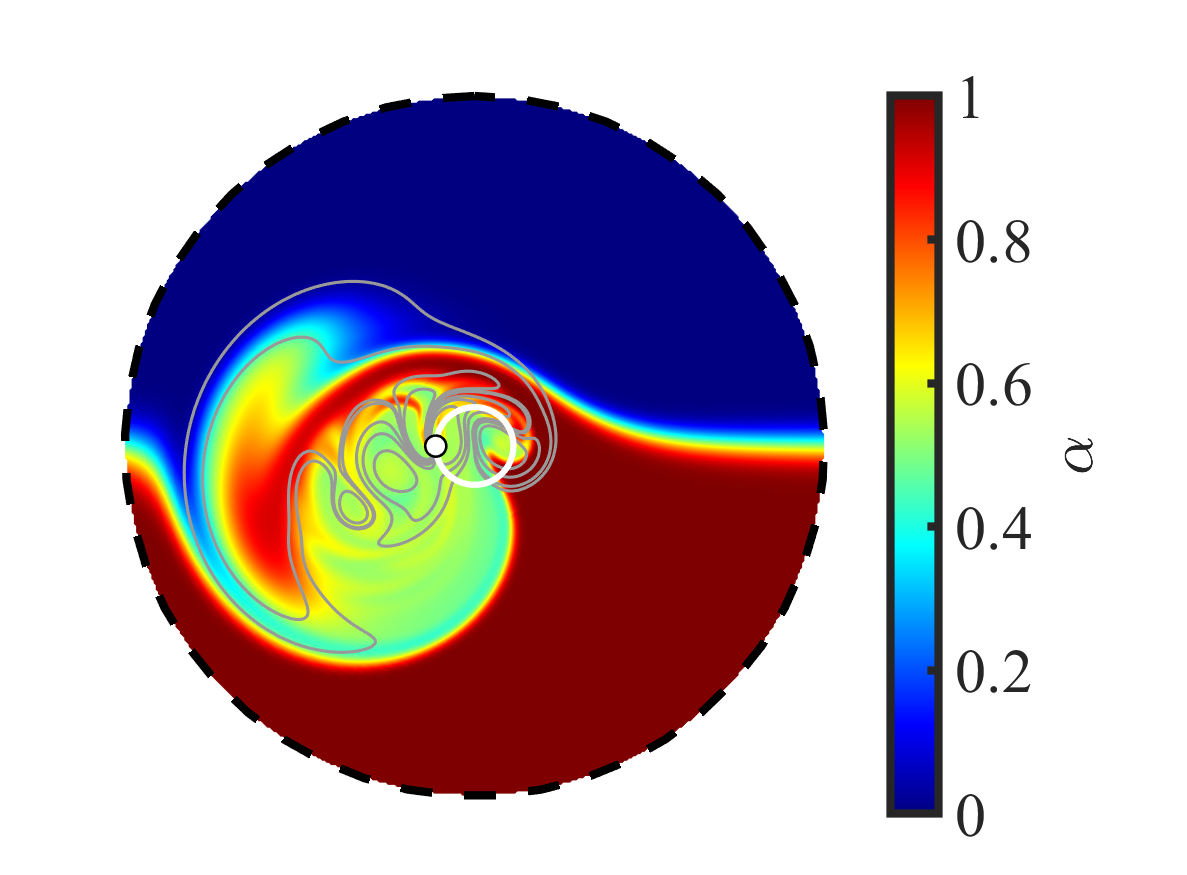}
	}~
	\hspace{-0.3cm}
	\subfloat[$R_{sd}=9$\label{fig:alpha_Re=50_F}]{
		\includegraphics[trim=2cm 0.5cm 5.5cm 1.25cm, clip=true,height=\subfigthreeB\textwidth]{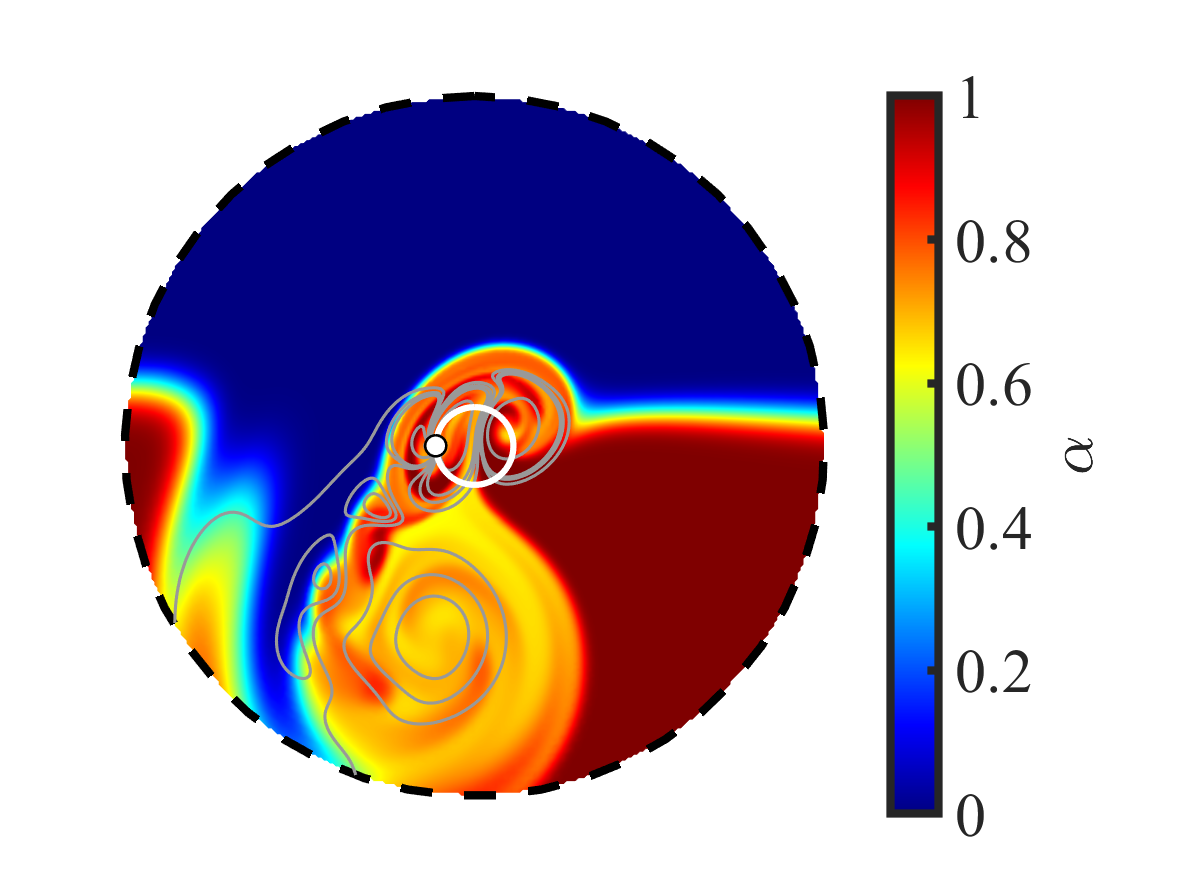}
	}~
	\hspace{-0.3cm}
	\subfloat[$R_{sd}=15$\label{fig:alpha_Re=50_G}]{
		\includegraphics[trim=2cm 0.5cm 5.5cm 1.25cm, clip=true,height=\subfigthreeB\textwidth]{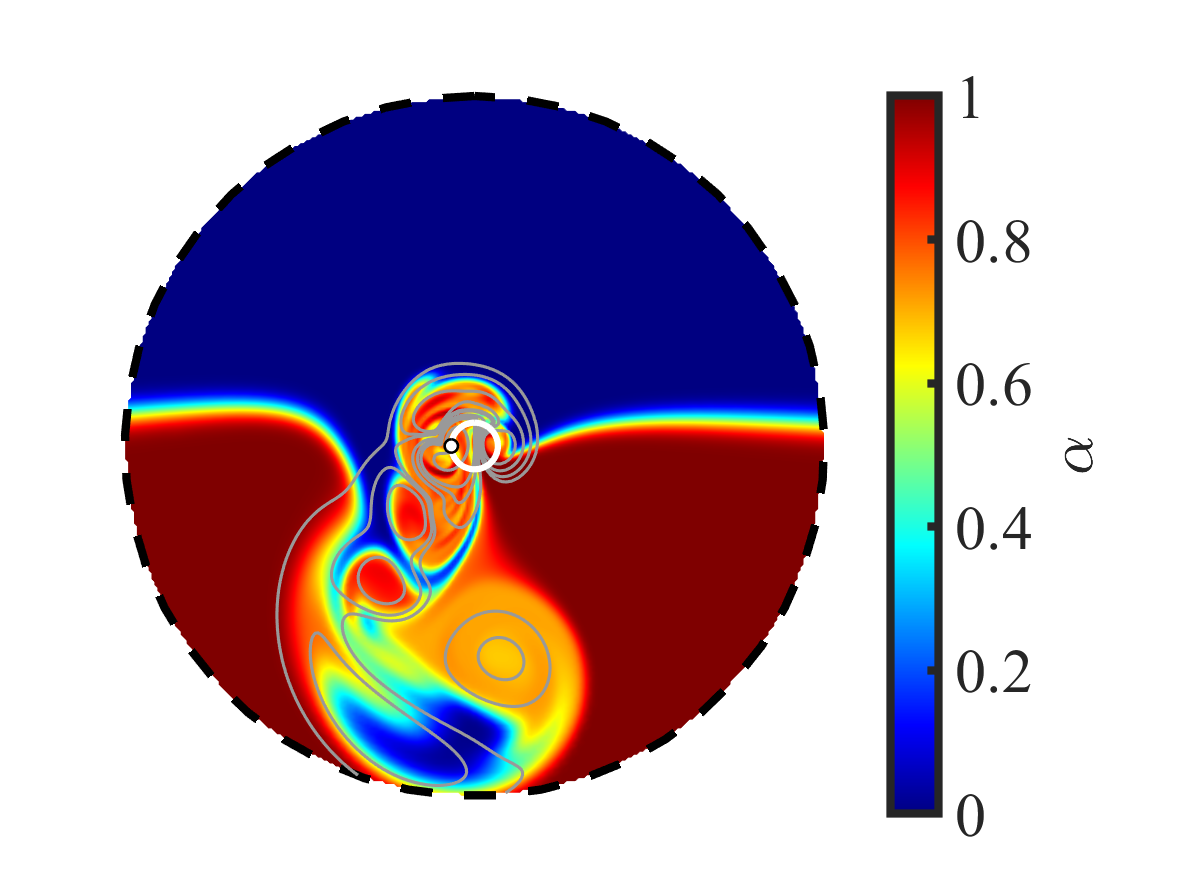}
	}~
	\hspace{-0.3cm}
	\subfloat[$R_{sd}=15$\label{fig:alpha_Re=50_H}]{
		\includegraphics[trim=2cm 0.5cm 1.5cm 1.25cm, clip=true,height=\subfigthreeB\textwidth]{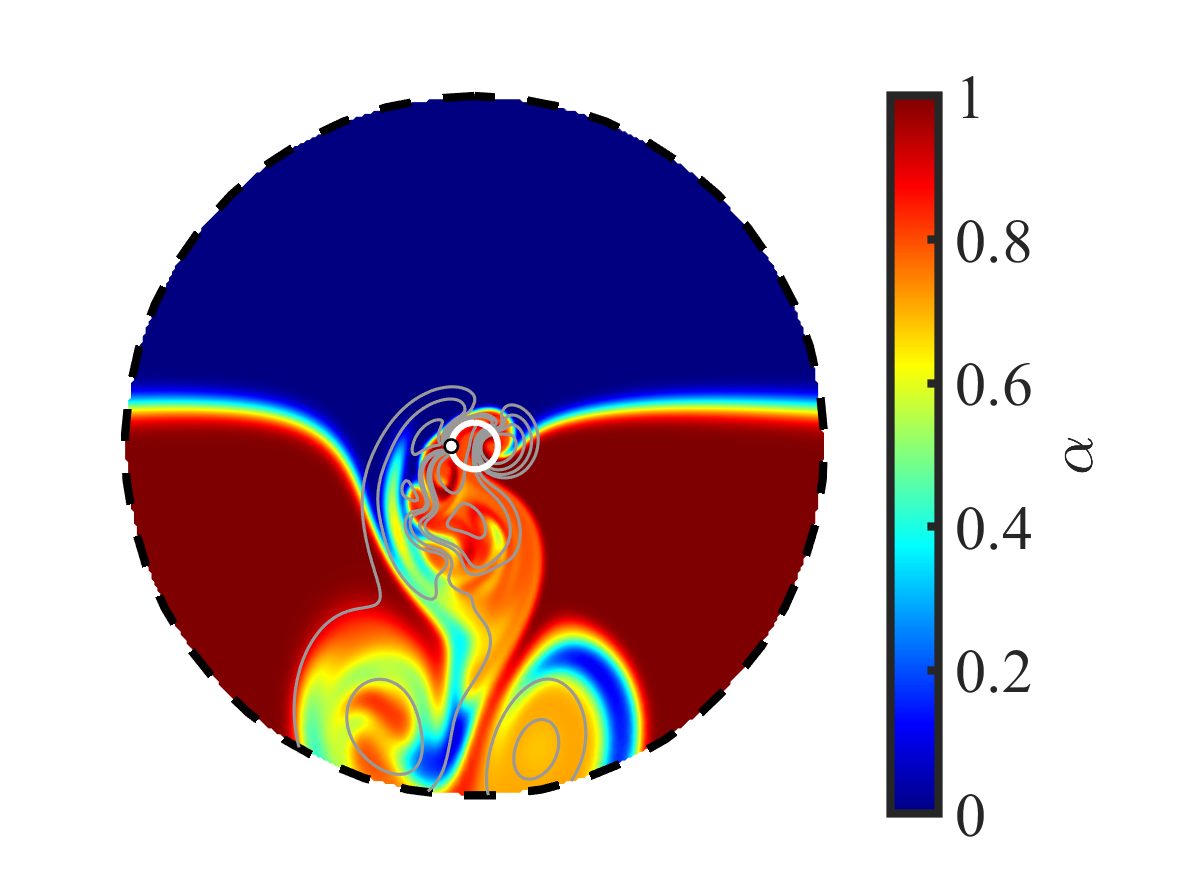}
	}
	\\
	\subfloat[$R_{sd}=3$\label{fig:vorticity_Re=50_A}]{
		\includegraphics[trim=2cm 0.5cm 5.5cm 1.25cm, clip=true,height=\subfigthreeB\textwidth]{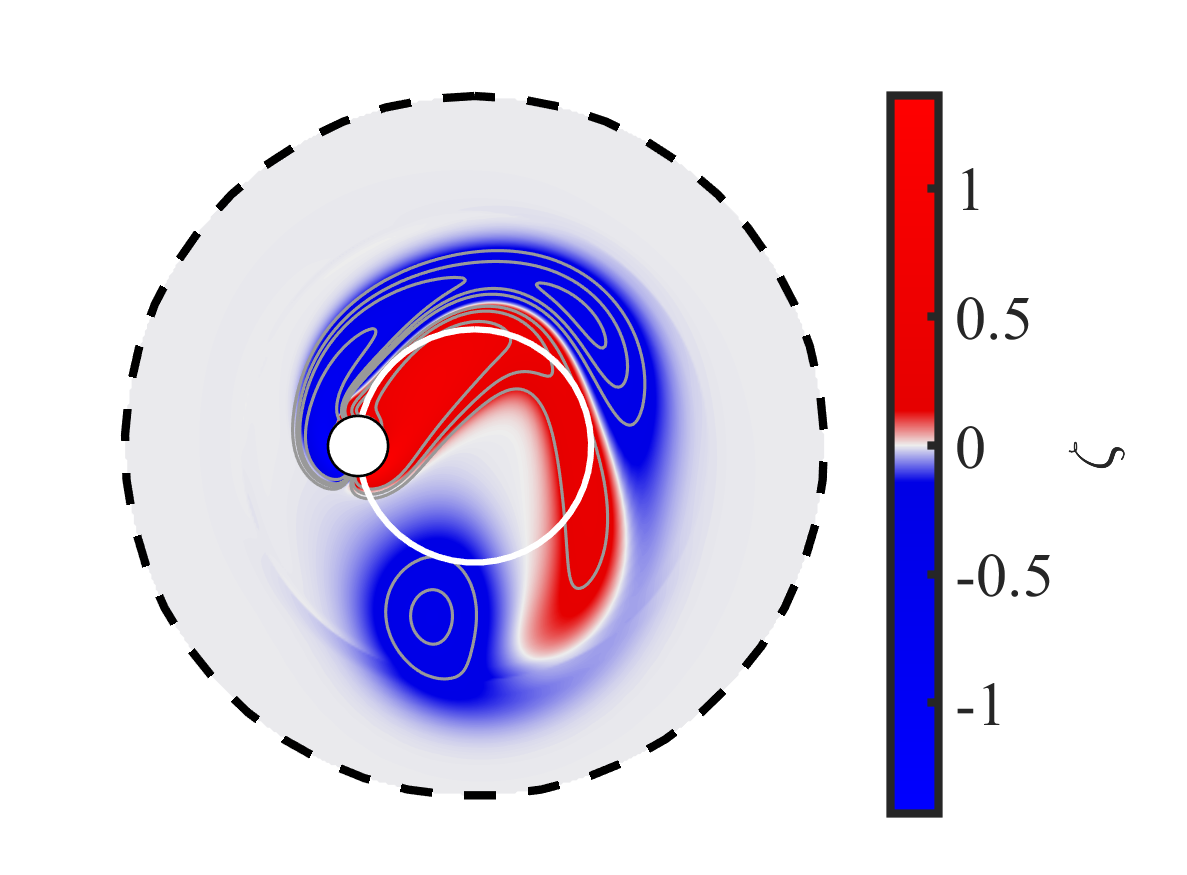}
	}~
	\hspace{-0.3cm}
	\subfloat[$R_{sd}=3$\label{fig:vorticity_Re=50_B}]{
		\includegraphics[trim=2cm 0.5cm 5.5cm 1.25cm, clip=true,height=\subfigthreeB\textwidth]{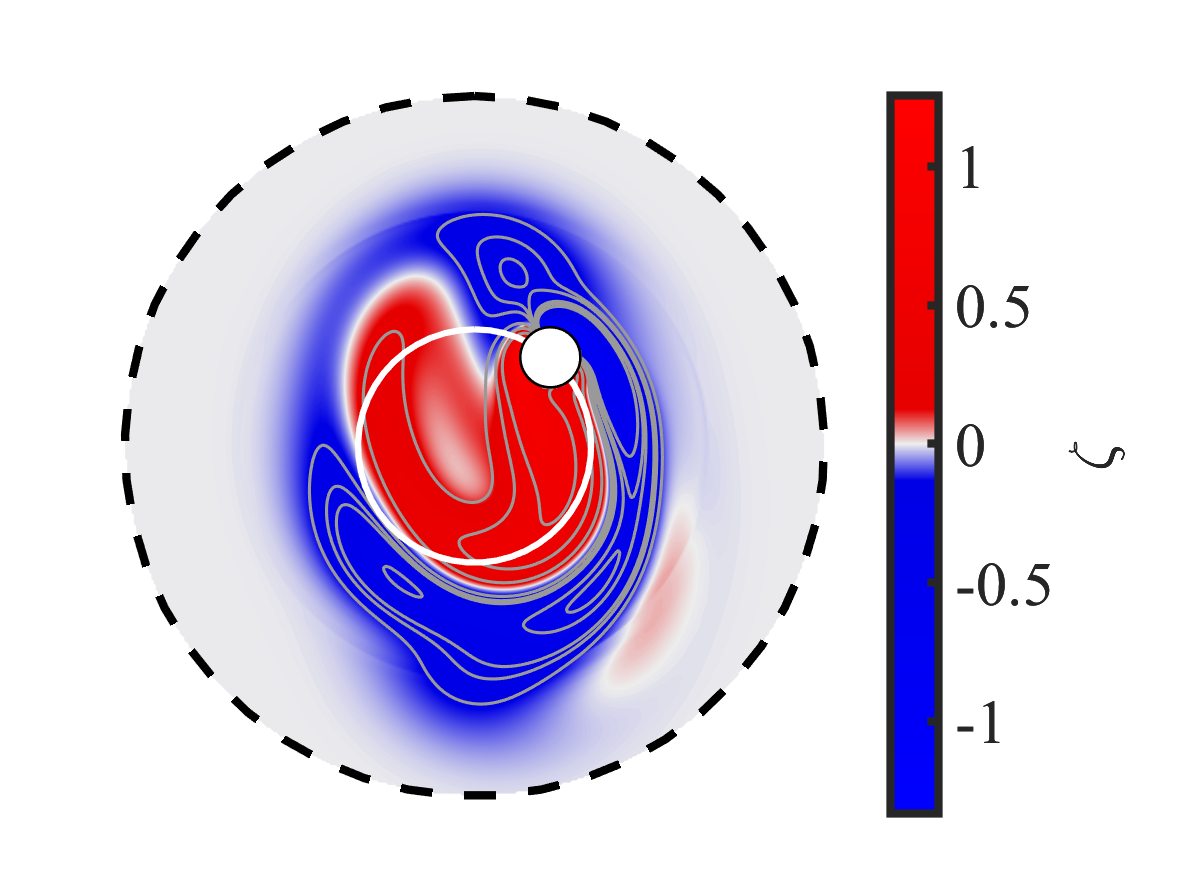}
	}~
	\hspace{-0.3cm}
	\subfloat[$R_{sd}=6$\label{fig:vorticity_Re=50_C}]{
		\includegraphics[trim=2cm 0.5cm 5.5cm 1.25cm, clip=true,height=\subfigthreeB\textwidth]{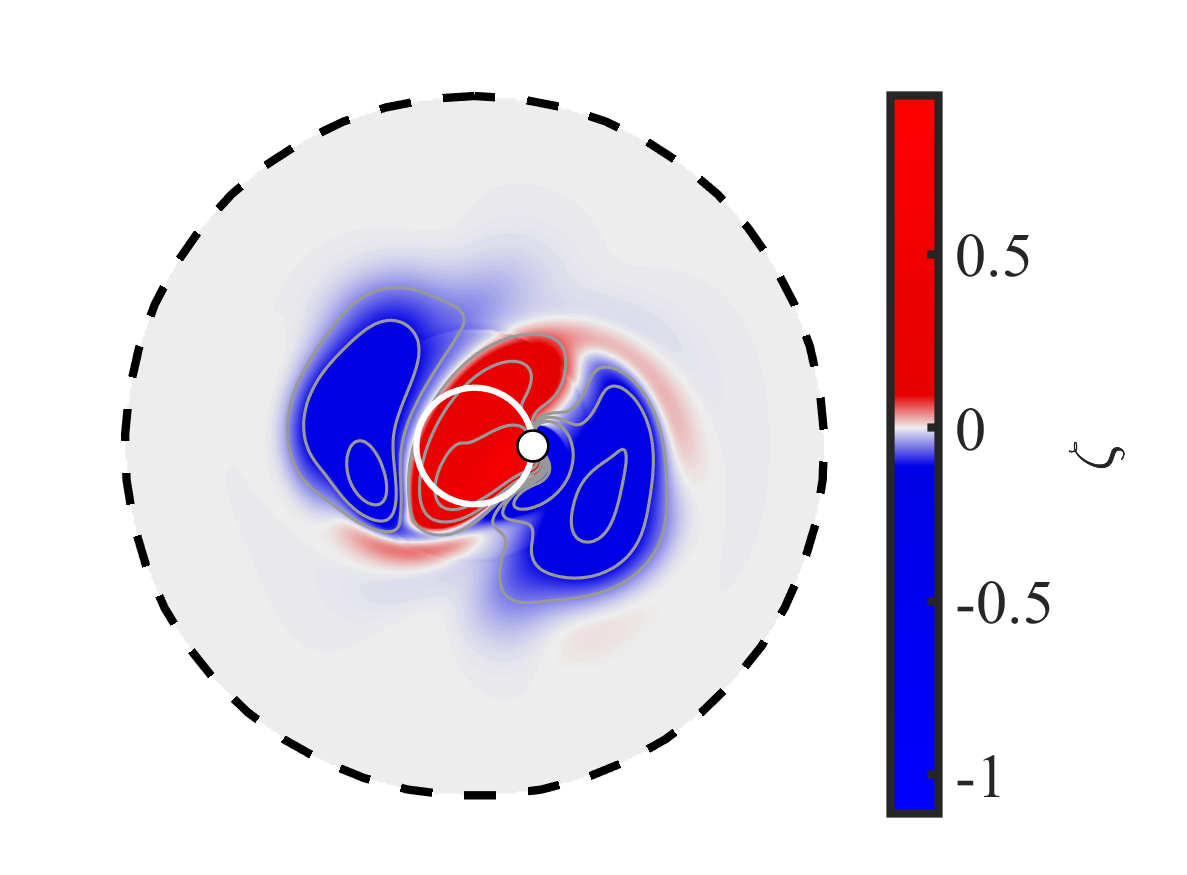}
	}~
	\hspace{-0.3cm}
	\subfloat[$R_{sd}=6$\label{fig:vorticity_Re=50_D}]{
		\includegraphics[trim=2cm 0.5cm 5.5cm 1.25cm, clip=true,height=\subfigthreeB\textwidth]{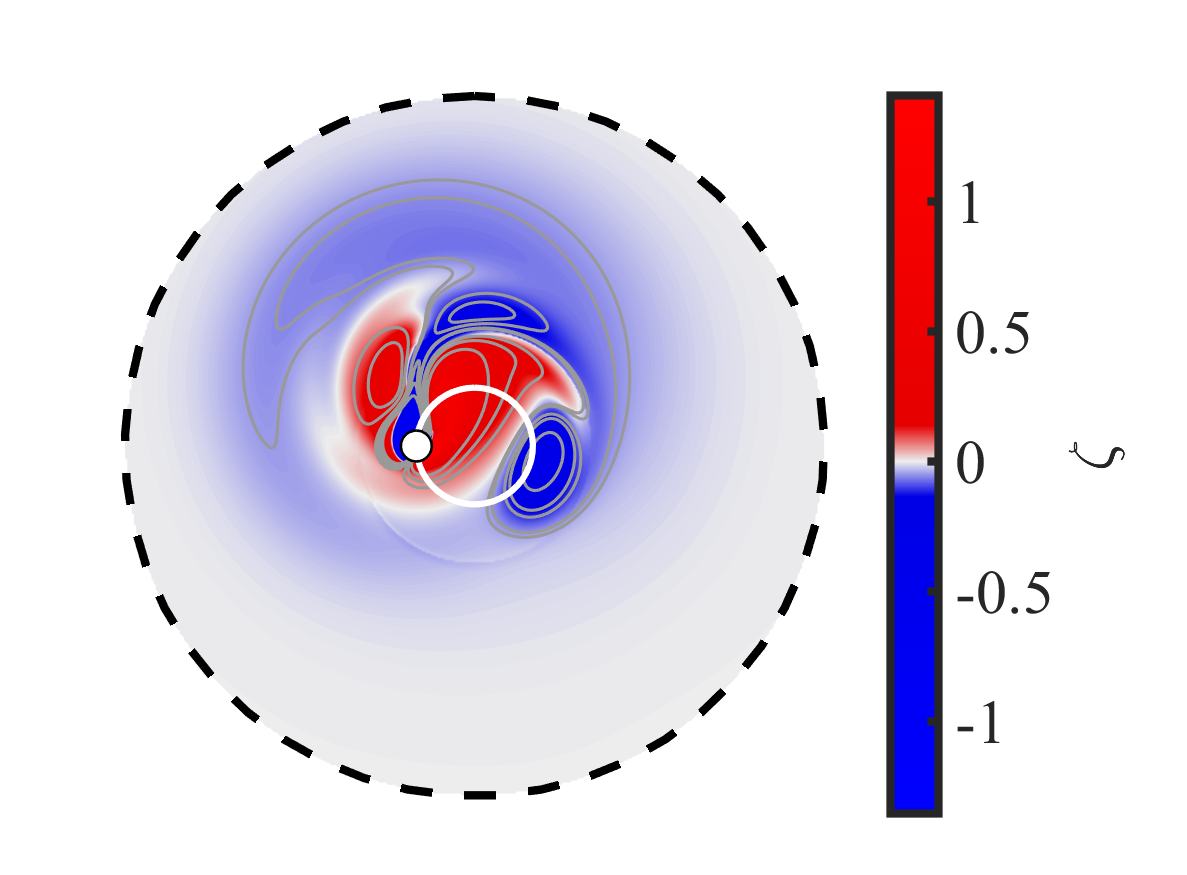}
	}\\
	\vspace{-0.3cm}
	\subfloat[$R_{sd}=6$\label{fig:vorticity_Re=50_E}]{
		\includegraphics[trim=2cm 0.5cm 5.5cm 1.25cm, clip=true,height=\subfigthreeB\textwidth]{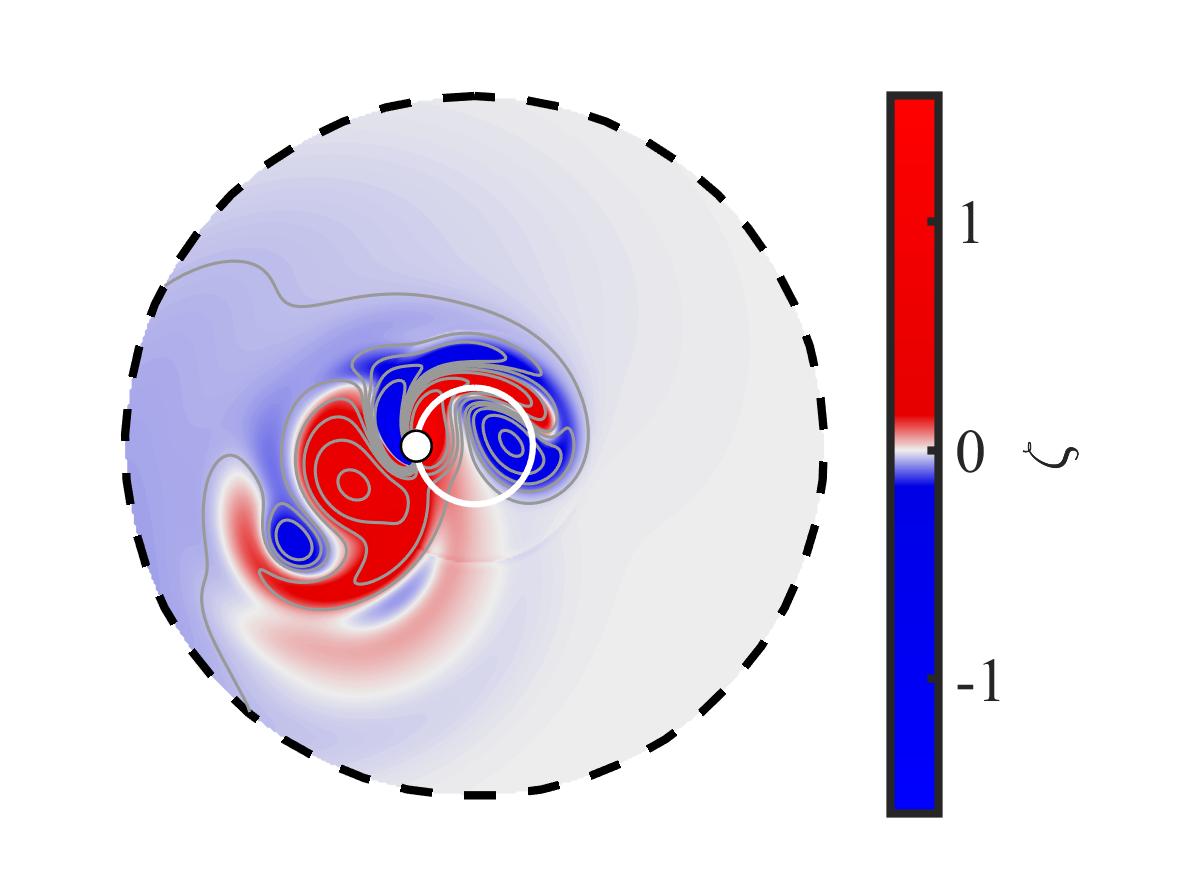}
	}~
	\hspace{-0.3cm}
	\subfloat[$R_{sd}=9$\label{fig:vorticity_Re=50_F}]{
		\includegraphics[trim=2cm 0.5cm 5.5cm 1.25cm, clip=true,height=\subfigthreeB\textwidth]{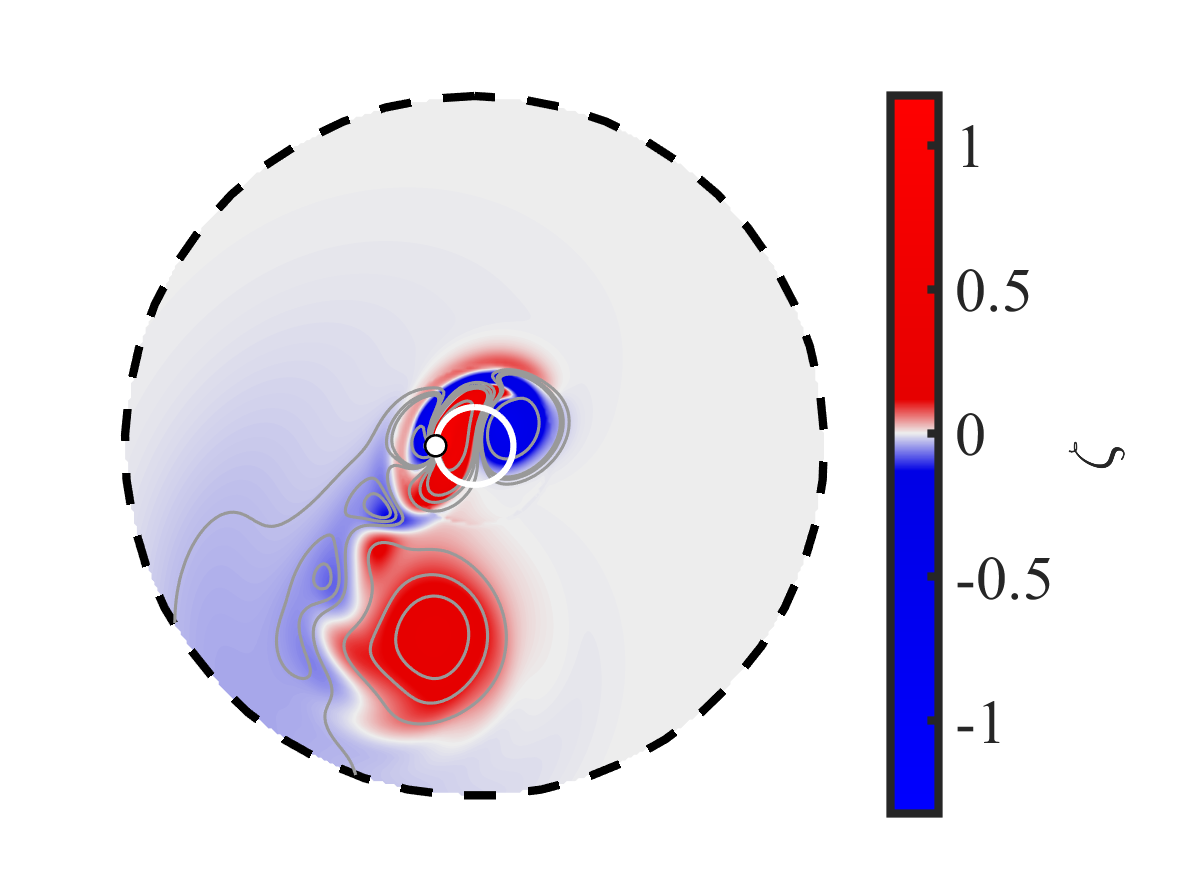}
	}~
	\hspace{-0.3cm}
	\subfloat[$R_{sd}=15$\label{fig:vorticity_Re=50_G}]{
		\includegraphics[trim=2cm 0.5cm 5.5cm 1.25cm, clip=true,height=\subfigthreeB\textwidth]{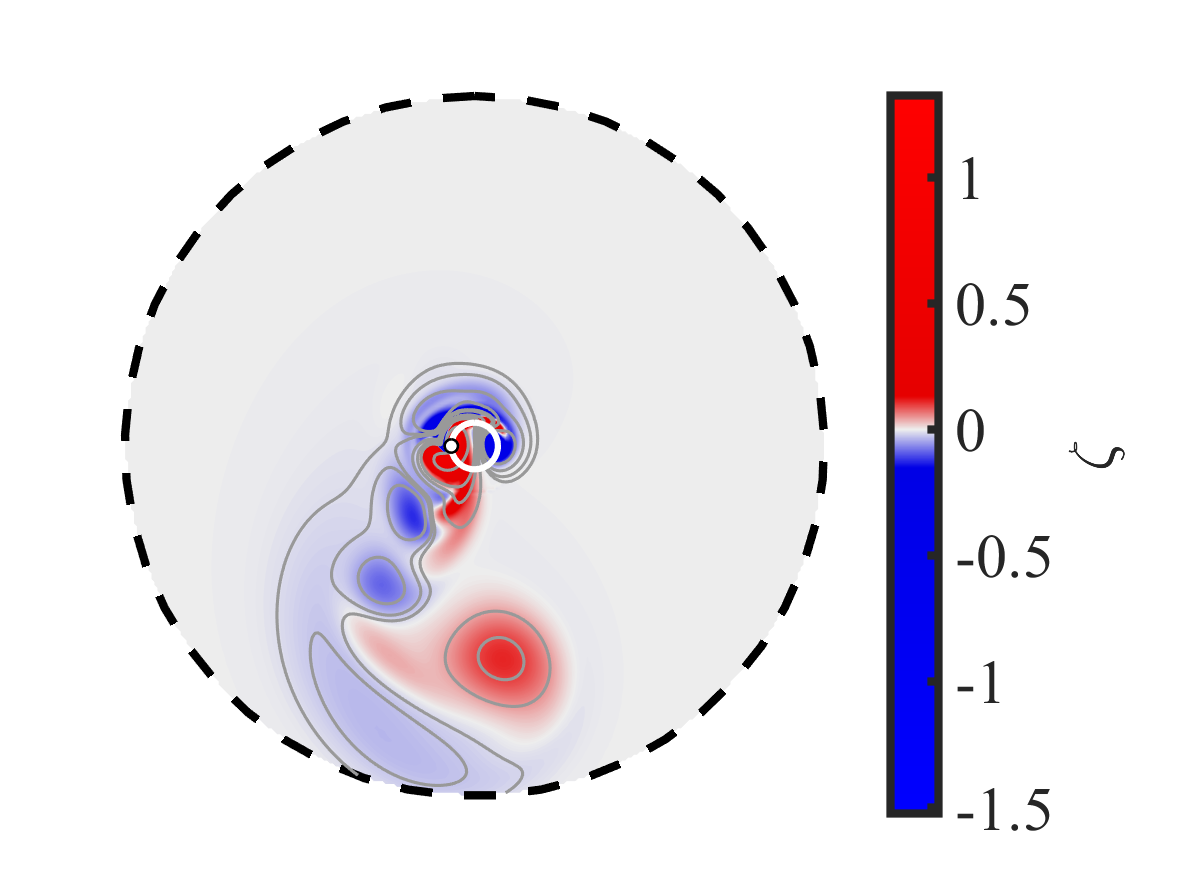}
	}~
	\hspace{-0.3cm}
	\subfloat[$R_{sd}=15$\label{fig:vorticity_Re=50_H}]{
		\includegraphics[trim=2cm 0.5cm 5.5cm 1.25cm, clip=true,height=\subfigthreeB\textwidth]{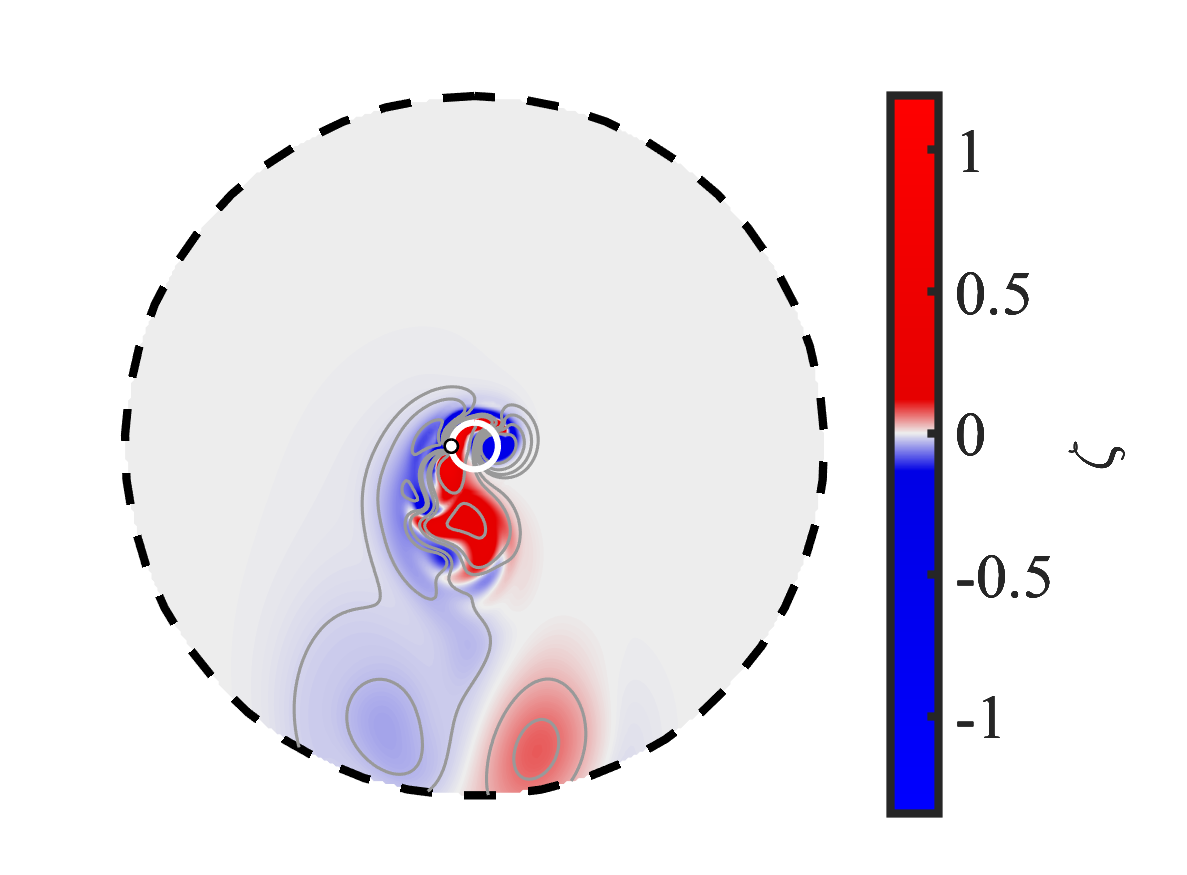}
	}
	\caption{{\color{black}Snapshots of the dye concentration field (a-h) and vorticity field (i-p) at times $T= 1, 1.65, 7.5, 20, 30, 40, 50, \text{and}~60$.  In the bottom two rows, the red, blue, and grey colors denote counterclockwise (CCW), clockwise (CW), and near-zero vorticity values, respectively. The white and grey lines indicate the stirrer’s path and the streamlines, respectively.} The radius of the field of view is provided in the caption. $Re=50${\color{black}}.
	}
	\label{fig:alpha_Re=50}
\end{figure}

\ida{The final advective transport mechanism identified in the present study} is illustrated in Figure \ref{fig:alpha_Re=50}, $Re=50$. Initial stretching of the interface and the formation of a striated dye pattern remains a common feature. However, even at early times the interface is stretched and deformed beyond the stirrer's path (see Figure \ref{fig:alpha_Re=50_C}), \ida{preceding} the formation of the well-mixed region at the center (Figure \ref{fig:alpha_Re=50_D}). The well-mixed region is soon advected away from the center (Figure \ref{fig:alpha_Re=50_E}), followed by extensive stretching and folding of the interface (Figures \ref{fig:alpha_Re=50_F}-\ref{fig:alpha_Re=50_H}).

The evolution of the vorticity field in Figure \ref{fig:alpha_Re=50} shows that, as expected, the attached vortices extend more rapidly, exceeding the stirrer's path (Figures \ref{fig:vorticity_Re=50_A}–\ref{fig:vorticity_Re=50_B}). In contrast to lower Reynolds numbers, where a ring of CW vorticity surrounds the stirrer's path for a few periods after the onset of stirring, here the flow field outside the stirrer's path is dominated by two CW vortices that rotate around the path at a lower speed than the stirrer itself. These two CW vortices are responsible for the significant deformation of the interface outside the stirrer's path during the early stages of mixing (Figure \ref{fig:alpha_Re=50_C}).
This flow structure is quickly disturbed by the growth of the central CCW vortex (Figure \ref{fig:vorticity_Re=50_D}). As CCW vortices are shed from the central region, the satellite CW vortices are carried away with them (Figures \ref{fig:vorticity_Re=50_E}–\ref{fig:vorticity_Re=50_H}). \ida{Consequently, the dye interface is stretched and deformed deep into the flow domain} (Figures \ref{fig:alpha_Re=50_E}–\ref{fig:alpha_Re=50_H}).

\begin{figure}
	\centering
	\subfloat[ $R_{sd}=3$\label{fig:locationVortex_Re=50_A}]{
		\includegraphics[trim=0cm 0cm 0cm 0cm, clip=true,height=.25\textwidth]{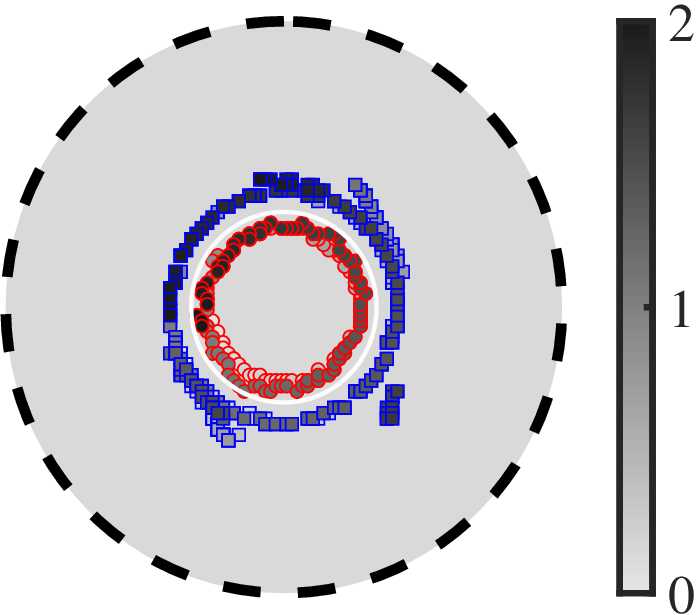}
	}~
	\subfloat[$R_{sd}=3$\label{fig:locationVortex_Re=50_B}]{
		\includegraphics[trim=0cm 0cm 0cm 0cm, clip=true,height=.25\textwidth]{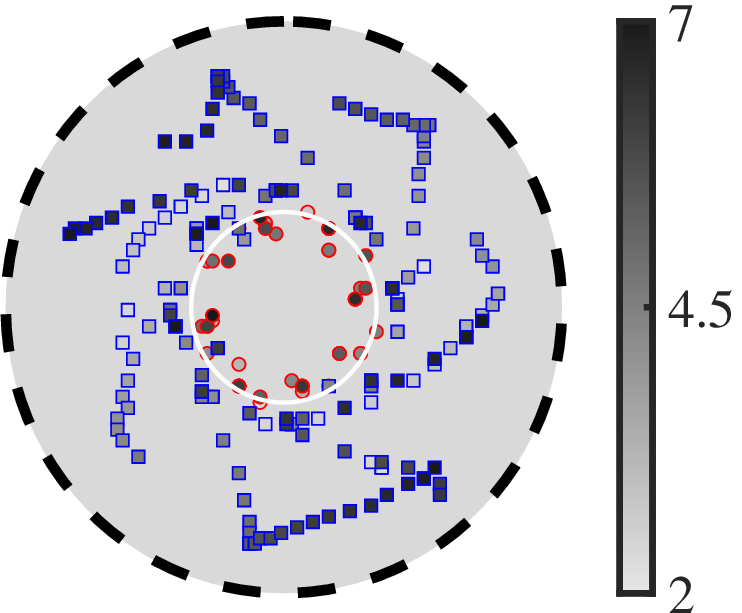}
	}
	\subfloat[$R_{sd}=16$\label{fig:locationVortex_Re=50_C}]{
		\includegraphics[trim=0cm 0cm 0cm 0cm, clip=true,height=.25\textwidth]{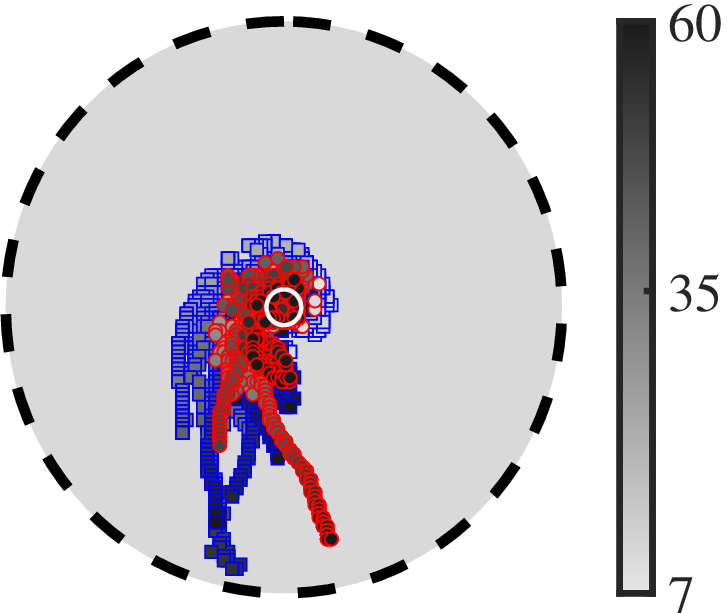}
	}
	\caption{Time evolution of the vortex centers at different time intervals. The blue square and red circle marks represent center of CW and CCW vortices, respectively. Lighter shades correspond to earlier times. The white solid line represents the stirrer's path and the black dashed lines denote the subdomain boundary. The radius of the subdomain is presented in the captions. $Re=50$.}
	\label{fig:locationVortex_Re=50}
	
\end{figure}

Figure \ref{fig:locationVortex_Re=50} illustrates the development of vortex centers. As in the previous cases, the formation of the attached vortices is apparent in Figure \ref{fig:locationVortex_Re=50_A}. Figure \ref{fig:locationVortex_Re=50_B} illustrates the slow azimuthal \ida{advection} of the satellite CW vortices around the stirrer's path and their  interaction with CW vortices shed from the stirrer. The shedding of vortices from the central region and their advection far into the flow domain is visualized in Figure \ref{fig:locationVortex_Re=50_C}.

\begin{figure}
	\centering
	\subfloat{
		\includegraphics[trim=0cm 1cm 1cm 0cm, clip=true,height=.45\textwidth]{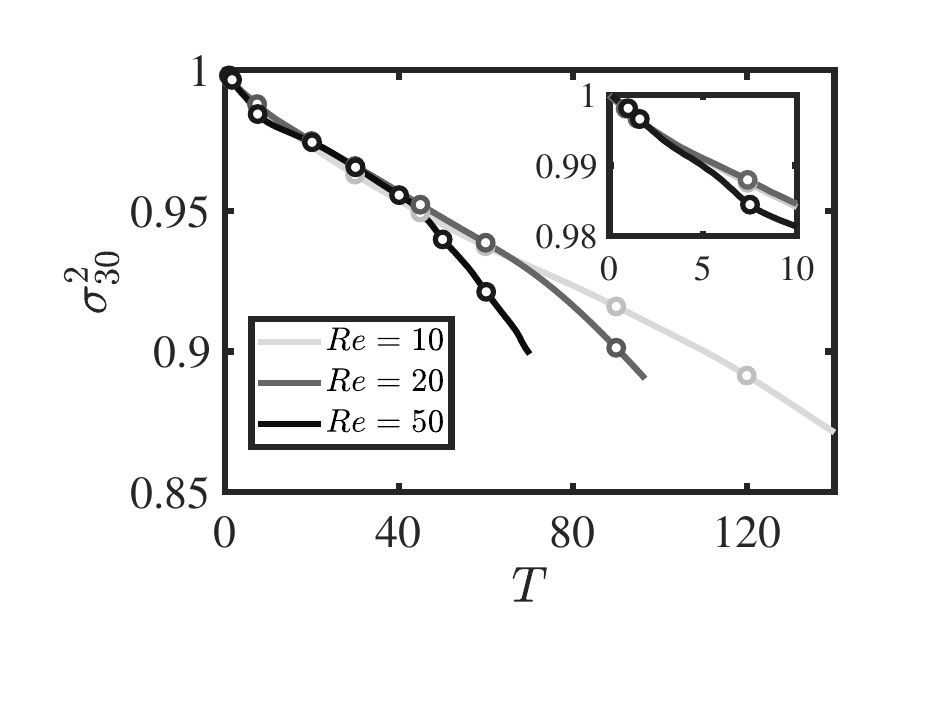}
	}
	
	\caption{Time evolution of normalized variance of dye concentration, $\sigma_{30}^2$, at different Reynolds numbers. The inset illustrates a magnified view over the range of $0\le T\le10$. }
	\label{fig:variance_regimeIII}
	
\end{figure}

Figure~\ref{fig:variance_regimeIII} shows the time evolution of the dye concentration variance, $\sigma_{30}^2$, comparing the mixing rate for $10 \leq Re \leq 50$. The markers indicate the time instances at which concentration and vorticity snapshots are presented in {\color{black}Figures \ref{fig:alpha_Re=10} to \ref{fig:alpha_Re=50}}.

The improved initial rapid decay at $Re = 50$ (see $T \lesssim 7$ in figure~\ref{fig:variance_regimeIII}) corresponds to the stretching and folding of the interface by the two satellite CW vortices. This is followed by a slower mixing phase during $7 \lesssim T \lesssim 40$, associated with the azimuthal movement of the well-mixed region. {\color{black}Overall, however, the variance decays similarly across all cases until the onset of radial vortex advection and extensive interface stretching (see, e.g., $T\approx 40$ and $T\approx 60$ for $Re=50$ and 20, respectively).}

\subsection{\ida{Mechanisms and regimes of laminar mixing}}

In the previous section, we distinguished between two primary mixing regimes: diffusion-dominated and advective. The mixing dynamics were analyzed in relation to the emergence and evolution of vortices. The \ida{resulting flow} shares several similarities with flow past a bluff body. First, two attached vortices form behind the bluff body. Second, vortex shedding is observed at sufficiently high $Re$. The key difference here is that the stirrer (i.e., the bluff body) does not follow a straight path, but instead moves along a circular trajectory.

As in the classical vortex-shedding problem behind a cylinder, the strength of the shed vortices and the shedding frequency increase with $Re$. \ida{The number and size of the shed vortices are governed by the interplay of three timescales:} the shedding period, the timescale of periodic interactions with the attached vortices, and viscous dissipation.

Here, vortex shedding occurs at a lower Reynolds number ($5 \lesssim Re$) than in the classical case, where shedding typically begins at $Re \approx 40$ (Roshko \cite{roshko1954development}). The shed vortices are, however, weak at low Reynolds numbers and dissipate rapidly. The asymmetry introduced by the stirrer’s circular path also prevents the shedding of CCW vortices; instead, \ida{only CW vortices are shed outside the stirrer's path.}

\textbf{Regime \NS~(Negligible Shedding)}: When $Re = O(1)$, vortex shedding is absent or very weak. Mixing is diffusion-dominated, and the dye pattern is characterized by the formation of a well-mixed region at the center and the development of a dye concentration field resembling a spiral pattern. \ida{The spiral expands in a self-similar manner over time through diffusion.}

\textbf{Regime \WS~(Weak Shedding, $5 \lesssim Re \lesssim 10$)}: The hallmark of the emergence of the advective regime is the extensive deformation of the mixing region beyond the stirrer’s path. The size of the attached vortices increases with $Re$, as expected. As they extend along the stirrer’s path, a new shedding mechanism \ida{emerges through} the stirrer’s interaction with the attached vortices. In this process, new vortices detach from the attached CCW vortex when the stirrer crosses its region of influence. At lower $Re$, the detached CCW vortices decay rapidly. When $Re \approx 10$, the periodic interaction between the stirrer and the attached CCW vortex becomes dominant over viscous dissipation, and the detached CCW vortex persists \ida{long enough to form, along with the shed CW vortex, a satellite vortex pair} that remains in the central region while slowly expanding radially. This leads to extensive, albeit slow and predominantly azimuthal, stretching of the interface.

\textbf{Regime \ES~(Escaping Vortices, $Re \gtrsim 20$)}: At $Re \approx 20$, as inertial effects become more dominant, the shed vortex pairs advect away from the central region. Beyond the central region, these vortices are no longer directly influenced by interactions with the attached vortices. Instead, they drift outward, gradually decaying due to viscous dissipation. The frequency of \ida{vortex-pair formation and detachment} increases with $Re$. As these vortex pairs separate from the central region, they induce significant deformation of the dye interface beyond the stirrer’s path, with disturbances propagating far into the domain.

	
	\subsection{\ida{Mixing rate and transport mechanisms}}
	
\begin{figure}
		\centering
		\subfloat[\label{fig:sigma30t}]{
			\includegraphics[trim=2.4cm 1.5cm 2.22cm 1cm, clip=true,width=.45\textwidth]{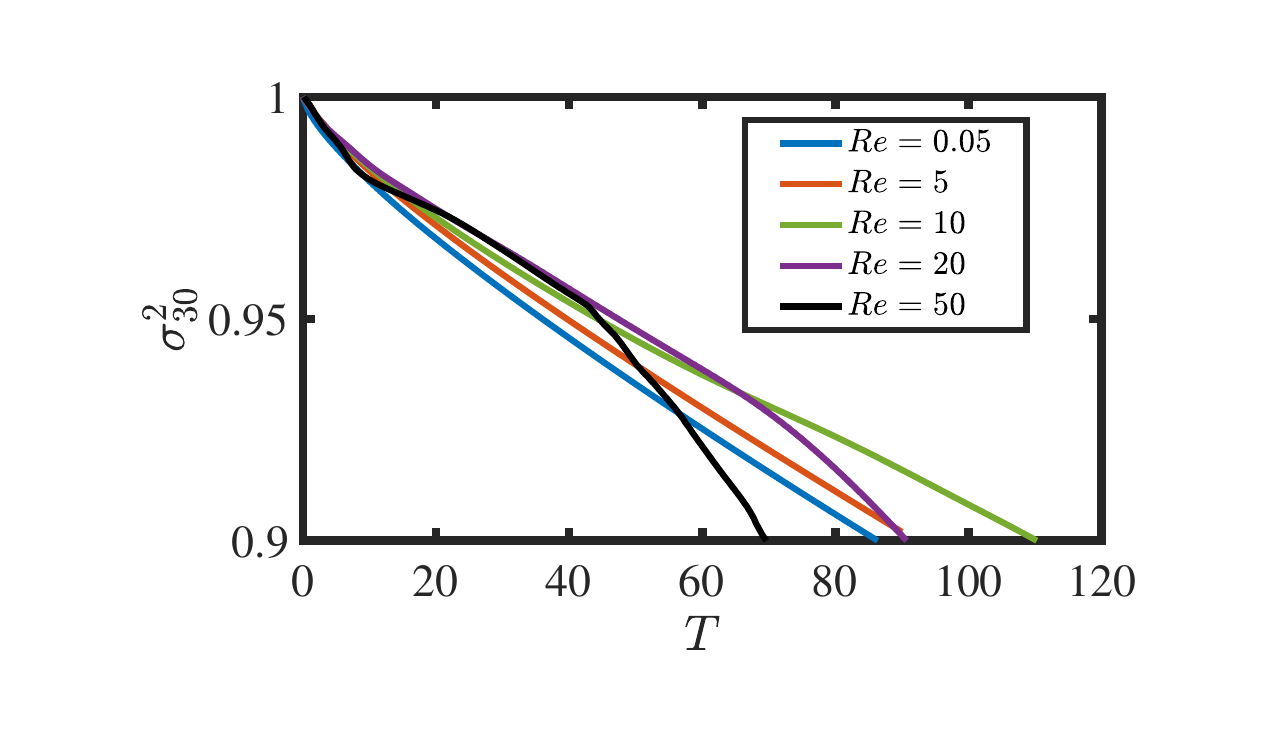}
		}
		\subfloat[\label{fig:sigma30ts}]{
			\includegraphics[trim=2.4cm 1.5cm 2.22cm 1cm, clip=true,width=.45\textwidth]{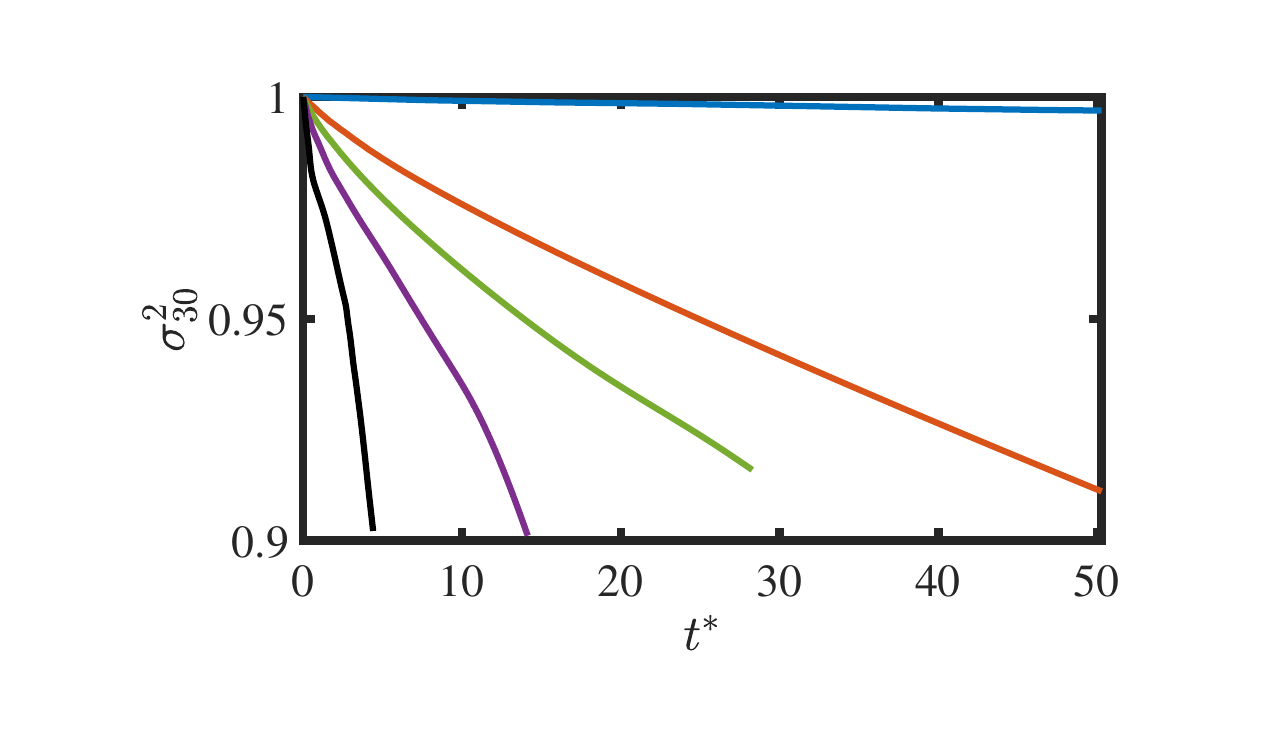}
		}
		\caption{Evolution of normalized variance of dye concentration, $\sigma^2_{30}$, versus (a) stirrer's period, $T$, and (b) dimensionless viscous time, $t^{*}$.  
		}
		\label{fig:variance-all}
	\end{figure}

	In the previous sections, \ida{we identified the transport mechanisms responsible for the mixing dynamics observed in the present flow. We now examine how these mechanisms influence the overall mixing rate}. Figure \ref{fig:sigma30t} shows $\sigma_{30}^2$ for all illustrative cases discussed in section \S\ref{sec:mixingEv}. At early times, the mixing rate is surprisingly similar across all $Re$ considered here. More significant differences emerge at later times, \ida{once vortices escape the central region and induce} extensive stretching of the interface.
	
	Notably, the timescale used for non-dimensionalization ($t = \hat{t}\hat{\Omega}$) differs among the cases shown. Figure \ref{fig:sigma30ts} presents the progress of mixing using a shared timescale, ${\color{black}t^*=\hat{t}\hat{\mu}/(\hat{\rho}\hat{r}_o^2)}$. Comparing Figures \ref{fig:sigma30t} and \ref{fig:sigma30ts} highlights that the primary effect of faster stirring is \ida{to complete} more periods per unit time. In other words, while interface deformation remains localized around the stirrer's path, the progress of mixing per stirrer period does not change significantly with the stirring rate. The rate of mixing per period is influenced by $Re$ only \ida{after} vortices escape the central region, stretching the interface far beyond the stirrer’s path.

\section{Summary}
\label{sec:III_summary}

We have investigated the mixing dynamics of a Newtonian fluid stirred by a cylindrical obstacle moving along a circular path in a two-dimensional, infinite domain. By systematically varying the stirring Reynolds number within the laminar range, we identified how transitions in flow topology, particularly the emergence, persistence, and transport of vortical structures, govern the evolution of mixing \ida{through successive transport mechanisms,} from diffusion-dominated to advection-dominated regimes.

Based on the observed vortex dynamics, three flow regimes can be differentiated: (i) In the negligible-shedding regime, vortex shedding is absent or weak and mixing remains diffusion-dominated. (ii) In the weak-shedding regime, vortices detach intermittently but remain confined near the stirrer’s path, leading primarily to azimuthal stretching of the concentration field. (iii) At higher Reynolds numbers, shed vortex pairs escape the central region and drift outward, producing strong deformation of the dye interface over a large portion of the domain. \ida{These regimes therefore represent transitions between distinct transport mechanisms governing scalar evolution, rather than merely different Reynolds-number ranges.}

\idaa{We have provided a mechanistic connection between scalar evolution and the evolving Eulerian flow field by identifying the flow features responsible for different advective transport mechanisms}. Advective mixing is driven by (i) the direct interaction of the interface and the stirrer as the stirrer travels along its path, \ida{followed by} (ii) interface deformation by shed vortices. Here, the vortices may be von K\'armán-like vortices traveling near the stirrer and trapped in the central region, or vortex pairs shed from the central region that advect far into the flow domain. The former enhance advective mixing during the early stages of the process, while the latter govern the mixing rate at later stages. We associate the escaping vortices with nonlocal transport and a significant enhancement of mixing.

The mixing rate was assessed using characteristic timescales based on both the stirring period and viscous diffusion. When vortical activity remains confined near the stirrer, the mixing achieved per rotation is found to be only weakly dependent on the stirring rate. Substantial enhancement of global mixing occurs only once vortices escape and transport scalar gradients away from the central region.

Overall, this study demonstrates that mixing performance in this system is governed not simply by stirring intensity, but by vortex–stirrer interactions and the ability of vortical structures to escape the central region, thereby enabling mixing beyond the central region. \ida{Rather than increasing continuously with Reynolds number, significant improvements in mixing arise when new transport mechanisms emerge.} \idaa{By connecting these transport mechanisms to identifiable Eulerian flow features, the present approach provides a route from the kinematic description of mixing toward its connection with the underlying fluid dynamics.}


\begin{acknowledgements}
We gratefully acknowledge financial support from Natural Sciences and Engineering Research Council (NSERC), Canada, through the Discovery Grant program. This research was enabled in part by support provided by Calcul Qu\'ebec (\url{www.calculquebec.ca}) and Compute Canada (\url{www.computecanada.ca}).
\end{acknowledgements}

	\bibliography{aipsamp.bib}
\end{document}